\documentclass[sigconf, nonacm]{acmart}

\usepackage{dcolumn}
\usepackage{xspace}
\usepackage{subcaption}

\usepackage{longtable}

\usepackage{acmart-taps}

\copyrightyear{2025}
\acmYear{2025}
\setcopyright{cc}
\setcctype{by-nc}
\acmConference[ICER 2025 Vol. 1]{ACM Conference on International Computing Education Research V.1}{August 3--6, 2025}{Charlottesville, VA, USA}
\acmBooktitle{ACM Conference on International Computing Education Research V.1 (ICER 2025 Vol. 1), August 3--6, 2025, Charlottesville, VA, USA}
\acmDOI{10.1145/3702652.3744201}
\acmISBN{979-8-4007-1340-8/2025/08}

\newcommand{\N}{\ensuremath{\mathbb{N}_0}}

\newcommand{\term}[1]{\emph{#1}}

\newcommand{\hypothesis}[2]{\aptLtoX[graphic=no, type=html]{}{\phantomsection}\label{hyp: #1}\textbf{\sffamily Hypothesis~#1 (#2):}}
\newcommand{\hypothesisref}[1]{\hyperref[hyp: #1]{\mbox{\sffamily#1}}}

\newcommand{\intervention}[1]{{\emph{#1}}}
\newcommand{\formula}{\intervention{verbose set notation}\xspace}
\newcommand{\nest}{\intervention{nesting}\xspace}
\newcommand{\natLang}{\intervention{natural language}\xspace}
\newcommand{\mult}{\intervention{multiplicities}\xspace}

\newcommand{\citeExtAppendix}[1]{#1}
\newcommand{\citeExtAppendixParen}[1]{(\citeExtAppendix{#1})}
\newcommand{\extAppendixTaskRegressionCFG}{Section~\ref{app:participants-tasklevel-cfg}}
\newcommand{\extAppendixTaskRegressionPDA}{Section~\ref{app:participants-tasklevel-pda}}

\newcommand{\extAppendixPooledRegressionPDA}{Section~\ref{app:participants-pooled}}
\newcommand{\extAppendixPanelRegressionCFG}{Section~\ref{app:participants-panel}}
\newcommand{\extAppendixPanelRegressionPDA}{Section~\ref{app:participants-panel}}
\newcommand{\extAppendixBoxPlots}{Section~\ref{app:participants-attempts_time}}
\newcommand{\extAppendixSessions}{Section~\ref{app:sessions}}
\newcommand{\extAppendixBHCorrection}{Section~\ref{app:bhCorrection}}

\newcommand*{\Iltis}{\textsc{Iltis}\xspace}

\let\epsilon\varepsilon

\newcommand{\enquote}[1]{``#1''}

\begin{document}

\title{Difficulty Generating Factors for Context-free Language Construction Assignments}

\author{Florian Schmalstieg}
\affiliation{\institution{Ruhr University Bochum}
  \city{Bochum}
  \country{Germany}}
\email{florian.schmalstieg@rub.de}
\orcid{0009-0001-2277-7502}

\author{Marko Schmellenkamp}
\affiliation{\institution{Ruhr University Bochum}
  \city{Bochum}
  \country{Germany}}
\email{marko.schmellenkamp@rub.de}
\orcid{0000-0003-3966-6590}

\author{Jakob Schwerter}
\affiliation{\institution{TU Dortmund University}
	\city{Dortmund}
	\country{Germany}}
\email{jakob.schwerter@tu-dortmund.de}
\orcid{0000-0001-5818-2431}

\author{Thomas Zeume}
\affiliation{\institution{Ruhr University Bochum}
  \city{Bochum}
  \country{Germany}}
\email{thomas.zeume@rub.de}
\orcid{0000-0002-5186-7507}

\renewcommand{\shortauthors}{Schmalstieg et al.}

\begin{abstract}
	Computer science students often struggle with abstract theoretical concepts, particularly in introductory courses on theoretical computer science. One such challenge is understanding context-free languages and their various representations.
	In this study we investigate factors that influence the difficulty of constructing context-free grammars and pushdown automata for context-free languages. {We propose two potential difficulty generating factors targeting how a language is presented to students: representation in natural language and as a verbose set notation. Furthermore, we propose two factors targeting the structure of the given context-free language: nesting of constructs and insertion of multiplicities.}

	We conducted a controlled experiment using within-subject randomization in an interactive learning system, testing the proposed difficulty factors for constructing context-free grammars and pushdown automata. Our results suggest that three of the four factors significantly influence students' objective performance in solving exercises for constructing context-free grammars, while students' perceived difficulties only partly align with the objective performance measures. {The findings for pushdown automata tasks differed markedly from those for context-free grammar tasks. Our variations either had negligible effects or, in some cases, even reduced difficulty.} Thus, no robust statistical conclusions can be made for pushdown automata tasks.

	The results lay foundations for learning systems that adaptively choose appropriate exercises for individual students.
\end{abstract}



\maketitle

\section{Introduction}	
\label{section:introduction}

Understanding context-free languages and their formal representations is a great challenge for undergraduate computer science students. These concepts form the basis for specifying and parsing the syntax of programming languages, and are thus covered in undergraduate computer science curricula (see, e.g., \cite{ACM2024,GI2016}). Students typically encounter formal languages and how they can be described in courses on \emph{theoretical computer science}. For the class of context-free languages, usual representations are context-free grammars, pushdown automata, and descriptions in set notation or natural language (also see \cite{HopcroftMU2014}). We refer to \autoref{fig:CFL} for an example of these four representations.

\begin{figure}[t]
		\small
		\begin{subcaptiongroup}
			\definecolor{boxcolor}{rgb}{0.93,0.93,0.93}%
			\newcommand{\boxskip}{\par\vspace*{1.55ex}}%
			\colorbox{boxcolor}{
			\begin{minipage}{.98\linewidth}
				{\small\enquote{We consider the language of all words over the alphabet \mbox{\(\Sigma = \{a,b,c\}\)}, which consist of three (possibly empty) blocks. The first block contains only \(a\)s, the second only \(b\)s and the third only \(c\)s. The number of \(a\)s and \(b\)s together is exactly equal to the number of \(c\)s.}}
				\subcaption{A description in natural language for the language.}
			\end{minipage}
			}
			\boxskip
			\begin{minipage}{1.021\linewidth}
				\colorbox{boxcolor}{
				\begin{minipage}[t][1.6cm][b]{.44\linewidth}
					\centering \vfill\mbox{$\{ a^n b^m c^{n+m} \mid n,m\in\N \}$}\vfill
					\subcaption{A description in set notation for the language.}
				\end{minipage}
				}
				\hfill
				\colorbox{boxcolor}{
				\begin{minipage}[t][1.6cm][b]{.45\linewidth}
					\vspace*{-2ex}
					\begin{align*}
						S \to aSc \mid B\\
						B \to bBc \mid \epsilon
					\end{align*}
					\vspace*{-4ex}
					\subcaption{A context-free grammar for the language.}
				\end{minipage}
				}
			\end{minipage}
			\boxskip
			\colorbox{boxcolor}{
			\begin{minipage}{.98\linewidth}
				\centering
				\scalebox{.9}{
					\includegraphics{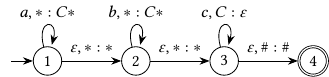}
				}
				\Description{The picture shows a pushdown automaton with four states which are chained with epsilon transitions. In the initial state, a's are read and collected on the stack. In the second state, b's are read and added to the stack. In the third state, c's are read and per c the stack is decreased by one symbol. If the stack is empty the automaton can accept in the fourth state.}
				\subcaption{A pushdown automaton for the language. The automaton reads words letter by letter, can store symbols on a stack, and accepts if the accepting state 4 is reached.}
			\end{minipage}
			}
		\end{subcaptiongroup}
		\setcounter{figure}{0}
		\captionof{figure}{Different representations of the same context-free language.}\label{fig:CFL}
\end{figure}

A key learning objective in these courses is the ability to construct formal representations---that is, designing context-free grammars or pushdown automata based on descriptions in natural language or set notation. Empirical evidence and high failure rates suggest that students often struggle with theoretical content \cite{Dougherty25,SollEA2023, Knobelsdorf2014}, and our teaching experience confirms that constructing formal representations is among the most difficult tasks. However, the specific cognitive and representational challenges that contribute to this difficulty remain largely unexplored. 
{Constructing formal models is also a recurring challenge in domains like algebra and programming \cite{Sfard95, GinatB17}. Such tasks frequently require recursive reasoning or symbolic abstraction, both of which are known to place heavy demands on cognitive resources \cite{poletiek_under_2018, fedor_semantics_2012}.}

The difficulty in the tasks can be understood through the lens of \emph{Cognitive Load Theory} \cite{Sweller1988,Sweller2011}, which distinguishes between two types of cognitive load: intrinsic load (inherent complexity of the material itself) and extraneous load (complexity introduced by the way the task is presented). 

When students construct context-free grammars or pushdown automata, they have to simultaneously understand the underlying language properties, translate informal descriptions into formal rules, and apply procedural knowledge. This imposes a high intrinsic cognitive load due to the abstract nature of formal grammars. Additionally, poorly structured tasks, ambiguous instructions, and unfamiliar representations can introduce extraneous cognitive load, complicating the learning process.
Thus, reducing unnecessary cognitive burden while preserving essential problem-solving complexity is crucial for improving student learning outcomes.

In this paper we start the exploration of the following question:
\begin{itemize}
 \item[] \emph{What makes constructing context-free grammars and pushdown automata difficult for students?} 
\end{itemize}

We address this question by exploring potential \emph{difficulty generating factors} that influence the difficulty of constructing formal representations. 
A principled map of difficulty generating factors is essential for addressing three pressing needs in computer science instruction. First, they provide a foundation for developing targeted instructional strategies. Future work can leverage these insights to design assignments that systematically scaffold student learning, ensuring that foundational concepts are well understood before introducing more complex constructs. Adaptive practice systems have to predict how small, surface-level changes to a formal language task alter its cognitive load, so that they can deliver exercises at a \enquote{just-right} level of challenge---promoting desirable difficulty during retrieval practice \citep{AgarwalEA2021}. This is especially valuable in large, theory-heavy courses, where students often struggle with self-regulation and procrastination \citep{WoltersB2021}, and where the uptake of effective strategies such as spaced self-testing remains low \citep{DunloskyR2015}. Personalized task generation that avoids both triviality and overwhelm may foster sustained engagement, better self-monitoring, and improved performance \citep{SchwerterDBM2022, SchwerterB2024}.
Second, they allow educators to refine exercises by eliminating extraneous cognitive barriers that increase task difficulty without supporting the intended learning objectives. 
Third, instructors frequently need to modify formal language tasks to deter plagiarism or generate exam questions. Without empirical guidance, such changes may unintentionally alter item difficulty and compromise score comparability. A fine-grained understanding of difficulty generating factors can thus support both adaptive tutoring and fair assessment, motivating the present study.

While difficulty generating factors have been studied extensively in didactics of mathematics (see, e.g., \cite{WilliamsC1997,Lehner2019} for a survey), we are not aware of any prior work in theoretical computer science education. {Therefore, this research builds on theoretical foundations from mathematics, drawing on cognitive models of task translation and model construction \cite{vanlehn_model_2013} to frame the observed student difficulties. These connections allow us to contextualize our hypotheses within a broader understanding of how learners engage with symbolic systems.}

\begin{figure}[t]
		\centering
		\small
		\includegraphics{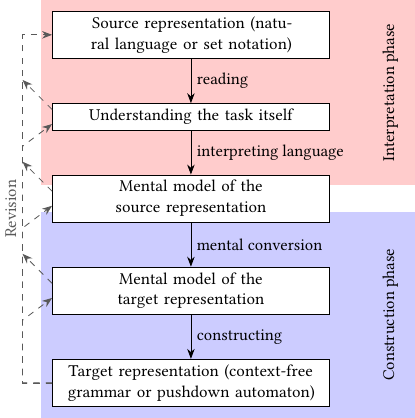}
		\captionof{figure}{An idealized workflow for how students approach context-free construction tasks.}\label{fig:cognitive model}
		\Description{The workflow describes how students can come from a source representation to a target representation, while going through a chain of steps, divided in, first, the interpretation phase and, second, the construction phase. At every point a revision to a previous step is possible. The steps are as follows: In the interpretation phase, from a source representation (natural language or set notation) by reading to an understanding of the task itself, and by interpreting the language to a mental model of the source representation. Then, in the construction phase form this mental model by a mental conversion to a mental model of the target representation, and by executing the construction to the target representation (context-free grammar or pushdown automaton).}
\end{figure}

To identify potential difficulty generating factors for context-free construction tasks, we adapt the four phases of problem solving by \citeauthor{Polya1945} \cite{Polya1945} {(\emph{Understanding the problem}, \emph{Devising a plan}, \emph{Carrying out the plan} and \emph{Looking back})}
to construction tasks for context-free languages: In an {interpretation phase}, learners read the task at hand and interpret the source representation (natural language or set notation) of the given formal language (\emph{Understanding the problem}). This phase yields a mental model of the source representation which can, for instance, include an informal verbalized description (e.g., \enquote{all words having first $a$s, then $b$s, and then as many $c$s as $a$s and $b$s combined}) or typical (counter-)examples and edge-cases. In a {construction phase}, learners then convert their mental model for the source representation into a mental model for the target representation (\emph{Devising a plan}) and finally construct the target representation (\emph{Carrying out the plan}). At every step, a revision of what has been done before might be necessary (\emph{Looking back}). These phases are visualized in \autoref{fig:cognitive model}.

This phase model suggests two main sources for difficulty of context-free construction tasks:
\term{Interpretation difficulty} that refers to the difficulty that stems from understanding the assignment, the given language, and the construction of a mental model, and \term{construction difficulty} that refers to the cognitive processes and content-specific knowledge required to convert the mental model into the target representation. The construction difficulty is also influenced by the complexity and size of the constructed solution (e.g., number of rules of a grammar, number of states/transitions of an automaton). We refer to Section~\ref{sec:relwork-hypo} for details.

In this study, we propose specific difficulty generating factors contributing to interpretation and construction difficulty (see Section~\ref{section:hypotheses}). We hypothesize that each factor (a) objectively increases task difficulty and (b) is perceived by students as making tasks more difficult. To test these hypotheses, we conduct a controlled experiment with within-subject randomization in an interactive learning system, systematically manipulating task difficulty and measuring both objective performance metrics and students' perceived difficulty.
Importantly, we focus on difficulty generating factors that are not inherent to the type of task (e.g., operational complexity) or to the phrasing of the assignment (e.g., linguistic complexity), but that depend solely on the specific subject of the task, i.e. the \enquote{values} used. 

\section{Background}

We will first give an overview over previous research {on formal languages in computer science education (Section~\ref{sec:relwork-FL-CSEd})}, on factors that influence the difficulty of tasks (Section~\ref{sec:relwork-task-diff}), and then describe the learning theoretical background for our hypotheses (Section~\ref{sec:relwork-hypo}).

\subsection{Formal languages in computer science education}\label{sec:relwork-FL-CSEd}

The difficulty of theoretical computer science, including formal languages and automata theory, is primarily attributed to the abstract nature of theoretical concepts (e.g., \cite{Chesnevar2004,Chuda2007}). It is widely assumed that this leads to low motivation among students  (e.g., \cite{Korte2007,Zingaro2008,Sigman2007}). However, recent research 
advocates for a more nuanced perspective (e.g., \cite{Knobelsdorf2016, Frede2021}).

To tackle some of the difficulties in teaching and learning formal languages and automata theory, multiple approaches have been proposed. \citet{Knobelsdorf2014} suggest a cognitive apprenticeship approach, advocating the teacher to demonstrate problem-solving methods in a hands-on way. Others emphasize the importance of problem-based and practical approaches \cite{Hamalainen2004, Habiballa2004, Maurer2021} and discovery learning \cite{Sigman2007}. However, most propositions aim to enable experimentation with immediate feedback, thus addressing limitations of the pen-and-paper approach \cite{Morazan2014,Bezakova2022}. These include integrating programming (e.g., \cite{Zingaro2008, Wermelinger2005}), game building (e.g., \cite{Korte2007}), interactive textbooks and exercises (e.g., \cite{Mohammed2021, Schmellenkamp2024, Dantoni2015}), and visualization techniques (e.g., \cite{Chuda2007, Hielscher2006, Rodger2006}, see \cite{Chakraborty2011} and \cite{Silva2024} for reviews).

{Despite the inherent challenges, there is only very little high-quality work on why specific assignments in the formal language domain are difficult.}
\citet{Sanders2015} found that, when solving construction tasks with finite automata, the most common errors made by their students were: using non-determinism when deterministic automata were required; adding unnecessary states; and failing to properly consume the remainder of a string after reaching an accepting state. {To mitigate difficulties and to support students to reason about automata, it has been found that simulating automata for specific words and therefore testing them for single inputs is a sensible first step \cite{Souza15, MorazanA14}.}

{\citet{Pillay2010} analyzed solutions of (among others) grammar construction tasks from weekly tutorials and mid-semester tests but did not report on specific errors. \citet{Schmellenkamp2024a} analyzed data from students in Year 12 of a grammar school. Utilizing procedural data provided by an interactive learning system, they identified seven distinct error types in the construction of context-free grammars. {They differentiated between syntactical, conceptual, and strategic difficulties and---in particular---highlighted common difficulties with recursive grammar rules.}
}

\subsection{Related work on task difficulty}\label{sec:relwork-task-diff}

Understanding what makes a task difficult has been studied in many fields of computer science, based on \citeauthor{campbell88}'s comprehensive review \cite{campbell88} and its extension by \citeauthor{Braarud2001} \cite{Braarud2001} (see, e.g., \cite{ahadi15,kasto13,Qian2017,sheard13}). For example, research has focused on programming tasks \cite{kasto13} and database queries \cite{ahadi15}, but we are not aware of any prior work examining theoretical computer science or, more specifically, construction tasks for context-free languages. In particular, \citeauthor{campbell88} and \citeauthor{Braarud2001}'s differentiation between objective and subjective complexity \cite{campbell88,Braarud2001} appears relevant here: while formal representations introduce inherent complexity, students' subjective perceptions of difficulty can vary significantly based on how tasks are presented. Further, research in computer science education has explored creating isomorphic tasks—i.e., tasks covering the same content and (ideally) having the same difficulty \cite{fowler24,parker22}—but again, we find no prior examination of isomorphic \emph{context-free language} tasks. However, there is growing evidence from related fields suggesting that symbolic reasoning tasks—especially those requiring rule construction or translation across representational formats—pose domain-general cognitive challenges \cite{koedinger_real_2004, landy_abstract_2014}. These studies show that surface features of symbolic tasks, such as spatial layout or syntactic conventions, can strongly affect learners' accuracy and strategy use.

Due to the limited research on theoretical computer science, we also look to mathematics education research, as both fields share important cognitive similarities and mathematics education research has studied task difficulty extensively. In particular, the symbolic reasoning, abstraction, and interpretation of structured representations required in theoretical computer science parallel those in mathematics. Hence, findings from mathematics problem-solving research can provide a useful foundation for analyzing the difficulty of context-free language construction tasks. 

For example, Williams and Clarke \cite{WilliamsC1997} categorized mathematical problem complexity into six dimensions—linguistic, contextual, representational, operational, conceptual, and intellectual—which were later adopted for programming education \cite{sheard13,Carbone2020}. Other categorizations focusing on content-specific knowledge, the cognitive processes involved, and formal factors in the assignment and solution are described in \cite{Prenzel2002,HussmannP07} (see also \cite{Lehner2019} for a survey). {Additionally, \citet{vanlehn_model_2013} proposes that model construction tasks require learners to generate and coordinate internal representations, a process that imposes considerable cognitive demands and benefits from explicit scaffolding.}

\subsection{Cognitive and learning theories behind task difficulty}\label{sec:relwork-hypo}

{Already in Section~\ref{section:introduction}, we described the phase model of problem solving by \citeauthor{Polya1945} \cite{Polya1945} which builds a foundation for our differentiation between interpretation and construction difficulty.}

Additionally, we focus on two well-known cognitive load effects \cite{Sweller1988,Sweller2011}. The \term{element interaction} describes how many different aspects must be held in working memory at the same time. In the context of context-free grammars and pushdown automata, high element interaction occurs when students have to track symbol dependencies, rule hierarchies, and recursive structures. This parallels challenges observed in algebra and logic instruction, where learners often misapply symbolic conventions due to shallow processing or representational mismatches \cite{landy_abstract_2014, dawkins_theos_2023}.
The \term{split-attention effect} refers to the load required to combine multiple sources of information that are separated from each other.

By analyzing task difficulty through \citeauthor{Polya1945}'s model and cognitive load theory, we can systematically identify and control factors that contribute to learners' challenges in constructing formal representations for context-free languages.

\section{Difficulty generating factors for context-free language construction tasks}
\label{section:hypotheses}
We next present our preregistered \cite{prereg} hypotheses of factors contributing to interpretation and construction difficulty. For each factor we hypothesize that it (a) increases the difficulty objectively, and that it (b) increases the perceived difficulty. Due to a limitation on the number of tasks we can pose during our study, we limit ourselves to four factors.

\subsection{Interpretation difficulty}

We focus on two aspects of how context-free languages are given in the assignment which likely contribute to interpretation difficulty: (1) natural language vs. set notation; (2) compact vs. verbose set notation.

Reading and understanding natural language descriptions might introduce cognitive load, as relations between parts of a word are not explicit and therefore more elements have to be held in working memory. For instance, the description:

\begin{itemize}
 \item[] \emph{The language of all words over the alphabet $\Sigma = \{a, b, c \}$, which consist of three (possibly empty) blocks. The first block contains only $a$s, the second only $b$s and the third only $c$s. The number of $a$s and $b$s is equal to the number of $c$s.} 
\end{itemize}
and the set notation $\{ a^n b^m c^{n+m} \mid n, m\in\N\}$ describe the same context-free language. While this is only an example, often set notation provides a clearer representation of the language structure. We hypothesize that therefore set notation facilitates the construction of a mental representation. 
\begin{itemize}
 \item[] \hypothesis{I-1}{natural language vs. set notation}\\(a) Using natural language instead of a set notation increases the objective difficulty of a task, and (b) the perceived difficulty of a task.
\end{itemize}

Context-free languages can often be written as set notations either compactly or more verbose, by spatially separating the blocks and the relation between them. For example, the  language  above can be written as $\{ a^n b^m c^{n+m} \mid n, m\in\N\}$ or in a more verbose way as $\{ a^n b^m c^\ell \mid n, m, \ell \in\N \text{ and }\ell =  n + m\}$. We hypothesize that the latter notation increases difficulty because of a higher cognitive load due to a split-attention effect.

\begin{itemize}
\item[] \hypothesis{I-2}{verbose vs. compact set notation}\\(a) Using verbose set notation instead of compact set notation increases the objective difficulty of a task, and (b) the perceived difficulty of a task.
\end{itemize}

\subsection{Construction difficulty}
The difficulty of constructing target representations is influenced by which concepts are required for the construction. Context-free languages in typical assignments only use few language constructs, which are combined and parameterized to yield concrete languages. Examples of such language constructs are (i) nesting vs. concatenation  (e.g., $a^nb^nc^md^m$ vs. $a^nb^mc^md^n$), and (ii) adding multiplicities (e.g., $a^nb^n$ vs. $a^nb^{2n}$). For constructing a target representation, a conceptual understanding of how to translate each occurring construct is required. Here we focus on the constructs (i) and (ii).

Cognitive load can vary depending on how language constructs are connected. If two constructs are concatenated (e.g., in the language $\{ a^n b^n c^m d^m \mid n,m\in\N\}$ the same type of construct, $a^n b^n$ and $c^m d^m$, is repeated twice), they can be processed independently when constructing a context-free grammar or a pushdown automaton. However, when constructs are nested (e.g., in $\{ a^n b^m c^m d^n \mid n,m\in\N\}$ the construct $b^m c^m$ is nested inside of $a^nd^n$), they  need to be kept in mind at the same time, probably leading to a higher element interaction between them and therefore higher cognitive load.

\begin{itemize}
 \item[] \hypothesis{C-1}{nesting vs. concatenation}\\(a) Nesting of language constructs is more difficult than concatenation, and (b) nesting is perceived as more difficult than concatenation.
\end{itemize}

Another source of cognitive load might be the incorporation of multiplicities in relationships between different parts of words of a language. An example of this are the languages $\{ a^n b^n \mid n\in\N \}$ and $\{ a^n b^{2n} \mid n\in\N\}$. When constructing a context-free grammar or a pushdown automaton,  a one-to-one mapping of letters (as in the first language) requires a different approach than a one-to-many mapping (as in the second language).
In addition to establishing a relationship between two parts of the word, students must also keep a counting criterion in mind. We therefore assume that difficulty increases due to higher element interactivity, as multiple conditions must be simultaneously satisfied.

\begin{itemize}
 \item[] \hypothesis{C-2}{multiplicities vs. no multiplicities}\\(a) Using multiplicities increases the objective difficulty of a task and (b) the perceived difficulty of a task.
\end{itemize}

Contrary to our preregistration \cite{prereg}, we dropped the hypotheses comparing the effects of \mult on context-free grammar and pushdown automaton tasks due to resource and space constraints.

We note that construction difficulty is likely also influenced by the number of interacting language constructs used in a description, which is typically correlated with the size of the target representation. Although it is likely that more constructs and larger size lead to a higher difficulty, this is not the primary focus of our investigation.

\section{Methodology}

\label{section:methodology}
We used a preregistered \cite{prereg} controlled experimental design study with within-subject randomization in an interactive learning system to test our hypotheses for objective and perceived task difficulty. We describe our study design and setting (Section~\ref{section:study-design}), which and how data was collected (Sections~\ref{section:metrics} and~\ref{section:data collection}), and the analysis methods (Section~\ref{section:analysis}).

\begin{table*}[t]
	\caption{Overview of the task groups and their individual tasks. Each task group has a baseline task, a representational variation and a conceptual variation. The natural language descriptions are translations from the \anon[phrasings in the original language]{original German phrasings}.}\label{tab:overview tasks}
	\begin{minipage}{\linewidth}
		\renewcommand{\arraystretch}{1.3}
		\centering
		\footnotesize
		\begin{tabular}{p{.6cm}p{3.8cm}p{5.9cm}p{3.8cm}}
			\toprule
			\small Task group&\normalsize Baseline&\small Representational variation\newline \scriptsize Task groups~1--3: \formula\newline Task groups~4--6: \natLang&\small Conceptual variation\newline  \scriptsize Task groups~1--3: \nest\newline Task groups~4--6: \mult\\\midrule
			\normalsize 1&$\{a^n b^{n+2} c^{m}\mid n,m \in \mathbb{N}_0 \}$&$\{a^n b^{m} c^{l}\mid n,m,l \in \mathbb{N}_0, m=n+2 \}$&$\{a^n b^{m} c^{n+2}\mid n,m \in \mathbb{N}_0 \}$\\
			\normalsize 2&$\{a^n b^{n} c^{m} b^m\mid  n,m \in \mathbb{N}_0, n\equiv_2 0 \}$&$\{a^n b^{m} c^{k} b^l\mid n,m,k,l \in \mathbb{N}_0, n=m, k=l, n\equiv_2 0 \}$&$\{a^n b^{m} c^{m} b^n\mid  n,m \in \mathbb{N}_0, n\equiv_2 0 \}$\\
			\normalsize 3&$\{uc^{|u|}a^n \mid  u\in \{a,b\}^*, n\in \mathbb{N}_0 \}$&$\{uc^na^m \mid u\in \{a,b\}^*, n,m\in \mathbb{N}_0, n={|u|} \}$&$\{ua^nc^{|u|} \mid u\in \{a,b\}^*, n\in \mathbb{N}_0 \}$\\
			\normalsize 4&$\{a^n b^{m}\mid n,m\in \mathbb{N}_0, n < m \}$& We consider the language of all words over the alphabet \mbox{\(\Sigma = \{a,b\}\)}, which consist of two (possibly empty) blocks. In the first block only \(a\)s occur, in the second only \(b\)s and the number of \(a\)s is genuinely less than the number of \(b\)s.
&$\{a^{2n} b^{m}\mid n,m\in \mathbb{N}_0, 2n < m \}$\\
			\normalsize 5&$\{a^n b^{m} c^{n+m}\mid n,m \in \mathbb{N}_0 \}$& We consider the language of all words over the alphabet \mbox{\(\Sigma = \{a,b,c\}\)}, which consist of three (possibly empty) blocks. The first block contains only \(a\)s, the second only \(b\)s and the third only \(c\)s. The number of \(a\)s and \(b\)s together is exactly equal to the number of \(c\)s.
&$\{a^n b^{m} c^{2(n+m)}\mid n,m \in \mathbb{N}_0 \}$\\
			\normalsize 6&$\{uc^n \mid u\in \{a,b\}^*, n={\#_a(u)} \}$& We consider the language \(L\) of all words over the alphabet \mbox{\(\Sigma = \{a,b,c\}\)} that begin with any word from \(\{a,b\}^\ast\) and have a block of \(c\)s behind it. This \(c\)-block consists of exactly as many \(c\)s as there are \(a\)s in the word before it.
& $\{uc^n \mid   u\in \{a,b\}^*, n=2\cdot\#_a(u) \}$\\\bottomrule
		\end{tabular}
	\end{minipage}
\end{table*}

\subsection{Study design}\label{section:study-design}
The goal of the study is to assess the Hypotheses \hypothesisref{I-1}, \hypothesisref{I-2}, \hypothesisref{C-1}, and \hypothesisref{C-2} by posing carefully manipulated tasks. Each hypothesis is tested for three tasks to ensure that effects are not task specific. 
For this, we designed six task groups, each containing (i) a baseline task, (ii) a representational variation, to assess the hypotheses regarding interpretation difficulty, and (iii) a conceptual variation, to assess the hypotheses regarding construction difficulty. To keep the number of tasks per student within a reasonable limit, we randomly split students into three subject groups and assign each group only one task per task group. To ensure fairness with respect to exam preparation, each subject group received every variation equally often.

\subsubsection{Tasks and task groups}\label{section:bounded} Each of the six task groups contains a baseline task, a representational variation, and a conceptual variation (see \autoref{tab:overview tasks}). The baseline tasks are given in compact set notation. Task groups~1--3 additionally use tasks with a more verbose set notation (as representational variation) and nesting (as conceptual variation). Task groups~4--6 use tasks with natural language (as representational variation) and multiplicities (as conceptual variation). 

As an example, the baseline task for Task group~1 is \linebreak$\{a^n b^{n+2} c^{m}\mid n,m \in \mathbb{N}_0 \}$. The representational variation task uses $\{a^n b^{m} c^{l}\mid n,m,l \in \mathbb{N}_0, m=n+2 \}$ with a more verbose set notation to test Hypothesis \hypothesisref{I-2} (verbose vs. compact set notation). The conceptual variation task uses $\{a^n b^{m} c^{n+2}\mid n,m \in \mathbb{N}_0 \}$ which nests the $b^m$ part between $a^n$ and $c^{n+2}$ to test Hypothesis C-1 (nesting vs. concatenation). Note that the representational variation does not change the language to be modeled while the conceptual variation does.

As is typical in many assignments for context-free languages (see, e.g., \cite{HopcroftMU2014}), all our tasks use languages that follow a block-like structure.
This makes the tasks easily accessible to students and allows for combining multiple language constructs in a simple way.
{It was also ensured that the languages---while still being natural and common (e.g., two of them appear similarly in \cite{HopcroftMU2014})---are not included as examples in the lecture or have easily accessible solutions.}

The languages used in Task groups~1, 2, 4, and~5 more specifically use \emph{bounded context-free languages}, that is,  languages with a fixed block structure where words of the language only differ by the frequency of the repetitions of the single blocks (see \cite{Ginsburg1964} for a formal definition). For example, $\{a^n b^{n} c^{m} b^m\mid n,m \in \mathbb{N}_0, n\equiv_2 0 \}$ is bounded as all words have the same block structure of first $a$s, then $b$s, then $c$s, and then $b$s again, but $\{uc^{|u|}a^n \mid u\in \{a,b\}^*, n\in \mathbb{N}_0 \}$ is not bounded as arbitrary words over $\{a,b\}$ can be substituted for $u$. Bounded context-free languages are very common in assignments on formal languages (see, e.g., \cite{HopcroftMU2014}).
The languages in Task groups~3 and 6 use non-bounded languages to assess whether the results differ between these classes of languages.

To reduce the effect of learned schemes on the students' performance, the languages of different task groups were chosen to all have an individual aspect distinguishing them from the languages of the other task groups. For example, an additional constant number of $b$s is used in Task group~1, a modulo constraint in Task group~2, and so on.

\subsubsection{Subject groups} With a total of 18 tasks, not every participating student could be expected to solve all of them. We therefore randomly distributed students into three subject groups. Each group was assigned one task of each task group for constructing context-free grammars (first week of the study) and one (different) task of each task group for constructing pushdown automata (second week of the study). 

The tasks from Task groups~1--3 were used as warm-up tasks with no further incentive, to assess the effect of the \formula and the \nest variation. The tasks from Task groups~4--6 enabled students to collect bonus points for the exam\footnote{Students could earn a bonus of up to 10\,\% on their final exam grade by achieving up to 80\,\% of all points in the assignment sheets of the first and second half of the course, each. Each incentivized assignment within the study was worth 9 out of 600 points achievable in the first half of the course.} and were used to assess the effect of the \natLang and the \mult variation. This split between warm-up tasks and tasks with bonus points was dictated by the course set-up. 

The same set of languages was used for both target representations (context-free grammars and pushdown automata). To prevent the results for assignments in the second week from being influenced by previous ones, the distribution of variations to the subject groups was permuted between both weeks. For limitations of this setup, we refer to Section~\ref{sec:discussion}.

\begin{table*}[t]
	\caption{Distribution of variations in the task groups. \formula: Verbose instead of compact set notation (representational), \natLang: Natural language instead of set notation (representational), \mult: Incorporating multiplicities in relationships between parts of a word (conceptual), \nest: Nesting instead of concatenation of language constructs (conceptual).}\label{tab:distribution variations}
	\begin{minipage}{\linewidth}
\centering
\begin{tabular}{p{1.7cm}lllll}
			\toprule
			Target &Incentive    & Task group  & Subject group 1          & Subject group 2          & Subject group 3          \\ \midrule
			context-free&warm-up      & 1 & baseline             & \formula          & \nest          \\
			grammar&& 2 & \nest          & baseline             & \formula          \\
			&& 3 & \formula          & \nest          & baseline             \\
			&bonus points & 4 & baseline             & \natLang & \mult   \\
			&& 5 & \mult   & baseline             & \natLang \\
			&& 6 & \natLang & \mult   & baseline             \\ \midrule
			pushdown &warm-up      & 1 & \formula          & \nest          & baseline             \\
			automaton&& 2 & baseline             & \formula          & \nest          \\
			&& 3 & \nest          & baseline             & \formula          \\
			&bonus points & 4 & \natLang & \mult   & baseline             \\
			&& 5 & baseline             & \natLang & \mult   \\
			&& 6 & \mult   & baseline             & \natLang \\ \bottomrule
		\end{tabular}
	\end{minipage}
\end{table*}

\subsubsection{Course integration}\label{section:data}
The study was executed in an introductory course on theoretical computer science during winter term 2024/2025 at \anon[a metropolitan university]{Ruhr University Bochum, Germany}. The course is mandatory in three different computer science-related undergraduate study programs and students typically attend it in their third semester. In the first half of the semester, the course featured formal languages (regular and context-free languages), then computability and complexity theory.

\begin{figure*}[tbh]
	\begin{minipage}[b]{.43\linewidth}
		\includegraphics{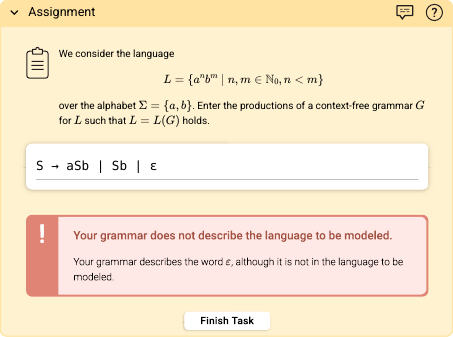}
	\end{minipage}
	\hfil
	\begin{minipage}[b]{.41\linewidth}
		\includegraphics{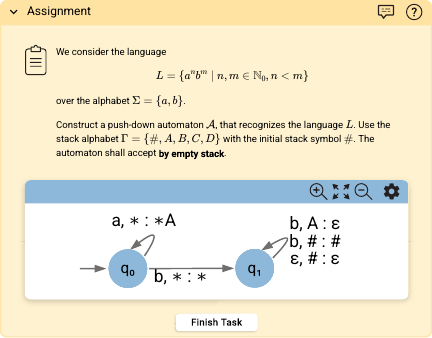}
	\end{minipage}
	\caption{Screenshots from two (translated) tasks in the \anon[interactive learning system]{\Iltis system}. Students could use the question mark icon in the upper right-hand corner to receive help on how to use the respective task. For grammar tasks, this included a brief summary of the syntax and conventions of specifying a grammar; for automata tasks, this included a brief tutorial on how to insert states and edges. In the left screenshot, an exemplary feedback generated by the system for the current attempt is shown.}\label{fig:screenshots}
	
	\Description{The first screenshot shows an example task for constructing context-free grammars. It is split in three vertical sections: the assigment is given at the top, the student can enter a grammar in a text field in the middle, and feedback for the student attempt is shown at the bottom. The assigment in this example reads: "We consider the language L = { a^n b^m | n,m in N_0, n < m } over the alphabet Sigma = { a, b }. Enter the productions of a context-free grammar G for L such that L = L(G) holds." For the grammar that has been entered in a text field (S -> aSb | Sb | epsilon) the shown feedback reads: "Your grammar does not describe the language to be modeled. Your grammar describes the word epsilon, although it is not in the language to be modeled."
	The second screenshot shows an example task for constructing a pushdown-automata for the same language. Here, instead of a text field, a canvas is shown where students can add states and draw transitions between them, in a graphical way.}
\end{figure*}

Each week, the course included two 90-minute lectures with all students and a 90-minute tutorial in groups of 20--30 students. Assignments consisted of (a) interactive web-based assignments provided in the \anon[interactive learning system]{\Iltis system  \cite{Schmellenkamp2024}}, some as warm-up and some incentivized by bonus points; and (b) incentivized paper-based assignments graded by student tutors.
Students had one week and unlimited attempts to solve the incentivized interactive assignments. After each failed attempt, students received feedback including syntactic and semantic checks on several levels. {The semantic feedback consisted of a witness word (see Figure~\ref{fig:screenshots}) and (if applicable) stated that the attempt grammar/automaton generates not a single word or only a finite number of words. For the grammar tasks, depending on the specific attempt, the system also provided explanations on a higher abstraction level: it stated that the attempt grammar generates words with an incorrect block structure or that the symbol frequency of the generated words is incorrect (e.g., \enquote{Your grammar generates words that contain twice as many $a$s than $b$s, however the words should contain twice as many $b$s than $a$s}). For some attempts, it could also provide a representation of the student attempt language in set notation.}

\begin{figure*}[t]
	\begin{minipage}[c]{.55\linewidth}
		\newcommand{\assignmentLength}{3.6cm}
		\newcommand{\tutorialLength}{1.8cm}
		\newcommand{\assignmentHeight}{0.5cm}
		\newcommand{\tutorialHeight}{0.5cm}
		\newcommand{\rasterHeight}{0.7cm}
		\includegraphics{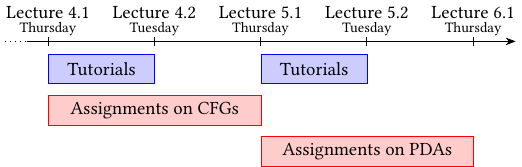}
	\end{minipage}
	\begin{minipage}[c]{.35\linewidth}
		\small
		\newcommand{\lecItem}[1]{\par\textbf{#1}}
			\lecItem{Lecture 4.1:} Introduction of context-free grammars (syntax, semantics, examples)
			\lecItem{Lecture 4.2:} Normal forms and correctness proofs for context-free grammars
			\lecItem{Lecture 5.1:} Introduction of pushdown automata\linebreak(syntax, semantics, examples)
			\lecItem{Lecture 5.2:} Pumping lemma, deterministic PDAs,\linebreak algorithms, closure and other properties
			\lecItem{Lecture 6.1:} Word problem and syntax analysis \end{minipage}
	\caption{A schematic overview of the timing of lectures and assignments: In Week~4 of the course, context-free languages were introduced in the lecture, then students had one week to work on the incentivized web-based and paper-based assignments. In Week~5, pushdown automata were introduced in the lecture and, again, students had one week for solving the incentivized assignments.}\label{fig:timings}
	\Description{The assignments on context-free grammars could be solved between Lectures 4.1 and 5.1, with tutorials after Lecture 4.1. The assignments on pushdown automata could be solved between Lectures 5.1 and 6.1, with tutorials after Lecture 5.1.}
\end{figure*}

The study was integrated into the interactive web-based assignments in two weeks of the first half of the course. {An overview of the timing is given in Figure~\ref{fig:timings}.}
From 371 students who engaged with the course at some point, 238 were active in these two weeks. Such interaction rates are common at many courses in our university as interaction with the assignments is optional and no strict schedule in which semester to attend a lecture or write an exam is enforced.

Students encountered the variations \formula and \nest only in non-incentivized warm-up tasks. As a result, fewer students worked on and completed these tasks. 	

\subsection{Metrics}\label{section:metrics}
To quantify objective task difficulty, we used several performance measures that are derived from data automatically collected by \anon[the interactive learning system]{the \Iltis system} and that are common in the literature (see, e.g., \cite{SchmitzH2023, SchwerterWG2022}). Specifically, we measured: (i) the proportion of students successfully completing the task, (ii) the proportion of students successfully completing the task on the first attempt (before receiving feedback from \anon[the learning system used]{the \Iltis system}), (iii) the average number of attempts until students either successfully complete the task or stop further interaction, and (iv) the average time spent on the task until students either successfully complete the task or stop further interaction. Time measurements were automatically recorded by the system from the moment the task was presented until the student submitted a correct solution or terminated interaction. We measured only the time spent actively working on the tasks. We excluded any periods of inactivity (no mouse movements, keystrokes, etc.) lasting longer than one minute.

In addition, after each successful completion of the task, participants were presented with a brief survey consisting of two questions designed to assess their perceived difficulty of the task. These questions separately addressed the difficulty of interpreting the given language and the difficulty of constructing a solution:\begin{description} 
	\item[Q1] How difficult did you find it to understand the language?
	\item[Q2] How difficult did you find it to convert the given language to a context-free grammar/pushdown automaton?
\end{description}
Students were asked to rate both questions on a 5-point Likert scale from 1 = very easy to 5 = very difficult. 	

Overall, participants achieved high completion rates across all but two tasks (>\,80\,\%, except for the \formula variation for context-free grammars and the \nest variation for pushdown automata, both in Task group~1 with $62\,\%$ and $75\,\%$, resp.).
This is likely influenced by the design of the \anon[interactive learning system]{\Iltis system}, which allowed students to make an unlimited number of attempts. Thus, the corresponding metric of the proportion of students who successfully completed the task does not provide sufficient granularity to meaningfully differentiate task difficulty. We, therefore, focus on the remaining performance measures \emph{number of attempts}, \emph{time spent}, and \emph{success on first attempt}. {Importantly, number of attempts and time spent are not restricted to first-attempt data. These metrics capture the full interaction with the task, regardless of whether participants solved it on their initial try.}

\subsection{Data collection}\label{section:data collection}

Data is collected by \anon[the interactive learning system]{the \Iltis system} anonymously on a session-basis. A session comprises all uninterrupted activity until a break of at least 30 minutes is recognized. We collected data on 818 sessions (473 sessions for context-free grammar tasks, 345 sessions for pushdown automata). Of all 238 active students during the two weeks of our study, 229 completed at least one of the investigated tasks. For 137 students who had previously consented to the use of their data for research purposes, we linked interactions with \anon[the interactive learning system]{the \Iltis system} across multiple sessions. This linked dataset provides more robust insights compared to the general anonymous session data, as it allows tracking of individual student performance across multiple sessions. 

{Data from sessions of students who did not provide explicit consent were not included in the linked dataset. However, in accordance with institutional review board approval (\anon{GEKTUDO2023-42}), we were permitted to include anonymous, unlinked session data from non-consenting students in aggregate analyses.} A t-test ($\alpha=0.05$) revealed no significant differences in neither the number of attempts nor the time spent between sessions of students who consented in linking their sessions and sessions of other students. To assess potential selection bias, we also repeated the regressions described in Section~\ref{section:analysis} with the full dataset of all sessions, including unlinked ones and found no substantial differences \citeExtAppendixParen{for details, see \extAppendixSessions}.

Therefore, from here on we focus exclusively on the linked participant data.

\subsection{Statistical analysis}\label{section:analysis}
Before performing statistical analyses, we identified and removed outliers separately for the variables \emph{number of attempts} and \emph{time spent} and each variation group. Outliers were defined using the standard criterion of values being more than 1.5 times the interquartile range (IQR) above the third quartile or below the first quartile. Detailed box plots illustrating the distribution of data and identified outliers are provided in \extAppendixBoxPlots.

To better interpret the results of the survey on perceived difficulty, we standardized the answers, so that the results describe the difference from the mean value in multiples of the standard deviation.

To assess the overall effects of the task variations, we used a fixed-effects panel regression model \cite{Angrist2009}. This regression model controls for (unobserved) individual differences by explicitly focusing on within-subject variation. In this type of analysis, we only look at how changes occur within each person over time. Because time-constant variables (such as a person's gender or prior knowledge) never change for a given individual, they don't help us understand what's causing changes within that individual. As a result, these variables don't improve the model's predictions or explanations and are therefore not included. Consequently, we only included participants who completed at least two tasks, as participants with fewer tasks do not provide within-subject variation. This criterion resulted in the exclusion of 7 participants in the context-free grammar tasks and 5 participants in the pushdown automaton tasks from this analysis.  To clearly separate general treatment effects (i.e. effects due to the experimental variation) from task-specific effects, we included task fixed-effects (i.e. dummies for the tasks) in our regression model. This allowed us to control for general differences between tasks. In addition, we accounted for repeated measurements from the same individuals by using heteroskedasticity-robust and individually clustered standard errors.

To assess task-specific effects and as a sensitivity analysis, we ran separate ordinary least squares (OLS) linear regressions for each individual task group. In case the fixed-effects panel regression show no general treatment effects for the multiple tasks, the task-specific linear regression can highlight whether this is due to heterogeneity within the tasks, that means whether our variations affected the tasks differently.

For this study, we chose a regression analysis rather than \term{item response theory} (IRT) because our design and data structure are better suited to this approach. Most outcomes are continuous or count variables that are collected repeatedly, making panel regression with fixed effects ideal for controlling for individual differences and capturing changes in performance across tasks. In contrast, IRT is typically applied to binary or ordinal data to estimate latent traits and item properties, and its assumptions (e.g., uni-dimensionality and local independence) may not hold in our multiple trial per task setting. As our focus is on assessing the causal effects of the experimental variations, rather than simply describing item properties, regression analysis was the more appropriate choice.

{Our hypotheses are theory-driven and pre-registered. Nevertheless, to further ensure the robustness of our findings, we conducted a Benjamini-Hochberg correction across the CFG and PDA panels and present the results in \extAppendixBHCorrection. Although 3 of our 19 significant results lose statistical significance, the overall pattern remains the same.} 	
	\section{Results}
\label{section:results}

{Table~\ref{tab:overview results} summarizes the main regression findings. For context-free grammar (CFG) tasks, three of the four difficulty generating factors—natural language, multiplicities, and nesting—were associated with lower performance and increased perceived difficulty. In contrast, pushdown automaton (PDA) tasks were less affected: only perceived difficulty rose in some cases, while some variations even improved first-attempt success. The following subsections detail these effects.}

\begin{table*}[bt]
	\caption{Overview over the results of the fixed-effects panel regressions regarding our hypotheses. The entries describe whether the corresponding factor overall influences the objective difficulty (a) or the perceived difficulty (b) regarding the target representation.}\label{tab:overview results}
	\begin{tabular}{lcccc}
		\toprule
		Target representation&\hypothesisref{I-1} (\natLang)&\hypothesisref{I-2} (\formula)&\hypothesisref{C-1} (\mult)&\hypothesisref{C-2} (\nest)\\\midrule
		context-free grammar&(a)&-&(a), (b)&(a), (b)\\\midrule
		pushdown automaton&-&(b)&(b)&-\\\bottomrule
	\end{tabular}
\end{table*}

\subsection{Results for tasks with context-free grammars as target representation}

\subsubsection{Interpretation difficulty}

\paragraph{Hypothesis \hypothesisref{I-1} (natural language vs. set notation)}

In the Task groups~\makebox{4--6}, natural language formulations reduced performance compared to set notation: students made more attempts, took longer, and were less likely to succeed on their first try (Table~\ref{tab: cfg plm}). These results support Hypothesis \hypothesisref{I-1} (a) that the \natLang variation increases a task's objective difficulty (for context-free grammar tasks), but do \emph{not} confirm Hypothesis \hypothesisref{I-1} (b) on perceived difficulty.

The task-level regression analysis \citeExtAppendixParen{\extAppendixTaskRegressionCFG} showed that the effect of the \natLang variation was strongest for bounded-language tasks (Task groups~4 and~5), where performance consistently declined and perceived difficulty increased. For Task group~4, students made more attempts and were less likely to succeed on the first try. In Task group~5, students took longer to solve the task and rated both interpretation and construction as more difficult. By contrast, for the unbounded-language task (Task group~6), performance was unaffected and perceived interpretation difficulty even decreased. These findings suggest that the impact of natural language varies by task context, particularly depending on the underlying structure of the target language.

\begin{table*}[tb]
	\caption{Results of the fixed-effects panel regression analysis of tasks with a \textbf{context-free grammar} as the target representation. For each variation and metric, (i)~the regression coefficient (i.e. how much the corresponding metric differs due to the variation) and (ii)~the standard error (in brackets) is given.
	Additionally, for each metric the number of included observations (i.e. the number of student-task combinations) and the determination coefficient ($R^2$) is indicated in the last two rows. The determination coefficients describe how well our independent variables (i.e. \emph{variation} and \emph{task}) predict the dependent outcome variable (i.e. measured metric). A version showing the task-dummy variables can be found in the appendix: \citeExtAppendix{\extAppendixPanelRegressionCFG}. 
	Significance levels: ***: $p<0.001$; **: $p<0.01$; *: $p<0.05$; +: $p<0.1$.}\label{tab: cfg plm}
	\centering
	\small
\begin{tabular}{l D{)}{)}{9)3} D{)}{)}{8)3} D{)}{)}{9)2} D{)}{)}{9)1} D{)}{)}{8)3}}
\toprule
 & \multicolumn{1}{p{1.7cm}}{Number of \newline attempts} & \multicolumn{1}{p{1.4cm}}{Time spent \newline (min)} & \multicolumn{1}{p{2.55cm}}{Probability of success \newline in first attempt (0--1)} & \multicolumn{1}{p{2.6cm}}{Perceived interpreta- \newline tion difficulty ($\times$SD)} & \multicolumn{1}{p{2.5cm}}{Perceived construc- \newline tion difficulty ($\times$SD) } \\
Baseline: & 3.56 & 7.74 & 0.306 & 2.27 & 3.09 \\ 
\midrule
\natLang & 1.83 \; (0.49)^{***} & 4.75 \; (1.15)^{***} & -0.10 \; (0.05)^{*}  & -0.05 \; (0.09)    & 0.11 \; (0.08)       \\
\formula & -0.25 \; (0.63)      & 0.79 \; (1.09)       & -0.13 \; (0.08)      & 0.07 \; (0.12)     & 0.08 \; (0.13)       \\
\nest    & 1.53 \; (0.80)^{+}   & 3.95 \; (1.56)^{*}   & -0.22 \; (0.08)^{**} & -0.02 \; (0.12)    & 0.61 \; (0.16)^{***} \\
\mult    & 2.78 \; (0.56)^{***} & 5.39 \; (1.02)^{***} & -0.09 \; (0.05)^{*}  & 0.20 \; (0.08)^{*} & 0.15 \; (0.09)^{+}   \\
\midrule
R$^2$                  & 0.25       & 0.24     & 0.13  & 0.08 & 0.16   \\
Observations               & 477        & 487   & 522  & 476 & 477  \\
\bottomrule
\end{tabular}
 \end{table*}

\paragraph{Hypothesis \hypothesisref{I-2} (verbose vs. compact set notation)}

Across Task groups~1--3, verbose instead of compact set notation (baseline) did not significantly impact aggregate performance or perception (Table~\ref{tab: cfg plm}). However, in Task group~3, students made more attempts, took longer, and were less likely to succeed on the first try. Though not statistically significant, perceived difficulty was also rated slightly higher. This stresses task-specific sensitivity to notation complexity.

\subsubsection{Construction difficulty}

\paragraph{Hypothesis \hypothesisref{C-1} (nesting vs. concatenation)}

We explored whether nesting blocks within formal languages (variation) increases difficulty compared to simple concatenation (baseline), using tasks from Groups 1--3. Nesting led to significantly worse performance and higher perceived construction difficulty
(see \autoref{tab: cfg plm}). These findings support both Hypothesis~\hypothesisref{C-1}~(a) and~(b). Task-level regressions aligned with these patterns for Groups~2 and~3, but not Group~1, suggesting variation in the effect's strength across tasks.

\paragraph{Hypothesis \hypothesisref{C-2} (multiplicities vs. no multiplicities)}

We used Task groups~4--6 to assess whether expressing relations between language blocks through multiplicities (variation) increases task difficulty compared to formulations without such dependencies (baseline). Table~\ref{tab: cfg plm} shows that multiplicities substantially impaired performance and increased both interpretation and construction difficulty
These effects were robust in Groups~4 and~5 but absent in Group~6, again indicating that task structure modulates the difficulty introduced by conceptual features \citeExtAppendixParen{\extAppendixTaskRegressionCFG}.

\subsection{Results for tasks with pushdown automata as the target representation}

Unlike CFG tasks,
PDA tasks yielded a more complex picture---showing weaker or even reversed effects.

\subsubsection{Interpretation difficulty}

\paragraph{Hypothesis \hypothesisref{I-1} (natural language vs. set notation)}

In contrast to CFG tasks, for which natural language significantly impaired performance, PDA tasks showed the opposite trend: students in the \natLang group were more likely to succeed on their first attempt, without affecting the other outcomes. Thus, Hypotheses~\hypothesisref{I-1}~(a) and~(b) were not supported for PDA tasks (see \autoref{tab: pda plm}).

Task-level analyses \citeExtAppendixParen{\extAppendixTaskRegressionPDA} showed mixed results. In Task group~6, performance improved: students made fewer attempts and spent less time. In Task group~5, students took more attempts and more time , yet were also more likely to succeed on their first attempt . No significant effects were observed for Task group~4. These results point to task-specific variability in how natural language formulations affect student performance, possibly reflecting differences in structural clarity or familiarity.

\paragraph{Hypothesis \hypothesisref{I-2} (verbose vs. compact set notation)}

We examined whether presenting language constraints more verbosely as formulas (variation) rather than using compact exponents (baseline) affected performance, across Task groups~1--3. Table~\ref{tab: pda plm} reveals no reduction in performance; instead, students in the \formula group were more likely to succeed on their first attempt, although this was not accompanied by improvements in time or attempts. Perceived interpretation difficulty was rated slightly higher, suggesting a possible dissociation between subjective and objective difficulty, though this trend was not confirmed in a pooled analysis \citeExtAppendixParen{\extAppendixPooledRegressionPDA}.

More specifically, only Task group~3 showed significant effects: students in the \formula variation used fewer attempts, were more likely to succeed on their first attempt, rated construction as less difficult, and spent less time. Surprisingly, the verbose formulation in Task group~3 both improved performance and reduced perceived difficulty---contradicting the pattern in CFG tasks and suggesting a possible facilitation effect unique to PDA construction.

\begin{table*}[tb]
	\caption{Results of the fixed-effects panel regression analysis of tasks with a \textbf{pushdown automaton} as the target representation. For each variation and metric, (i)~the regression coefficient (i.e. how much the corresponding metric differs due to the variation) and (ii)~the standard error (in brackets) is given.
	Additionally, for each metric the number of included observations (i.e. the number of student-task combinations) and the determination coefficient ($R^2$) is indicated in the last two rows. The determination coefficients describe how well our independent variables (i.e. \emph{variation} and \emph{task}) predict the dependent outcome variable (i.e. measured metric). A version showing the task-dummy variables can be found in the appendix: \citeExtAppendix{\extAppendixPanelRegressionPDA}. 
	Significance levels: ***: $p<0.001$; **: $p<0.01$; *: $p<0.05$; +: $p<0.1$.}\label{tab: pda plm}
	\centering
		\small
\begin{tabular}{l D{)}{)}{9)0} D{)}{)}{9)0} D{)}{)}{8)2} D{)}{)}{9)2} D{)}{)}{9)2}}
\toprule
& \multicolumn{1}{p{1.7cm}}{Number of \newline attempts} & \multicolumn{1}{p{1.4cm}}{Time spent \newline (min)} & \multicolumn{1}{p{2.55cm}}{Probability of success \newline in first attempt (0--1)} & \multicolumn{1}{p{2.6cm}}{Perceived interpreta- \newline tion difficulty ($\times$SD)} & \multicolumn{1}{p{2.5cm}}{Perceived construc- \newline tion difficulty ($\times$SD) } \\
Baseline: & 4.49 & 12.8 & 0.218 & 2.21 & 3.10 \\ 

\midrule
\natLang & 0.67 \; (0.54)  & -0.10 \; (1.49) & 0.10 \; (0.05)^{*}  & -0.02 \; (0.08)     & 0.09 \; (0.09)      \\
\formula & -0.49 \; (0.86) & 0.41 \; (1.99)  & 0.24 \; (0.08)^{**} & 0.20 \; (0.12)^{+}  & -0.06 \; (0.17)     \\
\nest    & -0.78 \; (0.75) & -1.30 \; (2.01) & 0.20 \; (0.10)^{*}  & 0.17 \; (0.15)      & -0.15 \; (0.15)     \\
\mult    & 0.28 \; (0.59)  & 0.30 \; (1.58)  & 0.11 \; (0.05)^{*}  & 0.24 \; (0.08)^{**} & 0.24 \; (0.08)^{**} \\
\midrule
R$^2$    & 0.28   & 0.36   & 0.15  & 0.13 & 0.33   \\
Observations   & 395    & 396    & 435  & 403 & 402  \\
\bottomrule
\end{tabular}
 \end{table*}

\pagebreak
\subsubsection{Construction difficulty}

\paragraph{Hypothesis \hypothesisref{C-1} (nesting vs. concatenation)}

We examined whether nesting language blocks (variation) affected performance compared to simple concatenation (baseline), using Task groups~1--3. Our analysis (Table~\ref{tab: pda plm}) revealed no consistent effects across performance or time, but students were more likely to succeed on their first attempt.

Task-level regressions \citeExtAppendixParen{\extAppendixTaskRegressionPDA} showed mixed estimation results. In Task group~1, perceived construction difficulty increased. In contrast, Task groups~2 and~3 showed improved performance and reduced difficulty: in Task group~2, students made fewer attempts, were more successful on their first-attempt, and rated construction to be easier; in Task group~3, students spent less time and had a higher first-attempt success rate. These results suggest that nesting, while potentially confusing in some contexts, may actually support performance under certain structural conditions.

\paragraph{Hypothesis \hypothesisref{C-2} (multiplicities vs. no multiplicities)}

To assess the impact of multiplicity relationships compared to simpler formulations, we analyzed Tasks Groups~4--6. Though performance improved for \mult,
students reported both interpretation and construction as more difficult, confirming Hypothesis~\hypothesisref{C-2}~(b) but not \hypothesisref{C-2}~(a) (see Table~\ref{tab: pda plm}).

Task-level regressions \citeExtAppendixParen{\extAppendixTaskRegressionPDA} revealed inconsistent patterns. In Task group~5, students spent more time but had a higher first-attempt success rate. In Task group~6, the time spent also increased, but no other effects were significant. No effects for the perceived difficulty were observed on task level, despite the overall trend in the panel regression, indicating high heterogeneity in how multiplicities impact PDA construction tasks.

Taken together, PDA tasks diverged sharply from CFGs in how students responded to representational complexity. Several variations that hindered CFG performance were neutral or even beneficial for PDA tasks. This asymmetry underscores important differences in how learners approach grammar versus state-machine construction.

\section{Discussion}\label{sec:discussion}

The primary goal of this study was to investigate how specific factors---namely, the use of natural language descriptions instead of set notation (\natLang), the use of more verbose expressions for relationships between blocks in set notation (\formula), the addition of multiplicities to these relationships (\mult), and the nesting of language constructs (\nest)---influence the difficulty of tasks involving the construction of context-free grammars and pushdown automata. Understanding these difficulty generating factors is essential to support the adaptive selection of tasks tailored to students' individual needs, and ultimately to promote effective engagement with theoretical concepts.

Our results suggest that several of the proposed difficulty generating factors significantly influence context-free grammar task difficulty, aligning with cognitive load theory \cite{Sweller2011}. {The findings for pushdown automata tasks differed markedly from those for context-free grammar tasks. Contrary to our expectations, our variations either had negligible effects or, in some cases, even reduced difficulty as indicated by higher success rates on the first attempt.} 

{In the following, we first discuss our main findings on objective and perceived task difficulty (Sections~\ref{sec:discussion-objective} and~\ref{sec:discussion-perceived}). After that, for context-free grammars as target representation---where our data seems to be more robust---we evaluate our findings in more detail for the single task groups (Section~\ref{sec:discussion-additional}). Possible reasons for the divergent findings on pushdown automata are discussed in Section~\ref{sec:discussion-pushdown}. Lastly, we present limitations of our study in Section~\ref{sec:discussion-limitations}.}

\subsection{Objective task difficulty}\label{sec:discussion-objective}
{Our findings suggest that the factors involving natural language descriptions (\natLang), multiplicity (\mult), and nesting (\nest) are \emph{difficulty generating} for tasks with a context-free grammar as target representation. Specifically, we observed a significant decrease in performance for these variations, corresponding to Hypotheses \hypothesisref{I-1} (a), \hypothesisref{C-1} (a), and \hypothesisref{C-2} (a), respectively, which predicted increased objective task difficulty due to these factors.} 

{The results for the \mult and the \nest variations are in line with observations, that increasing the number of interacting elements (e.g., through more intricate constraints like multiplicities or nested structures) raises the cognitive load of a task \cite{Sweller1988, Sweller2011}.} 
{The differing findings for pushdown automata might suggest that the stack-based reasoning intrinsic to pushdown automata provides clearer structural cues to deal with these variations.}  

{The findings for the \natLang variation align with studies showing that symbolic representations---though more compact---can be easier to manipulate than natural language when syntactic relations are explicit \cite{koedinger_real_2004, landy_abstract_2014}. In contrast, natural language can obscure structural dependencies, increasing working memory demands due to unresolved reference and implicit relations.}
{That this is not the case for the pushdown automata tasks might be due to learning effects from the previously posed context-free grammar tasks. Even though no student received the same task for constructing both a grammar and a pushdown automaton, their experience from encountering these variations for other languages might have diminished any noticeable differences in objective difficulty.}

\subsection{Perceived task difficulty}\label{sec:discussion-perceived}
{Regarding perceived difficulty, our results for context-free grammar tasks indicate significant effects for the variation \mult in both the construction and interpretation categories, and for \nest in the construction category only. This confirms hypotheses \hypothesisref{C-1}~(b) and \hypothesisref{C-2}~(b). The absence of significant effects in the interpretation category for \nest suggests that nesting primarily impacts the construction phase.}

{These results for context-free grammars as target representation are in line with prior research showing that recursive or nested structures pose special challenges for novice learners across domains \cite{poletiek_under_2018, wendebourg_semantic_2025}. Similarly, the fact that introducing multiplicities leads to increased difficulty supports theories that emphasize relational complexity as a key source of cognitive load \cite{Sweller1988, vanlehn_model_2013}. Multiplicities require learners to track both symbolic repetition and interdependent quantities---conditions that increase element interactivity and reduce cognitive bandwidth for construction planning.}

{For pushdown automata, we see no statistically significant changes in perceived difficulty for the \nest variation, underlining that stack-based reasoning might provide clearer structural cues. Interestingly, there is an effect for the \mult variation, that does not match the results for objective difficulty. It is conceivable, that tasks perceived to be more difficult lead to an activation of students, who were more careful and therefore performed better.}

\subsection{Task group specific findings for context-free grammar tasks}\label{sec:discussion-additional}

{The inconsistent results for the different task groups might indicate subtle differences in the way specific variations change the necessary cognitive processes. Additional research focusing on the actual solving process might be necessary.}

\paragraph{Bounded vs.\ non-bounded languages.}
A key takeaway from our task-specific analyses is that the influence of both the \natLang and \mult variations differed markedly between bounded and non-bounded languages. Bounded languages (Task groups~4 and~5, see Section~\ref{section:bounded}) exhibited increased objective and perceived difficulty under these variations. In contrast, the same modifications did not increase difficulty in non-bounded languages (Task group~6) and the \natLang variation of Task group~6 even  significantly decreased the perceived interpretation difficulty.

The latter result could be a result of bounded languages---due to their rigid block structure---often having a more verbose natural language description that is more difficult to parse for students. In our tasks, that is in particular true for the language of Task group~5 which used three blocks, compared to two blocks for Task groups~4 and~6.
{This also resonates with work in mathematics and logic education, where students struggle more with abstract rule translation when the underlying structure is less visible or harder to simulate mentally \cite{dawkins_theos_2023}.}
 
A major implication is that language classes 
matter to effectively calibrate difficulty levels of tasks.

\paragraph{Verbose set notation}{The inconsistent effects of verbose set notation may reflect task-specific differences in how split-attention affects processing. Similar results have been observed in artificial grammar learning, where added symbolic complexity only harms performance when it increases element interactivity, not just surface length \cite{fedor_semantics_2012}. This nuance suggests that \emph{how} complexity is encoded may matter more than how much is encoded.}

\paragraph{Increased perceived construction difficulty for natural-language tasks.}
An unexpected result was the increase in perceived construction difficulty when languages were presented in purely textual form (\natLang). Intuitively, one might expect that such a representation would mainly affect the {interpretation} rather than the construction process itself. However, students may experience additional cognitive overhead when translating more verbose statements into formal machinery. This is consistent with the fact that research in mathematics and programming has identified linguistic complexity as a dimension of complexity \cite{WilliamsC1997, sheard13}.

\paragraph{High variance in Task group~1.}
Finally, Task group~1's muted response to both the \formula and \nest variations may reflect a general learning effect: for many students, this was likely the first time they encountered an assignment to construct a context-free grammar, so the baseline overhead of recalling how context-free grammars work may have overshadowed any variation-specific impact.

Future work could use repeated or scaffolded exposures to clarify how learners adapt once initial familiarity is established.

\subsection{Divergent findings for pushdown automata}\label{sec:discussion-pushdown}

{Students who had previously solved context-free grammar tasks for similar languages might have developed mental schemes that supported their pushdown automata task performance, mitigating the anticipated difficulty increase. Since the construction of pushdown automata involves different cognitive processes, it should not introduce systematic bias. However students had additional lectures and exercises between the assignments, which might have strengthened their general understanding of context-free languages, reducing differences between the variations.}

{Another difference is the interactive learning environment's feedback mechanisms. It is conceivable that students found it easier to debug pushdown automata than context-free grammars, thus neutralizing expected increases in difficulty.
Feedback timing and guidance have been shown to modulate cognitive load
\cite{FyfeDR15,AttaliK17}, which suggests that future research should carefully consider feedback types and timing as moderating factors. } 

{As the creation of automata is time consuming, students might have thought more about their first attempts, especially when they perceived it as more difficult, thus leading to higher success rates in the first attempt. Additionally students might have been more familiar with the procedural approach of automata in contrast to the more declarative approach of grammars. This could have made it easier for them to adapt to the given variations.}

\subsection{Limitations}\label{sec:discussion-limitations}
This study has several limitations that arise from both the real teaching setting 
and design choices made for the study.

\paragraph{Generalizability}

The results of this study are based on one specific course at one university. Therefore, these findings might not generalize to different theory courses or other educational settings. However, as the tasks used in our study align well with those in popular textbooks (e.g., \cite{HopcroftMU2014}), we are confident that they are typical for many courses.

{While the tasks in this study were carefully designed to differ substantially from lecture examples and exercises, they of course involved similar concepts. Also, the order of the tasks may have influenced the results, with initial tasks of each type potentially having lowered performance as students needed time to adjust to the structure and interface of the task.}

\paragraph{Realistic educational setting}

As a consequence of collecting data in a real educational setting, we could not control for students' interactions, such as potential collaboration or the use of external resources. {Solutions to our tasks were not readily available in online resources {though students might well have found solutions to similar tasks.}} 

{Using large language models (LLMs) is popular among students. Though the broadly available LLMs can interpret both languages in set notation and natural language descriptions, we experienced that they do not consistently generate correct solutions for all our languages. In contrast to automata, grammars can easily be pasted from other sources. However, we have no evidence to suggest that this happened on a large scale.}

\paragraph{Student engagement}

The difference in student engagement between warm-up tasks and incentivized tasks may have introduced selection bias, with certain types of students systematically completing more tasks. {The same applies to students who drop out of the course between the assignments on context-free grammars and pushdown automata. However, we have no evidence to suggest that this influenced the main findings of this paper.}

{Moreover, the multiple attempt setting does not incentivize students to correctly solve a task in the first attempt, which could have led to an increase of careless mistakes or the adoption of a trial and error approach.} Still, because most students solved the tasks in the end and feedback is provided by the system after the first attempt, we used the rate of success on the first attempt as one of our performance metrics.

\paragraph{Study design and methodological considerations}

Because of a limited number of exercises we could pose for our study, we were only able to consider four difficulty generating factors, even though more factors could be of interest. 

Analyzing the number of attempts and the time spent on the tasks introduces an additional methodological consideration: We cannot directly infer the reasons why students abandoned their attempts. It is conceivable that a particularly difficult task might lead to fewer attempts if students became frustrated and abandoned the task prematurely. However, given the high success rates observed, this potential confound is unlikely to have a substantial effect.

To measure perceived difficulty, we conducted a survey among the students. However, students might not accurately attribute their difficulties to the \emph{interpretation} or \emph{construction} phase.

This issue is compounded by our design choice to ask only one question per difficulty type to reduce the survey's complexity and increase response rates.

Our use of a fixed-effects panel regression controlled effectively for all stable individual-specific characteristics, including prior knowledge and cognitive traits. Consequently, explicit control or measurement of these stable student characteristics was not necessary within our study's within-subject design. Nevertheless, future studies utilizing between-subject or mixed-method designs might benefit from explicitly measuring these cognitive characteristics to further examine their influence on performance across varying conditions or populations.

{A last limitation concerns the number of statistical tests conducted across related models, which increases the risk of false positives. By conducting a Benjamini-Hochberg correction \citeExtAppendixParen{\extAppendixBHCorrection}, some findings lose or only retain marginal significance, suggesting limited statistical power and warranting cautious interpretation of effect robustness.}

\section{Conclusion and Outlook}
\label{section:conclusion}

{We found indications that several factors contribute to interpretation and construction difficulty in context-free grammar tasks}, in particular for bounded languages.
Our results provide a basis for designing adaptive systems for tasks on constructing context-free grammars that are tailored to individual educational needs of students: teachers can manage difficulty of these tasks by determining whether tasks use natural language or set notation, whether language constructs are nested or concatenated, and whether they involve multiplicities. 
For pushdown automata, we found high heterogeneity between tasks for the same experimental variations, highlighting that changing the difficulty for context-free grammar tasks is more straightforward than for pushdown automata.

To determine \emph{how} these factors influence the difficulty of tasks and, especially, why they do not seem to influence unbounded languages and pushdown automaton construction tasks in the same way, further research is necessary. To this end, we plan to conduct think-aloud interviews. By observing students' approaches, answers, and problems on selected tasks, we hope to mitigate the expert blind spots.

In this study we have not focused on the phrasing of the assignments, but solely on the formal languages to be modeled by students and their representations.
As is well-known in various settings, the linguistic complexity of an assignment influences its difficulty (see, e.g., \cite{WilliamsC1997,Carbone2020,sheard13}). This can be an interesting additional future perspective.

{Beyond theoretical computer science, these results echo a broader pattern across STEM: students struggle not just with what to learn, but how it is represented. Effective instruction should therefore consider not only the formal correctness of task design but also its cognitive accessibility \cite{vanlehn_model_2013, landy_abstract_2014}. Future interventions might draw from strategies used in functional programming education \cite{winter_using_2019} or set-theoretic logic \cite{dawkins_theos_2023}, where visual and semantic scaffolds help learners build and refine internal models of recursive or symbolic systems.}

In the long term, this line of research aims to enable the design of adaptive tutoring systems that automatically generate practice exercises. Such systems can address common students' issues with self-organization and procrastination \cite{WoltersB2021} by providing appropriately challenging tasks. Given that students seldom engage voluntarily with practice exercises during the semester \cite{DunloskyR2015}---despite its proven association with higher exam performance \cite{SchwerterDBM2022, SchwerterB2024}---adaptively designed exercises may encourage regular use, potentially improving students' self-monitoring and academic success.

\begin{acks}
	We thank Jan Vahrenhold for insightful discussions and feedback. This work has been supported by the \grantsponsor{DFG}{Deutsche Forschungsgemeinschaft (DFG, German Research Foundation)}{}, grant \grantnum{DFG}{448468041}, by the \grantsponsor{RC Trust}{Research Center Trustworthy Data Science and Security}{https://rc-trust.ai} (\url{https://rc-trust.ai}), one of the Research Alliance centers within the UA Ruhr (\url{https://uaruhr.de}), and by \grantsponsor{Fair}{\enquote{From Prediction to Agile Interventions in the Social Sciences (FAIR, PROFILNRW-2020-068)}}{}.
	It has benefited from Dagstuhl Seminar \href{https://www.dagstuhl.de/24251}{24251} \enquote{Teaching Support Systems for Formal Foundations of Computer Science}. The sole responsibility for the content of this publication lies with the authors.
\end{acks}
	
\bibliographystyle{ACM-Reference-Format}
\bibliography{bibliography}

\appendix
\onecolumn
\raggedbottom

\makeatletter
\def\@subsubsecfont{\large\sffamily\bfseries}
\makeatother

	\section{Statistics and regressions regarding study
		participants}\label{app:participants}

	\subsection{Rate of success and of success in first
		attempt}\label{app:participants-success}
		
		Bar charts for all task groups, grouped by variation (base: baseline, verb: verbose set notation, nest: nesting, mult: multiplicities, \linebreak nat: natural
		language). Only participants are taken into account.

\begin{figure}[H]
	\centering
	\includegraphics{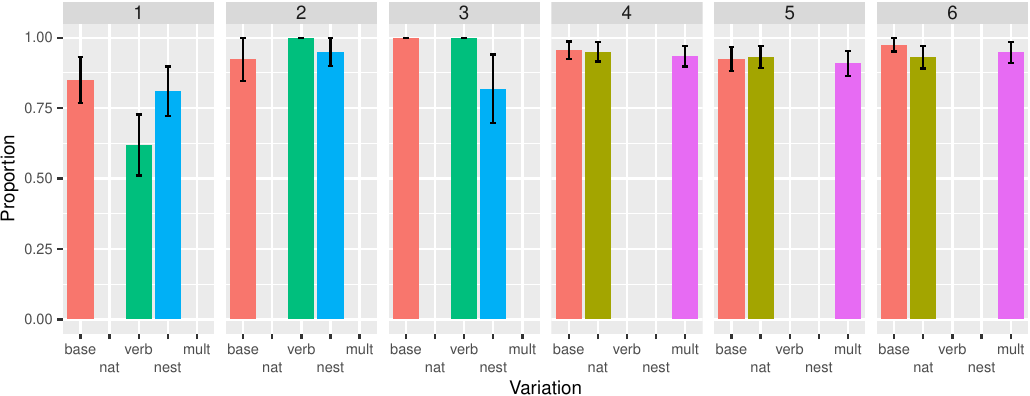}
	\caption{Rate of success for tasks with a context-free grammar as the
		target representation. Only study participants are considered.}
\end{figure}

\begin{figure}[H]
	\centering
	\includegraphics{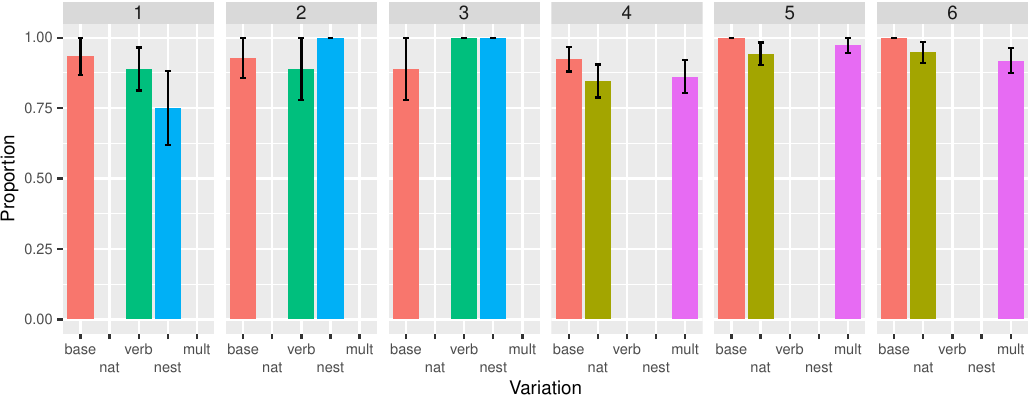}
	\caption{Rate of success for tasks with a pushdown automaton as the
		target representation. Only study participants are considered.}
\end{figure}

\begin{figure}[H]
	\centering
	\includegraphics{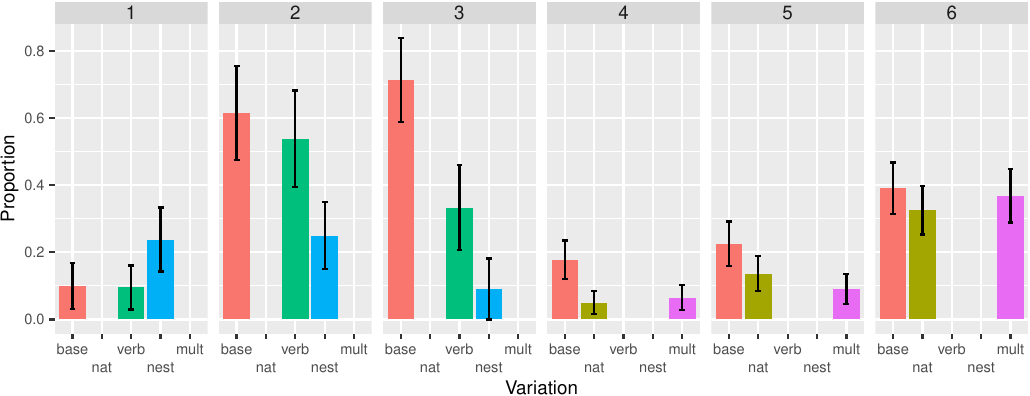}
	\caption{Rate of success in the first attempt for tasks with a
		context-free grammar as the target representation. Only study
		participants are considered.}
\end{figure}

\begin{figure}[H]
	\centering
	\includegraphics{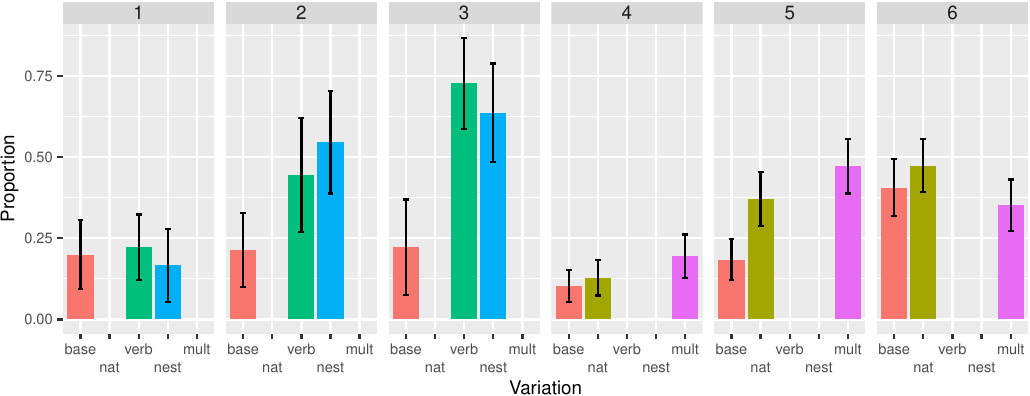}
	\caption{Rate of success in the first attempt for tasks with a pushdown
		automaton as the target representation. Only study participants are
		considered.}
\end{figure}

\clearpage
	\subsection{Needed attempts and time
		spent}\label{app:participants-attempts_time}

Box plots for all task groups, grouped by variation (base: baseline, verb: verbose set notation, nest: nesting, mult: multiplicities, \linebreak nat: natural
language). Mean values
are represented by the diamonds. Only participants are taken into account.

	\subsubsection{Attempts with outliers}\label{app:participants-attempts-outliers}\hphantom{}

\begin{figure}[H]
	\centering
	\includegraphics{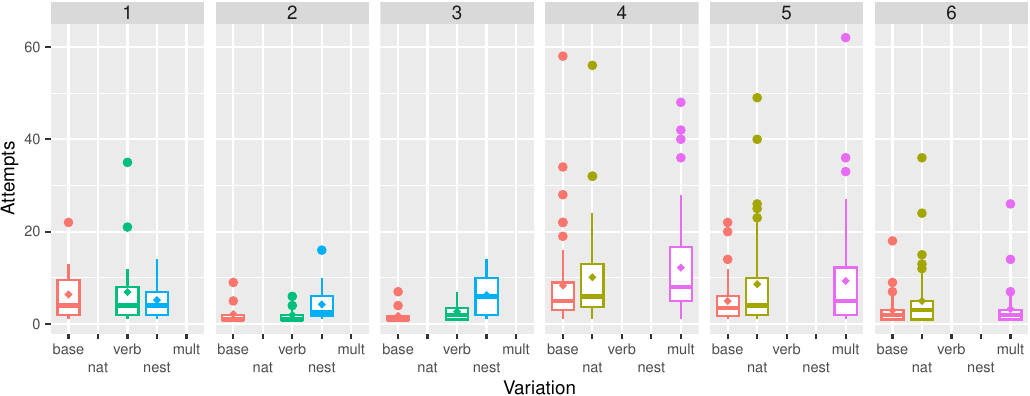}
	\caption{Box plot of the needed attempts for tasks with a context-free
		grammar as the target representation. Only study participants (including
		outliers)  are taken into account. The outliers are represented as dots}
\end{figure}

\begin{figure}[H]
	\centering
	\includegraphics{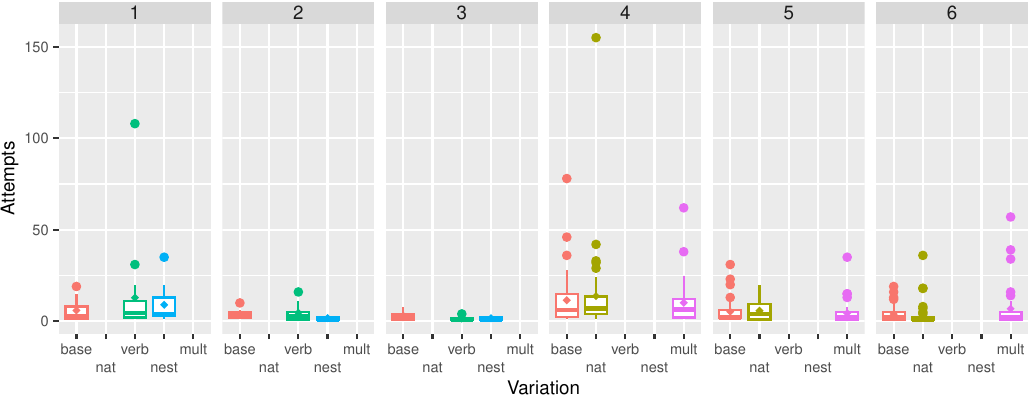}
	\caption{Box plot of the needed attempts for tasks with a pushdown
		automaton as the target representation. Only study participants
		(including outliers)  are taken into account. The outliers are represented as dots}
\end{figure}

\clearpage

	\subsubsection{Attempts without
		outliers}\label{app:participants-attempts-without-outliers}\hphantom{}

\begin{figure}[H]
	\centering
	\includegraphics{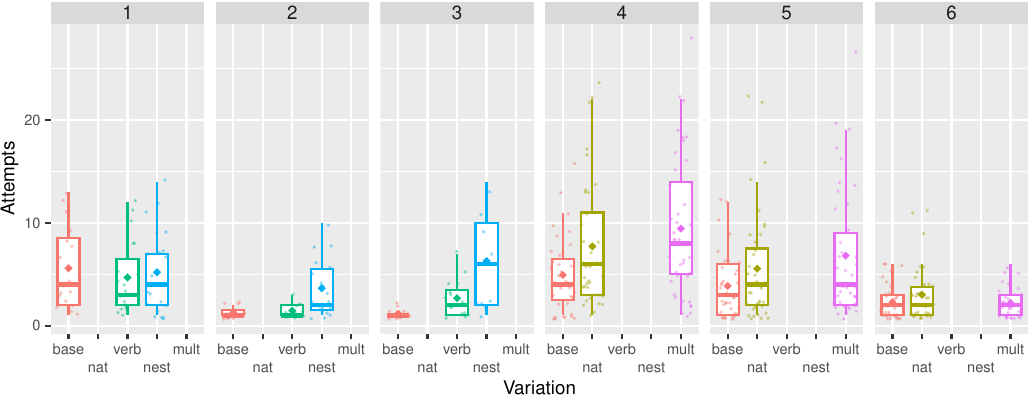}
	\caption{Box plot of the needed attempts for tasks with a context-free
		grammar as the target representation. Only study participants after the
		removal of outliers are taken into account. Additionally every participant is
		represented by a dot.}
\end{figure}

\begin{figure}[H]
	\centering
	\includegraphics{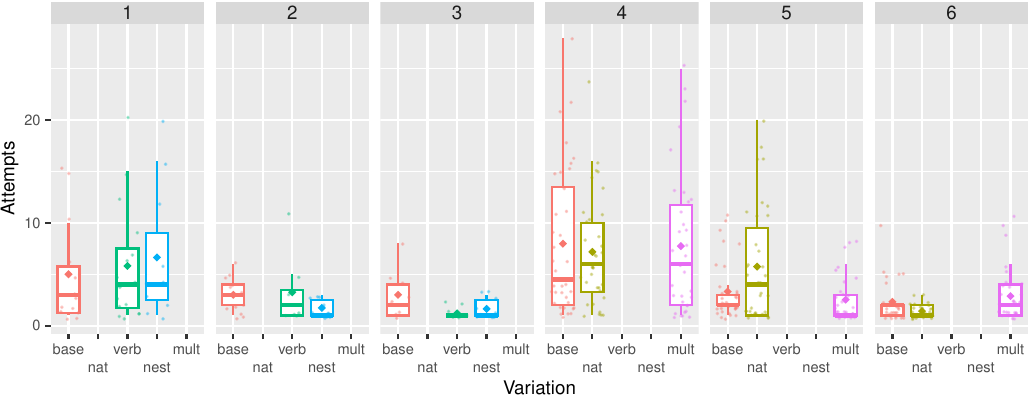}
	\caption{Box plot of the needed attempts for tasks with a pushdown
		automaton as the target representation. Only study participants after
		the removal of outliers are taken into account. Additionally every participant is
		represented by a dot.}
\end{figure}

\clearpage

	\subsubsection{Time with outliers}\label{app:participants-time-outliers}\hphantom{}

\begin{figure}[H]
	\centering
	\includegraphics{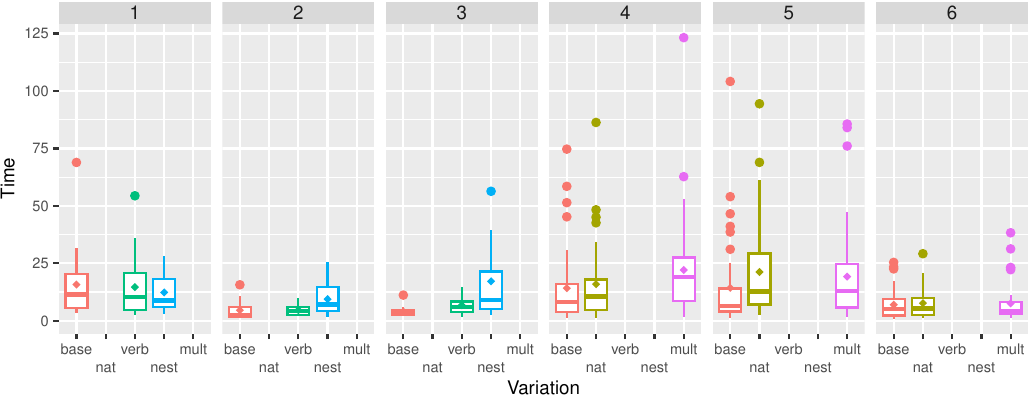}
	\caption{Box plot of time spent on tasks with a context-free grammar as
		the target representation. Only study participants (including outliers)
		are taken into account. The outliers are represented as dots}
\end{figure}

\begin{figure}[H]
	\centering
	\includegraphics{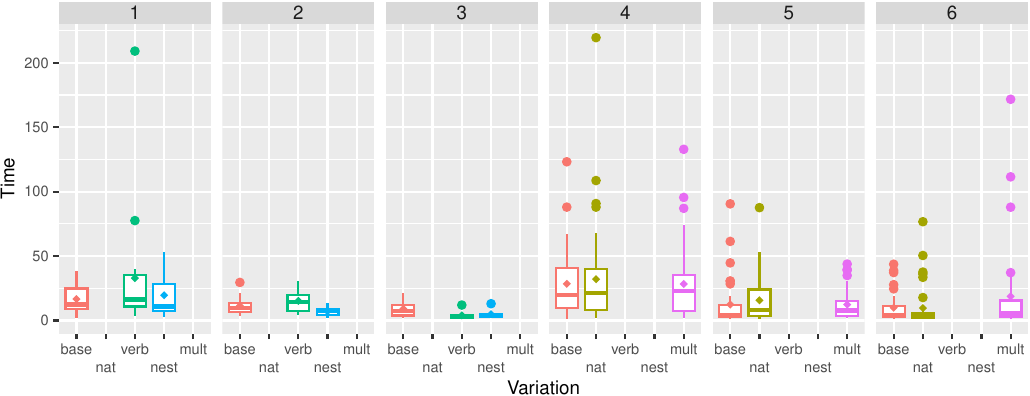}
	\caption{Box plot of time spent on tasks with a pushdown automaton as
		the target representation. Only study participants (including outliers)
		are taken into account. The outliers are represented as dots}
\end{figure}

\clearpage

	\subsubsection{Time without outliers}\label{app:participants-time-without-outliers}\hphantom{}

\begin{figure}[H]
	\centering
	\includegraphics{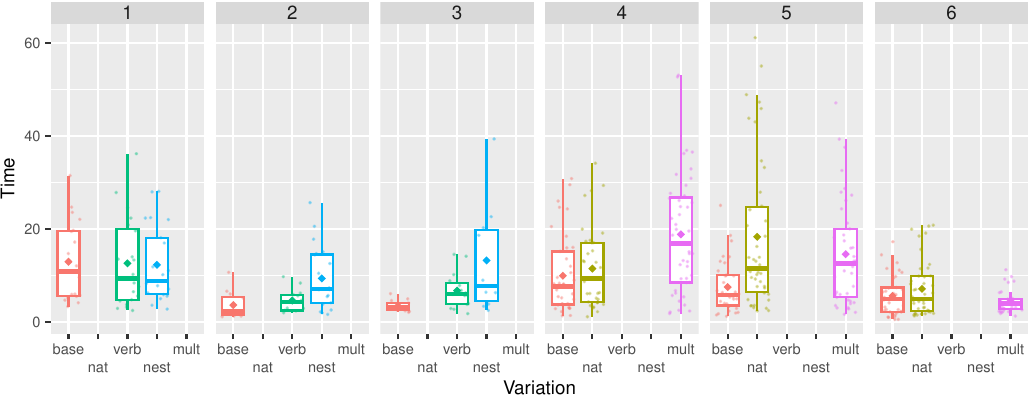}
	\caption{Box plot of time spent on tasks with a context-free grammar as
		the target representation. Only study participants after the removal of
		outliers are taken into account. Additionally every participant is represented by a
		dot.}
\end{figure}

\begin{figure}[H]
	\centering
	\includegraphics{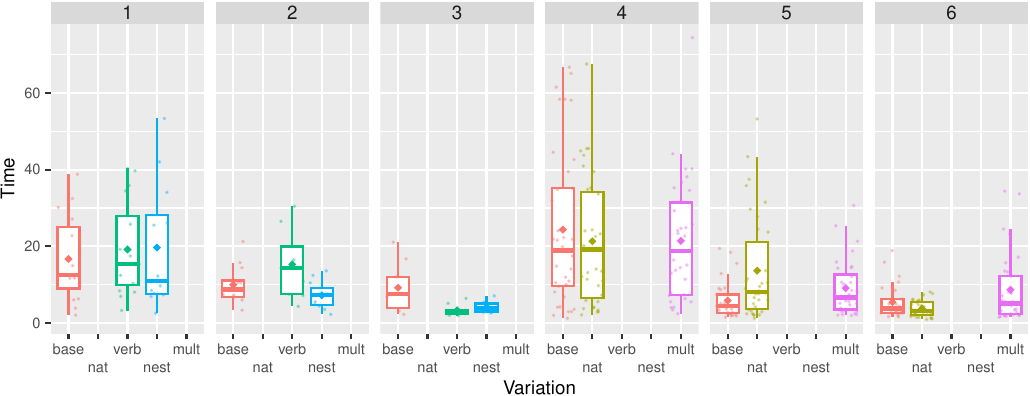}
	\caption{Box plot of time spent on tasks with a pushdown automaton as
		the target representation. Only study participants after the removal of
		outliers are taken into account. Additionally every participant is represented by a
		dot.}
\end{figure}

\clearpage

\subsection{Perceived
	difficulty}\label{app:participants-perceived}
	
	Distribution of the perceived difficulty grouped by variation (base: baseline, verb: verbose set notation, nest: nesting, mult: multiplicities, nat: natural
	language). Only participants are taken into account. The number of observations is noted below the corresponding bar.
	
	\subsubsection{Perceived interpretation
		difficulty}\label{app:participants-perceived-interpretation}\hphantom{}

\begin{figure}[H]
	\centering
{\includegraphics[keepaspectratio]{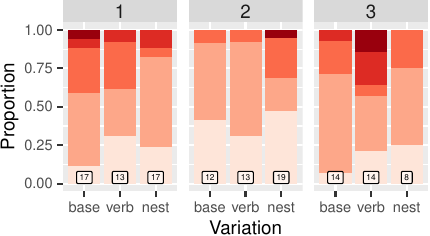}}
	\hfill
	\includegraphics[keepaspectratio]{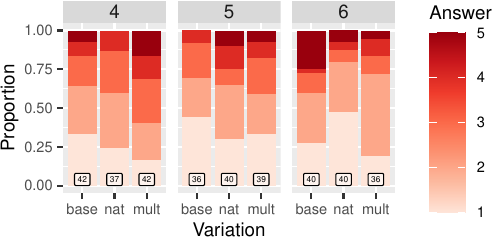}
	\caption{Distribution of the perceived interpretation difficulty (1 = very easy to 5 = very difficult) for the tasks with a context free grammar as the
		target representation.}
\end{figure}
	
\begin{figure}[H]
	\centering
{\includegraphics[keepaspectratio]{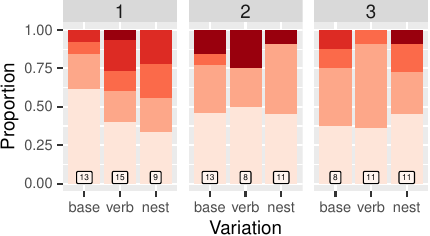}}
\hfill
{\includegraphics[keepaspectratio]{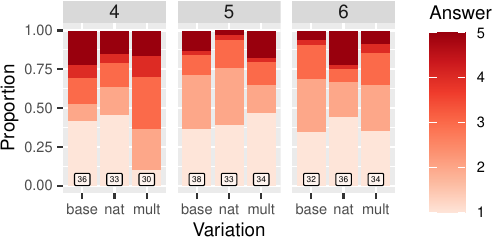}}
	\caption{Distribution of the perceived interpretation difficulty (1 = very easy to 5 = very difficult) for the tasks with a pushdown automaton as the
		target representation.}
\end{figure}

\clearpage

\subsubsection{Perceived construction
	difficulty}\label{app:participants-perceived-construction}\hphantom{}

\begin{figure}[H]
	\centering
{\includegraphics[keepaspectratio]{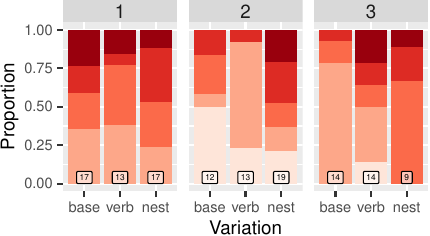}}
\hfill
{\includegraphics[keepaspectratio]{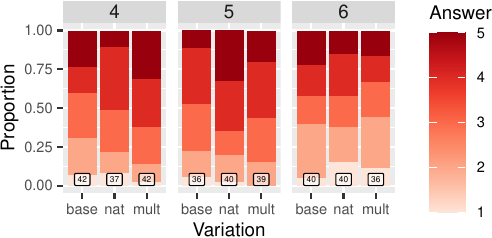}}
	\caption{Distribution of the perceived construction difficulty (1 = very easy to 5 = very difficult) for the tasks with a context free grammar as the
		target representation.}
\end{figure}

\begin{figure}[H]
	\centering
{\includegraphics[keepaspectratio]{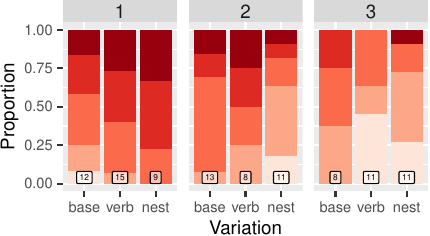}}
\hfill
{\includegraphics[keepaspectratio]{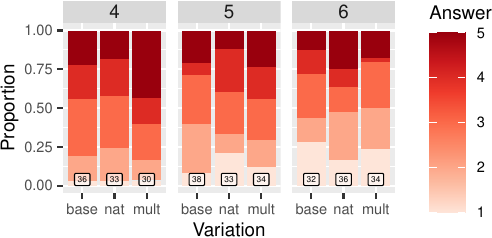}}
	\caption{Distribution of the perceived construction difficulty (1 = very easy to 5 = very difficult) for the tasks with a pushdown automaton as the
		target representation.}
\end{figure}

\clearpage
	\subsection{Task level linear OLS regression
		analysis}\label{app:participants-tasklevel}

Results of the task level linear regression analysis. The rows present
the regression coefficients for each variation, i.e.~how much the
corresponding metric (column) differs due to the variation. Below
each coefficient the standard error is specified. Additionally to the
variations we included task-fixed effects to control for general
differences between tasks, which we omitted here for readability. For
each metric the number of included observations and the coefficient of
determination (\(R^2\)) are presented. The \(R^2\) value describes how
good our independent variables (variations) predict the dependent
outcome variable (measured metric).

	\subsubsection{Context-free grammars as the target
		representation}\label{app:participants-tasklevel-cfg}\hphantom{}

\begin{table}[H]
	\caption{Task level linear regression on the participant data for Task group 1 regarding context-free grammars as the target representation.}
	\begin{center}
		\begin{tabular}{l c c c c c}
			\toprule
			& \parbox{1.5cm}{Number of attempts} & \parbox{1.5cm}{Time spent (min)} & \parbox{3.2cm}{Probability of success in first attempt (0--1)} & \parbox{3.2cm}{Perceived interpretation difficulty ($\times$SD)} & \parbox{3.2cm}{Perceived construction difficulty ($\times$SD)} \\
			\midrule
			(Intercept)          & $5.58^{***}$ & $12.97^{***}$ & $0.10$   & $0.16$   & $0.10$   \\
			& $(0.93)$     & $(1.95)$      & $(0.07)$ & $(0.19)$ & $(0.23)$ \\
			Verbose set notation & $-0.89$      & $-0.33$       & $-0.00$  & $-0.25$  & $-0.23$  \\
			& $(1.22)$     & $(2.91)$      & $(0.10)$ & $(0.29)$ & $(0.33)$ \\
			Nesting              & $-0.39$      & $-0.65$       & $0.14$   & $-0.32$  & $0.05$   \\
			& $(1.29)$     & $(2.64)$      & $(0.12)$ & $(0.26)$ & $(0.30)$ \\
			\midrule
			R$^2$                & $0.01$       & $0.00$        & $0.04$   & $0.04$   & $0.02$   \\
			Num. obs.            & $59$         & $60$          & $62$     & $47$     & $47$     \\
			\bottomrule
			\multicolumn{6}{l}{\scriptsize{Note: *** p<0.001, ** p<0.01, * p<0.05, + p<0.1.}}
		\end{tabular}
		
	\end{center}
\end{table}

\begin{table}[H]
	\caption{Task level linear regression on the participant data for Task group 2 regarding context-free grammars as the target representation.}
	\begin{center}
		\begin{tabular}{l c c c c c}
			\toprule
			& \parbox{1.5cm}{Number of attempts} & \parbox{1.5cm}{Time spent (min)} & \parbox{3.2cm}{Probability of success in first attempt (0--1)} & \parbox{3.2cm}{Perceived interpretation difficulty ($\times$SD)} & \parbox{3.2cm}{Perceived construction difficulty ($\times$SD)} \\
			\midrule
			(Intercept)          & $1.27^{***}$ & $3.66^{***}$ & $0.62^{***}$ & $-0.48^{**}$ & $-0.87^{**}$ \\
			& $(0.14)$     & $(0.84)$     & $(0.14)$     & $(0.15)$     & $(0.28)$     \\
			Verbose set notation & $0.18$       & $0.98$       & $-0.08$      & $0.08$       & $-0.13$      \\
			& $(0.25)$     & $(1.08)$     & $(0.20)$     & $(0.20)$     & $(0.33)$     \\
			Nesting              & $2.36^{**}$  & $5.73^{**}$  & $-0.37^{*}$  & $0.22$       & $0.81^{*}$   \\
			& $(0.66)$     & $(1.75)$     & $(0.17)$     & $(0.25)$     & $(0.39)$     \\
			\midrule
			R$^2$                & $0.26$       & $0.23$       & $0.11$       & $0.02$       & $0.17$       \\
			Num. obs.            & $41$         & $45$         & $46$         & $44$         & $44$         \\
			\bottomrule
			\multicolumn{6}{l}{\scriptsize{Note: *** p<0.001, ** p<0.01, * p<0.05, + p<0.1.}}
		\end{tabular}
		
	\end{center}
\end{table}

\begin{table}[H]
	\caption{Task level linear regression on the participant data for Task group 3 regarding context-free grammars as the target representation.}
	\begin{center}
		\begin{tabular}{l c c c c c}
			\toprule
			& \parbox{1.5cm}{Number of attempts} & \parbox{1.5cm}{Time spent (min)} & \parbox{3.2cm}{Probability of success in first attempt (0--1)} & \parbox{3.2cm}{Perceived interpretation difficulty ($\times$SD)} & \parbox{3.2cm}{Perceived construction difficulty ($\times$SD)} \\
			\midrule
			(Intercept)          & $1.17^{***}$ & $3.47^{***}$ & $0.71^{***}$  & $0.01$   & $-0.70^{***}$ \\
			& $(0.11)$     & $(0.33)$     & $(0.13)$      & $(0.15)$ & $(0.13)$      \\
			Verbose set notation & $1.50^{**}$  & $3.30^{**}$  & $-0.38^{*}$   & $0.34$   & $0.51$        \\
			& $(0.47)$     & $(1.05)$     & $(0.18)$      & $(0.34)$ & $(0.33)$      \\
			Nesting              & $5.11^{**}$  & $9.79^{*}$   & $-0.62^{***}$ & $-0.22$  & $0.92^{***}$  \\
			& $(1.46)$     & $(3.72)$     & $(0.15)$      & $(0.26)$ & $(0.23)$      \\
			\midrule
			R$^2$                & $0.36$       & $0.27$       & $0.26$        & $0.07$   & $0.18$        \\
			Num. obs.            & $38$         & $38$         & $40$          & $36$     & $37$          \\
			\bottomrule
			\multicolumn{6}{l}{\scriptsize{Note: *** p<0.001, ** p<0.01, * p<0.05, + p<0.1.}}
		\end{tabular}
		
	\end{center}
\end{table}

\begin{table}[H]
	\caption{Task level linear regression on the participant data for Task group 4 regarding context-free grammars as the target representation.}
	\begin{center}
		\begin{tabular}{l c c c c c}
			\toprule
			& \parbox{1.5cm}{Number of attempts} & \parbox{1.5cm}{Time spent (min)} & \parbox{3.2cm}{Probability of success in first attempt (0--1)} & \parbox{3.2cm}{Perceived interpretation difficulty ($\times$SD)} & \parbox{3.2cm}{Perceived construction difficulty ($\times$SD)} \\
			\midrule
			(Intercept)      & $4.92^{***}$ & $9.97^{***}$ & $0.18^{**}$ & $-0.01$    & $0.07$     \\
			& $(0.56)$     & $(1.27)$     & $(0.06)$    & $(0.15)$   & $(0.16)$   \\
			Natural language & $2.78^{*}$   & $1.54$       & $-0.13^{+}$ & $0.03$     & $0.05$     \\
			& $(1.17)$     & $(1.90)$     & $(0.07)$    & $(0.20)$   & $(0.21)$   \\
			Multiplicities   & $4.51^{***}$ & $8.91^{***}$ & $-0.11$     & $0.51^{*}$ & $0.40^{+}$ \\
			& $(1.18)$     & $(2.33)$     & $(0.07)$    & $(0.22)$   & $(0.21)$   \\
			\midrule
			R$^2$            & $0.10$       & $0.13$       & $0.04$      & $0.06$     & $0.04$     \\
			Num. obs.        & $118$        & $121$        & $131$       & $121$      & $121$      \\
			\bottomrule
			\multicolumn{6}{l}{\scriptsize{Note: *** p<0.001, ** p<0.01, * p<0.05, + p<0.1.}}
		\end{tabular}
		
	\end{center}
\end{table}

\begin{table}[H]
	\caption{Task level linear regression on the participant data for Task group 5 regarding context-free grammars as the target representation.}
	\begin{center}
		\begin{tabular}{l c c c c c}
			\toprule
			& \parbox{1.5cm}{Number of attempts} & \parbox{1.5cm}{Time spent (min)} & \parbox{3.2cm}{Probability of success in first attempt (0--1)} & \parbox{3.2cm}{Perceived interpretation difficulty ($\times$SD)} & \parbox{3.2cm}{Perceived construction difficulty ($\times$SD)} \\
			\midrule
			(Intercept)      & $3.86^{***}$ & $7.54^{***}$  & $0.22^{**}$ & $-0.26^{+}$ & $0.11$     \\
			& $(0.51)$     & $(0.95)$      & $(0.07)$    & $(0.13)$    & $(0.14)$   \\
			Natural language & $1.67^{+}$   & $10.80^{***}$ & $-0.09$     & $0.36^{+}$  & $0.35^{+}$ \\
			& $(1.00)$     & $(2.70)$      & $(0.08)$    & $(0.21)$    & $(0.20)$   \\
			Multiplicities   & $2.94^{**}$  & $7.08^{***}$  & $-0.13^{+}$ & $0.31$      & $0.25$     \\
			& $(1.10)$     & $(2.02)$      & $(0.08)$    & $(0.21)$    & $(0.19)$   \\
			\midrule
			R$^2$            & $0.05$       & $0.12$        & $0.02$      & $0.03$      & $0.03$     \\
			Num. obs.        & $117$        & $117$         & $128$       & $115$       & $115$      \\
			\bottomrule
			\multicolumn{6}{l}{\scriptsize{Note: *** p<0.001, ** p<0.01, * p<0.05, + p<0.1.}}
		\end{tabular}
		
	\end{center}
\end{table}

\begin{table}[H]
	\caption{Task level linear regression on the participant data for Task group 6 regarding context-free grammars as the target representation.}
	\begin{center}
		\begin{tabular}{l c c c c c}
			\toprule
			& \parbox{1.5cm}{Number of attempts} & \parbox{1.5cm}{Time spent (min)} & \parbox{3.2cm}{Probability of success in first attempt (0--1)} & \parbox{3.2cm}{Perceived interpretation difficulty ($\times$SD)} & \parbox{3.2cm}{Perceived construction difficulty ($\times$SD)} \\
			\midrule
			(Intercept)      & $2.29^{***}$ & $5.64^{***}$ & $0.39^{***}$ & $0.30$      & $0.02$   \\
			& $(0.26)$     & $(0.68)$     & $(0.08)$     & $(0.19)$    & $(0.16)$ \\
			Natural language & $0.71$       & $1.51$       & $-0.06$      & $-0.57^{*}$ & $-0.12$  \\
			& $(0.50)$     & $(1.14)$     & $(0.11)$     & $(0.24)$    & $(0.23)$ \\
			Multiplicities   & $-0.09$      & $-1.13$      & $-0.02$      & $-0.27$     & $-0.20$  \\
			& $(0.35)$     & $(0.80)$     & $(0.11)$     & $(0.24)$    & $(0.23)$ \\
			\midrule
			R$^2$            & $0.03$       & $0.05$       & $0.00$       & $0.05$      & $0.01$   \\
			Num. obs.        & $111$        & $113$        & $122$        & $116$       & $116$    \\
			\bottomrule
			\multicolumn{6}{l}{\scriptsize{Note: *** p<0.001, ** p<0.01, * p<0.05, + p<0.1.}}
		\end{tabular}
		
	\end{center}
\end{table}

\subsubsection{Pushdown automata as the target
		representation}\label{app:participants-tasklevel-pda}\hphantom{}

\begin{table}[H]
	\caption{Task level linear regression on the participant data for Task group 1 regarding pushdown automata as the target representation.}
	\begin{center}
		\begin{tabular}{l c c c c c}
			\toprule
			& \parbox{1.5cm}{Number of attempts} & \parbox{1.5cm}{Time spent (min)} & \parbox{3.2cm}{Probability of success in first attempt (0--1)} & \parbox{3.2cm}{Perceived interpretation difficulty ($\times$SD)} & \parbox{3.2cm}{Perceived construction difficulty ($\times$SD)} \\
			\midrule
			(Intercept)          & $5.00^{***}$ & $16.70^{***}$ & $0.20^{+}$ & $-0.52^{*}$ & $0.06$     \\
			& $(1.32)$     & $(2.90)$      & $(0.11)$   & $(0.21)$    & $(0.28)$   \\
			Verbose set notation & $0.81$       & $2.48$        & $0.02$     & $0.57$      & $0.44$     \\
			& $(1.93)$     & $(4.23)$      & $(0.15)$   & $(0.36)$    & $(0.34)$   \\
			Nesting              & $1.64$       & $2.98$        & $-0.03$    & $0.57$      & $0.68^{+}$ \\
			& $(2.33)$     & $(5.52)$      & $(0.15)$   & $(0.38)$    & $(0.35)$   \\
			\midrule
			R$^2$                & $0.01$       & $0.01$        & $0.00$     & $0.08$      & $0.11$     \\
			Num. obs.            & $41$         & $43$          & $45$       & $37$        & $36$       \\
			\bottomrule
			\multicolumn{6}{l}{\scriptsize{Note: *** p<0.001, ** p<0.01, * p<0.05, + p<0.1.}}
		\end{tabular}
		
	\end{center}
\end{table}

\begin{table}[H]
	\caption{Task level linear regression on the participant data for Task group 2 regarding pushdown automata as the target representation.}
	\begin{center}
		\begin{tabular}{l c c c c c}
			\toprule
			& \parbox{1.5cm}{Number of attempts} & \parbox{1.5cm}{Time spent (min)} & \parbox{3.2cm}{Probability of success in first attempt (0--1)} & \parbox{3.2cm}{Perceived interpretation difficulty ($\times$SD)} & \parbox{3.2cm}{Perceived construction difficulty ($\times$SD)} \\
			\midrule
			(Intercept)          & $3.00^{***}$ & $9.98^{***}$ & $0.21^{+}$ & $-0.15$  & $0.17$      \\
			& $(0.47)$     & $(1.35)$     & $(0.11)$   & $(0.32)$ & $(0.19)$    \\
			Verbose set notation & $0.25$       & $5.31$       & $0.23$     & $0.14$   & $0.09$      \\
			& $(1.29)$     & $(3.26)$     & $(0.21)$   & $(0.58)$ & $(0.38)$    \\
			Nesting              & $-1.27^{*}$  & $-2.76$      & $0.33^{+}$ & $-0.20$  & $-0.74^{*}$ \\
			& $(0.54)$     & $(1.74)$     & $(0.19)$   & $(0.42)$ & $(0.35)$    \\
			\midrule
			R$^2$                & $0.10$       & $0.24$       & $0.09$     & $0.01$   & $0.17$      \\
			Num. obs.            & $32$         & $33$         & $34$       & $32$     & $32$        \\
			\bottomrule
			\multicolumn{6}{l}{\scriptsize{Note: *** p<0.001, ** p<0.01, * p<0.05, + p<0.1.}}
		\end{tabular}
		
	\end{center}
\end{table}

\begin{table}[H]
	\caption{Task level linear regression on the participant data for Task group 3 regarding pushdown automata as the target representation.}
	\begin{center}
		\begin{tabular}{l c c c c c}
			\toprule
			& \parbox{1.5cm}{Number of attempts} & \parbox{1.5cm}{Time spent (min)} & \parbox{3.2cm}{Probability of success in first attempt (0--1)} & \parbox{3.2cm}{Perceived interpretation difficulty ($\times$SD)} & \parbox{3.2cm}{Perceived construction difficulty ($\times$SD)} \\
			\midrule
			(Intercept)          & $3.00^{***}$ & $9.19^{***}$ & $0.22$     & $-0.21$  & $-0.24$     \\
			& $(0.78)$     & $(2.12)$     & $(0.15)$   & $(0.29)$ & $(0.23)$    \\
			Verbose set notation & $-1.80^{*}$  & $-5.88^{*}$  & $0.51^{*}$ & $-0.21$  & $-0.77^{*}$ \\
			& $(0.79)$     & $(2.17)$     & $(0.20)$   & $(0.33)$ & $(0.32)$    \\
			Nesting              & $-1.36$      & $-5.04^{*}$  & $0.41^{+}$ & $0.00$   & $-0.55$     \\
			& $(0.83)$     & $(2.17)$     & $(0.21)$   & $(0.42)$ & $(0.36)$    \\
			\midrule
			R$^2$                & $0.23$       & $0.34$       & $0.18$     & $0.02$   & $0.14$      \\
			Num. obs.            & $30$         & $29$         & $31$       & $30$     & $30$        \\
			\bottomrule
			\multicolumn{6}{l}{\scriptsize{Note: *** p<0.001, ** p<0.01, * p<0.05, + p<0.1.}}
		\end{tabular}
		
	\end{center}
\end{table}

\begin{table}[H]
	\caption{Task level linear regression on the participant data for Task group 4 regarding pushdown automata as the target representation.}
	\begin{center}
		\begin{tabular}{l c c c c c}
			\toprule
			& \parbox{1.5cm}{Number of attempts} & \parbox{1.5cm}{Time spent (min)} & \parbox{3.2cm}{Probability of success in first attempt (0--1)} & \parbox{3.2cm}{Perceived interpretation difficulty ($\times$SD)} & \parbox{3.2cm}{Perceived construction difficulty ($\times$SD)} \\
			\midrule
			(Intercept)      & $7.97^{***}$ & $24.37^{***}$ & $0.10^{*}$ & $0.25$   & $0.22$   \\
			& $(1.19)$     & $(3.32)$      & $(0.05)$   & $(0.21)$ & $(0.15)$ \\
			Natural language & $-0.80$      & $-3.04$       & $0.03$     & $-0.24$  & $-0.09$  \\
			& $(1.50)$     & $(4.35)$      & $(0.07)$   & $(0.30)$ & $(0.21)$ \\
			Multiplicities   & $-0.24$      & $-2.96$       & $0.09$     & $0.33$   & $0.31$   \\
			& $(1.68)$     & $(4.32)$      & $(0.08)$   & $(0.28)$ & $(0.23)$ \\
			\midrule
			R$^2$            & $0.00$       & $0.01$        & $0.01$     & $0.04$   & $0.03$   \\
			Num. obs.        & $104$        & $105$         & $114$      & $99$     & $99$     \\
			\bottomrule
			\multicolumn{6}{l}{\scriptsize{Note: *** p<0.001, ** p<0.01, * p<0.05, + p<0.1.}}
		\end{tabular}
		
	\end{center}
\end{table}

\begin{table}[H]
	\caption{Task level linear regression on the participant data for Task group 5 regarding pushdown automata as the target representation.}
	\begin{center}
		\begin{tabular}{l c c c c c}
			\toprule
			& \parbox{1.5cm}{Number of attempts} & \parbox{1.5cm}{Time spent (min)} & \parbox{3.2cm}{Probability of success in first attempt (0--1)} & \parbox{3.2cm}{Perceived interpretation difficulty ($\times$SD)} & \parbox{3.2cm}{Perceived construction difficulty ($\times$SD)} \\
			\midrule
			(Intercept)      & $3.32^{***}$ & $5.86^{***}$ & $0.18^{**}$ & $-0.05$  & $-0.12$  \\
			& $(0.48)$     & $(0.84)$     & $(0.06)$    & $(0.17)$ & $(0.16)$ \\
			Natural language & $2.42^{*}$   & $7.78^{**}$  & $0.19^{+}$  & $-0.21$  & $-0.04$  \\
			& $(1.05)$     & $(2.50)$     & $(0.10)$    & $(0.22)$ & $(0.25)$ \\
			Multiplicities   & $-0.79$      & $3.21^{*}$   & $0.29^{**}$ & $0.04$   & $0.19$   \\
			& $(0.62)$     & $(1.53)$     & $(0.11)$    & $(0.27)$ & $(0.24)$ \\
			\midrule
			R$^2$            & $0.12$       & $0.11$       & $0.06$      & $0.01$   & $0.01$   \\
			Num. obs.        & $101$        & $98$         & $109$       & $105$    & $105$    \\
			\bottomrule
			\multicolumn{6}{l}{\scriptsize{Note: *** p<0.001, ** p<0.01, * p<0.05, + p<0.1.}}
		\end{tabular}
		
	\end{center}
\end{table}

\begin{table}[H]
	\caption{Task level linear regression on the participant data for Task group 6 regarding pushdown automata as the target representation.}
	\begin{center}
		\begin{tabular}{l c c c c c}
			\toprule
			& \parbox{1.5cm}{Number of attempts} & \parbox{1.5cm}{Time spent (min)} & \parbox{3.2cm}{Probability of success in first attempt (0--1)} & \parbox{3.2cm}{Perceived interpretation difficulty ($\times$SD)} & \parbox{3.2cm}{Perceived construction difficulty ($\times$SD)} \\
			\midrule
			(Intercept)      & $2.32^{***}$ & $5.45^{***}$ & $0.41^{***}$ & $-0.11$  & $-0.39^{*}$ \\
			& $(0.39)$     & $(0.85)$     & $(0.09)$     & $(0.16)$ & $(0.19)$    \\
			Natural language & $-0.90^{*}$  & $-1.74^{+}$  & $0.07$       & $0.19$   & $0.23$      \\
			& $(0.40)$     & $(0.93)$     & $(0.12)$     & $(0.26)$ & $(0.27)$    \\
			Multiplicities   & $0.55$       & $3.22^{+}$   & $-0.05$      & $0.09$   & $-0.03$     \\
			& $(0.62)$     & $(1.77)$     & $(0.12)$     & $(0.23)$ & $(0.27)$    \\
			\midrule
			R$^2$            & $0.09$       & $0.11$       & $0.01$       & $0.01$   & $0.01$      \\
			Num. obs.        & $91$         & $92$         & $107$        & $102$    & $102$       \\
			\bottomrule
			\multicolumn{6}{l}{\scriptsize{Note: *** p<0.001, ** p<0.01, * p<0.05, + p<0.1.}}
		\end{tabular}
		
	\end{center}
\end{table}

	\subsection{Full fixed-effects panel regression (including task dummy
		variables)}\label{app:participants-panel}

Results of the fixed-effects panel regression analysis. The rows present
the regression coefficients for each variation, i.e.~how much the
corresponding metric (column) differs due to the variation. Below
each coefficient the standard error is specified. Additionally to the
variations we included task-fixed effects to control for general
differences between tasks. For each metric the number of included
observations and the coefficient of determination (\(R^2\)) are
presented. The \(R^2\) value describes how good our independent
variables (variations and task) predict the dependent outcome
variable (measured metric).

\begin{table}[H]
	\caption{Fixed-effects panel regression on the participant data regarding context-free grammars as the target representation.}
	\begin{center}
		\begin{tabular}{l c c c c c}
			\toprule
			& \parbox{1.5cm}{Number of attempts} & \parbox{1.5cm}{Time spent (min)} & \parbox{3.2cm}{Probability of success in first attempt (0--1)} & \parbox{3.2cm}{Perceived interpretation difficulty ($\times$SD)} & \parbox{3.2cm}{Perceived construction difficulty ($\times$SD)} \\
			\midrule
			Natural language     & $1.83^{***}$  & $4.75^{***}$  & $-0.10^{*}$  & $-0.05$     & $0.11$        \\
			& $(0.49)$      & $(1.15)$      & $(0.05)$     & $(0.09)$    & $(0.08)$      \\
			Verbose set notation & $-0.25$       & $0.79$        & $-0.13$      & $0.07$      & $0.08$        \\
			& $(0.63)$      & $(1.09)$      & $(0.08)$     & $(0.12)$    & $(0.13)$      \\
			Nesting              & $1.53^{+}$    & $3.95^{*}$    & $-0.22^{**}$ & $-0.02$     & $0.61^{***}$  \\
			& $(0.80)$      & $(1.56)$      & $(0.08)$     & $(0.12)$    & $(0.16)$      \\
			Multiplicities       & $2.78^{***}$  & $5.39^{***}$  & $-0.09^{*}$  & $0.20^{*}$  & $0.15^{+}$    \\
			& $(0.56)$      & $(1.02)$      & $(0.05)$     & $(0.08)$    & $(0.09)$      \\
			Task group 2         & $-2.27^{***}$ & $-4.84^{***}$ & $0.27^{**}$  & $-0.30^{*}$ & $-0.73^{***}$ \\
			& $(0.67)$      & $(1.25)$      & $(0.09)$     & $(0.12)$    & $(0.14)$      \\
			Task group 3         & $-1.02$       & $-2.91^{+}$   & $0.17^{*}$   & $0.13$      & $-0.23$       \\
			& $(0.68)$      & $(1.54)$      & $(0.07)$     & $(0.15)$    & $(0.15)$      \\
			Task group 4         & $0.44$        & $-1.93$       & $-0.07$      & $-0.04$     & $-0.03$       \\
			& $(0.89)$      & $(1.73)$      & $(0.09)$     & $(0.16)$    & $(0.22)$      \\
			Task group 5         & $-1.42$       & $-2.04$       & $-0.02$      & $-0.24$     & $0.10$        \\
			& $(0.95)$      & $(1.79)$      & $(0.09)$     & $(0.15)$    & $(0.20)$      \\
			Task group 6         & $-4.18^{***}$ & $-9.74^{***}$ & $0.19^{*}$   & $-0.20$     & $-0.30$       \\
			& $(0.87)$      & $(1.76)$      & $(0.09)$     & $(0.15)$    & $(0.21)$      \\
			\midrule
			R$^2$                & $0.25$        & $0.24$        & $0.13$       & $0.08$      & $0.16$        \\
			Num. obs.            & $477$         & $487$         & $522$        & $476$       & $477$         \\
			\bottomrule
			\multicolumn{6}{l}{\scriptsize{Note: *** p<0.001, ** p<0.01, * p<0.05, + p<0.1.}}
		\end{tabular}
		
	\end{center}
\end{table}

\begin{table}[H]
	\caption{Fixed-effects panel regression on the participant data regarding pushdown automata as the target representation.}
	\begin{center}
		\begin{tabular}{l c c c c c}
			\toprule
			& \parbox{1.5cm}{Number of attempts} & \parbox{1.5cm}{Time spent (min)} & \parbox{3.2cm}{Probability of success in first attempt (0--1)} & \parbox{3.2cm}{Perceived interpretation difficulty ($\times$SD)} & \parbox{3.2cm}{Perceived construction difficulty ($\times$SD)} \\
			\midrule
			Natural language     & $0.67$        & $-0.10$        & $0.10^{*}$   & $-0.02$     & $0.09$        \\
			& $(0.54)$      & $(1.49)$       & $(0.05)$     & $(0.08)$    & $(0.09)$      \\
			Verbose set notation & $-0.49$       & $0.41$         & $0.24^{**}$  & $0.20^{+}$  & $-0.06$       \\
			& $(0.86)$      & $(1.99)$       & $(0.08)$     & $(0.12)$    & $(0.17)$      \\
			Nesting              & $-0.78$       & $-1.30$        & $0.20^{*}$   & $0.17$      & $-0.15$       \\
			& $(0.75)$      & $(2.01)$       & $(0.10)$     & $(0.15)$    & $(0.15)$      \\
			Multiplicities       & $0.28$        & $0.30$         & $0.11^{*}$   & $0.24^{**}$ & $0.24^{**}$   \\
			& $(0.59)$      & $(1.58)$       & $(0.05)$     & $(0.08)$    & $(0.08)$      \\
			Task group 2         & $-2.53^{**}$  & $-7.36^{**}$   & $0.16^{+}$   & $0.01$      & $-0.43^{*}$   \\
			& $(0.93)$      & $(2.31)$       & $(0.09)$     & $(0.13)$    & $(0.17)$      \\
			Task group 3         & $-3.23^{***}$ & $-11.16^{***}$ & $0.27^{**}$  & $-0.07$     & $-1.04^{***}$ \\
			& $(0.82)$      & $(1.97)$       & $(0.09)$     & $(0.15)$    & $(0.15)$      \\
			Task group 4         & $-0.38$       & $-1.85$        & $0.19^{*}$   & $0.20$      & $-0.83^{***}$ \\
			& $(1.10)$      & $(2.91)$       & $(0.09)$     & $(0.16)$    & $(0.22)$      \\
			Task group 5         & $-3.81^{***}$ & $-12.62^{***}$ & $0.38^{***}$ & $-0.23$     & $-1.24^{***}$ \\
			& $(1.00)$      & $(2.51)$       & $(0.10)$     & $(0.14)$    & $(0.21)$      \\
			Task group 6         & $-5.58^{***}$ & $-16.43^{***}$ & $0.44^{***}$ & $-0.12$     & $-1.45^{***}$ \\
			& $(1.05)$      & $(2.50)$       & $(0.10)$     & $(0.15)$    & $(0.22)$      \\
			\midrule
			R$^2$                & $0.28$        & $0.36$         & $0.15$       & $0.13$      & $0.33$        \\
			Num. obs.            & $395$         & $396$          & $435$        & $403$       & $402$         \\
			\bottomrule
			\multicolumn{6}{l}{\scriptsize{Note: *** p<0.001, ** p<0.01, * p<0.05, + p<0.1.}}
		\end{tabular}
		
	\end{center}
\end{table}

\clearpage
	\subsection{Full pooled linear regression (including task dummy
		variables)}\label{app:participants-pooled}

Results of the pooled linear regression analysis. The rows present the
regression coefficients for each variation, i.e.~how much the
corresponding metric (column) differs due to the variation. Below
each coefficient the standard error is specified. Additionally to the
variations we included task-fixed effects to control for general
differences between tasks. For each metric the number of included
observations and the coefficient of determination (\(R^2\)) are
presented. The \(R^2\) value describes how good our independent
variables (variations and task) predict the dependent outcome
variable (measured metric).

\begin{table}[H]
	\caption{Pooled linear regression on the participant data regarding context-free grammars as the target representation.}
	\begin{center}
		\begin{tabular}{l c c c c c}
			\toprule
			& \parbox{1.5cm}{Number of attempts} & \parbox{1.5cm}{Time spent (min)} & \parbox{3.2cm}{Probability of success in first attempt (0--1)} & \parbox{3.2cm}{Perceived interpretation difficulty ($\times$SD)} & \parbox{3.2cm}{Perceived construction difficulty ($\times$SD)} \\
			\midrule
			(Intercept)          & $4.47^{***}$  & $10.87^{***}$ & $0.27^{***}$ & $-0.01$      & $-0.17$       \\
			& $(0.68)$      & $(1.40)$      & $(0.07)$     & $(0.12)$     & $(0.16)$      \\
			Natural language     & $1.73^{***}$  & $4.70^{***}$  & $-0.09^{*}$  & $-0.06$      & $0.09$        \\
			& $(0.49)$      & $(1.11)$      & $(0.05)$     & $(0.09)$     & $(0.08)$      \\
			Verbose set notation & $0.07$        & $1.09$        & $-0.14^{+}$  & $0.04$       & $0.03$        \\
			& $(0.57)$      & $(1.25)$      & $(0.08)$     & $(0.13)$     & $(0.14)$      \\
			Nesting              & $1.85^{*}$    & $4.01^{**}$   & $-0.23^{**}$ & $-0.08$      & $0.56^{***}$  \\
			& $(0.73)$      & $(1.44)$      & $(0.08)$     & $(0.12)$     & $(0.16)$      \\
			Multiplicities       & $2.54^{***}$  & $5.16^{***}$  & $-0.09^{*}$  & $0.18^{*}$   & $0.15^{+}$    \\
			& $(0.55)$      & $(0.99)$      & $(0.04)$     & $(0.09)$     & $(0.09)$      \\
			Task group 2         & $-2.93^{***}$ & $-6.47^{***}$ & $0.30^{***}$ & $-0.32^{**}$ & $-0.64^{***}$ \\
			& $(0.60)$      & $(1.20)$      & $(0.08)$     & $(0.12)$     & $(0.14)$      \\
			Task group 3         & $-1.80^{**}$  & $-5.00^{**}$  & $0.25^{***}$ & $0.11$       & $-0.27^{+}$   \\
			& $(0.65)$      & $(1.55)$      & $(0.07)$     & $(0.15)$     & $(0.15)$      \\
			Task group 4         & $1.48^{+}$    & $-0.48$       & $-0.11$      & $0.14$       & $0.31$        \\
			& $(0.84)$      & $(1.62)$      & $(0.08)$     & $(0.16)$     & $(0.19)$      \\
			Task group 5         & $-0.48$       & $-0.47$       & $-0.06$      & $-0.06$      & $0.39^{*}$    \\
			& $(0.85)$      & $(1.73)$      & $(0.08)$     & $(0.15)$     & $(0.18)$      \\
			Task group 6         & $-3.36^{***}$ & $-8.25^{***}$ & $0.15^{+}$   & $-0.01$      & $0.00$        \\
			& $(0.75)$      & $(1.53)$      & $(0.08)$     & $(0.15)$     & $(0.18)$      \\
			\midrule
			R$^2$                & $0.21$        & $0.17$        & $0.11$       & $0.03$       & $0.10$        \\
			Num. obs.            & $484$         & $494$         & $529$        & $479$        & $480$         \\
			\bottomrule
			\multicolumn{6}{l}{\scriptsize{Note: *** p<0.001, ** p<0.01, * p<0.05, + p<0.1.}}
		\end{tabular}
		
	\end{center}
\end{table}

\begin{table}[H]
	\caption{Pooled linear regression on the participant data regarding pushdown automata as the target representation.}
	\begin{center}
		\begin{tabular}{l c c c c c}
			\toprule
			& \parbox{1.5cm}{Number of attempts} & \parbox{1.5cm}{Time spent (min)} & \parbox{3.2cm}{Probability of success in first attempt (0--1)} & \parbox{3.2cm}{Perceived interpretation difficulty ($\times$SD)} & \parbox{3.2cm}{Perceived construction difficulty ($\times$SD)} \\
			\midrule
			(Intercept)          & $5.87^{***}$  & $18.40^{***}$  & $0.06$       & $-0.26$     & $0.49^{*}$    \\
			& $(0.96)$      & $(2.22)$       & $(0.08)$     & $(0.17)$    & $(0.19)$      \\
			Natural language     & $0.29$        & $0.91$         & $0.10^{+}$   & $-0.09$     & $0.04$        \\
			& $(0.62)$      & $(1.75)$       & $(0.05)$     & $(0.09)$    & $(0.09)$      \\
			Verbose set notation & $-0.15$       & $0.95$         & $0.22^{**}$  & $0.20$      & $-0.06$       \\
			& $(0.90)$      & $(2.17)$       & $(0.08)$     & $(0.15)$    & $(0.18)$      \\
			Nesting              & $-0.21$       & $-1.06$        & $0.21^{*}$   & $0.14$      & $-0.19$       \\
			& $(0.76)$      & $(2.04)$       & $(0.09)$     & $(0.15)$    & $(0.15)$      \\
			Multiplicities       & $-0.15$       & $1.04$         & $0.11^{*}$   & $0.14$      & $0.15^{+}$    \\
			& $(0.63)$      & $(1.62)$       & $(0.05)$     & $(0.09)$    & $(0.08)$      \\
			Task group 2         & $-3.14^{***}$ & $-7.80^{***}$  & $0.20^{*}$   & $-0.02$     & $-0.47^{**}$  \\
			& $(0.92)$      & $(2.29)$       & $(0.08)$     & $(0.17)$    & $(0.18)$      \\
			Task group 3         & $-3.84^{***}$ & $-12.94^{***}$ & $0.34^{***}$ & $-0.15$     & $-1.11^{***}$ \\
			& $(0.81)$      & $(1.98)$       & $(0.09)$     & $(0.16)$    & $(0.15)$      \\
			Task group 4         & $1.72$        & $3.40$         & $0.02$       & $0.51^{**}$ & $-0.26$       \\
			& $(1.19)$      & $(3.17)$       & $(0.09)$     & $(0.19)$    & $(0.21)$      \\
			Task group 5         & $-2.01^{+}$   & $-9.45^{***}$  & $0.22^{*}$   & $0.14$      & $-0.61^{**}$  \\
			& $(1.03)$      & $(2.57)$       & $(0.09)$     & $(0.19)$    & $(0.21)$      \\
			Task group 6         & $-3.71^{***}$ & $-13.09^{***}$ & $0.28^{**}$  & $0.23$      & $-0.86^{***}$ \\
			& $(1.03)$      & $(2.46)$       & $(0.09)$     & $(0.19)$    & $(0.22)$      \\
			\midrule
			R$^2$                & $0.20$        & $0.25$         & $0.09$       & $0.03$      & $0.10$        \\
			Num. obs.            & $399$         & $400$          & $440$        & $405$       & $404$         \\
			\bottomrule
			\multicolumn{6}{l}{\scriptsize{Note: *** p<0.001, ** p<0.01, * p<0.05, + p<0.1.}}
		\end{tabular}
		
	\end{center}
\end{table}

\clearpage

	\section{Statistics and regressions regarding all anonymous sessions}\label{app:sessions}

	\subsection{Rate of success and of success in first attempt}\label{app:sessions-success}
		Bar charts for all task groups, grouped by variation (base: baseline, verb: verbose set notation, nest: nesting, mult: multiplicities, \linebreak nat: natural
		language). All anonymous sessions are taken into account.

\begin{figure}[H]
	\centering
	\includegraphics{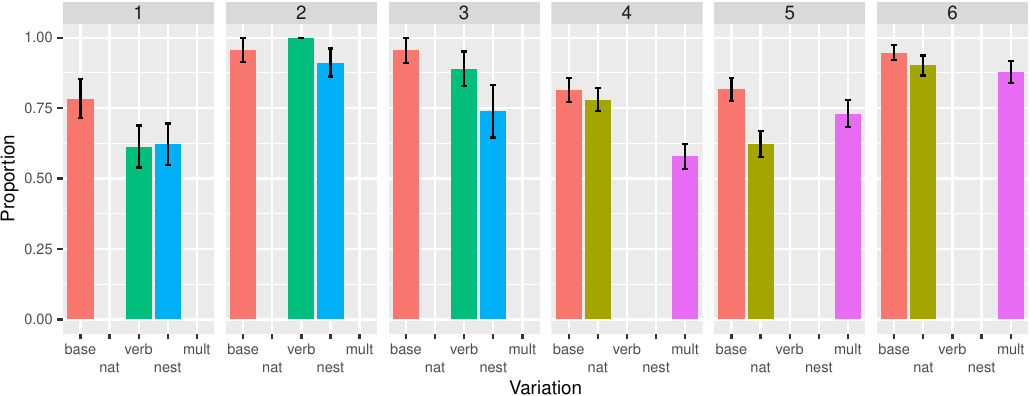}
	\caption{Rate of success for tasks with a context-free grammar as the
		target representation. All anonymous sessions are considered.}
\end{figure}

\begin{figure}[H]
	\centering
	\includegraphics{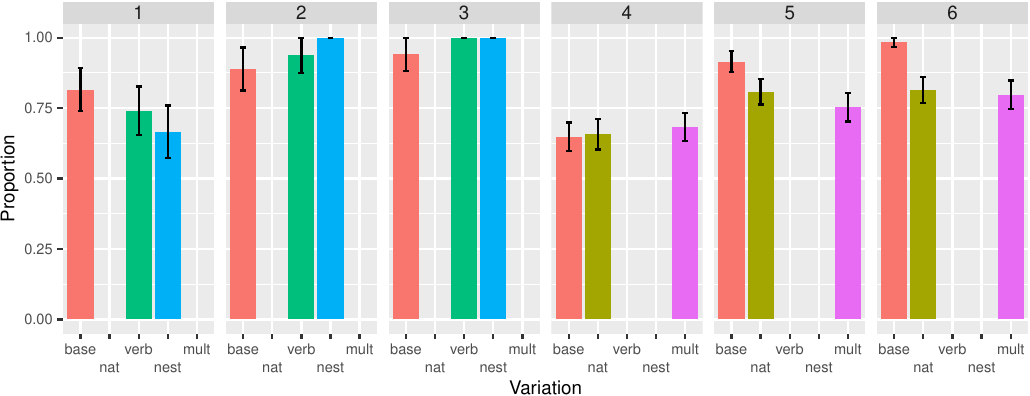}
	\caption{Rate of success for tasks with a pushdown automaton as the
		target representation. All anonymous sessions are considered.}
\end{figure}

\begin{figure}[H]
	\centering
	\includegraphics{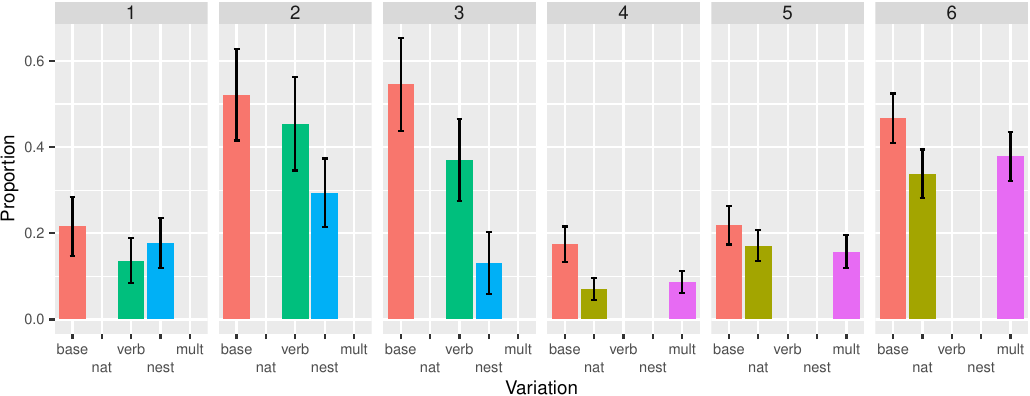}
	\caption{Rate of success in the first attempt for tasks with a
		context-free grammar as the target representation. All anonymous
		sessions are considered.}
\end{figure}

\begin{figure}[H]
	\centering
	\includegraphics{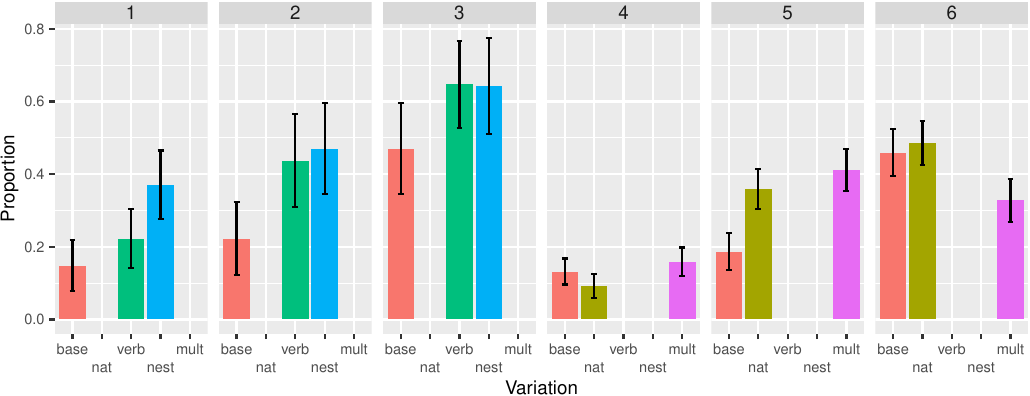}
	\caption{Rate of success in the first attempt for tasks with a pushdown
		automaton as the target representation. All anonymous sessions are
		considered.}
\end{figure}

\clearpage
	\subsection{Needed attempts and time
		spent}\label{app:sessions-attempts_time}

Box plots for all task groups, grouped by variation (base: baseline, verb: verbose set notation, nest: nesting, mult: multiplicities, \linebreak nat: natural
language). Mean values
are represented by the diamonds. All anonymous sessions are taken into account.

\subsubsection{Attempts with outliers}\hphantom{}

{\begin{figure}[H]
	\centering
	\includegraphics{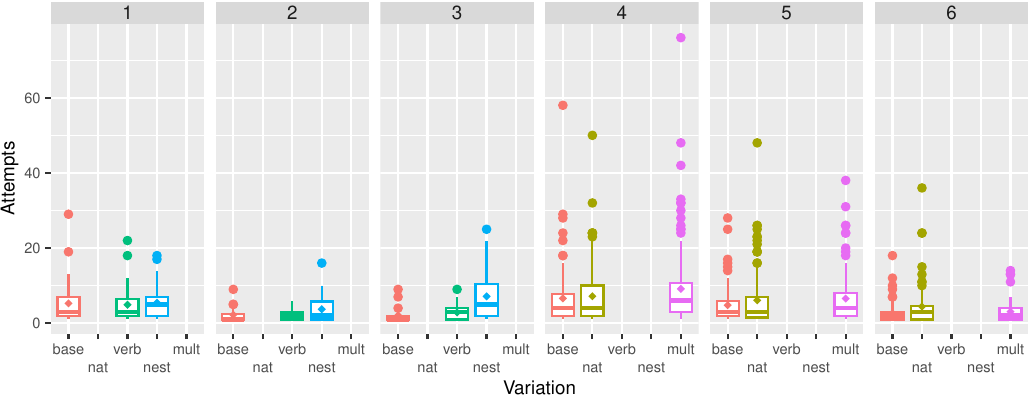}
	\caption{Box plot of time spent on tasks with a context-free grammar as
		the target representation. All anonymous sessions (including outliers)
		are taken into account. The outliers are represented as dots}
\end{figure}

\begin{figure}[H]
	\centering
	\includegraphics{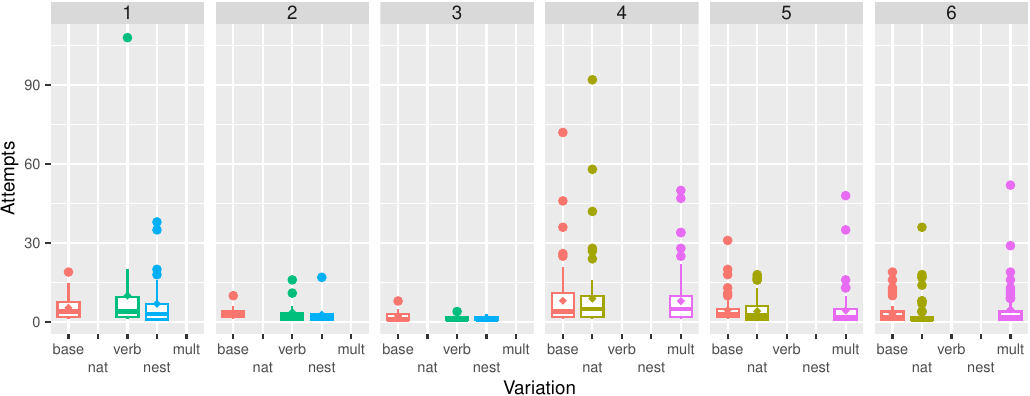}
	\caption{Box plot of time spent on tasks with a pushdown automaton as
		the target representation. All anonymous sessions (including outliers)
		are taken into account. The outliers are represented as dots}
\end{figure}}

\clearpage
	\subsubsection{Attempts without
		outliers}\hphantom{}

\begin{figure}[H]
	\centering
	\includegraphics{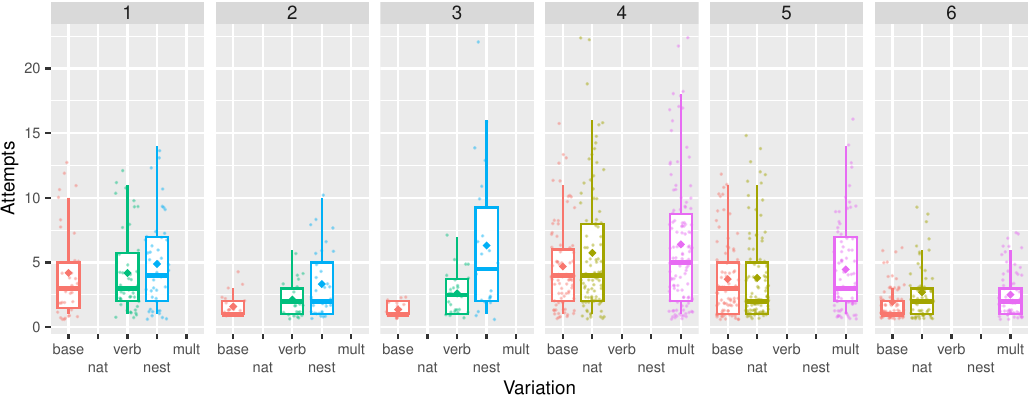}
	\caption{All anonymous sessions after the removal of outliers are
		visible. Additionally every participant is represented by a dot.}
\end{figure}

\begin{figure}[H]
	\centering
	\includegraphics{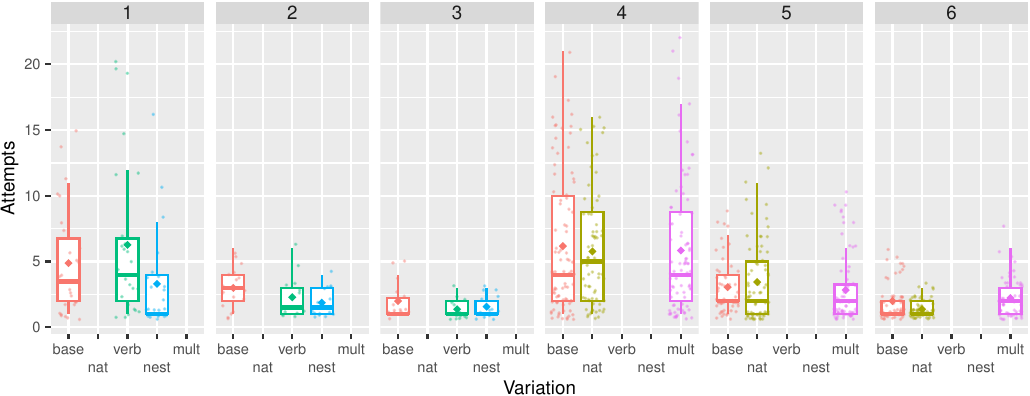}
	\caption{All anonymous sessions after the removal of outliers are
		visible. Additionally every participant is represented by a dot.}
\end{figure}

\clearpage
	\subsubsection{Time with outliers}\hphantom{}

\begin{figure}[H]
	\centering
	\includegraphics{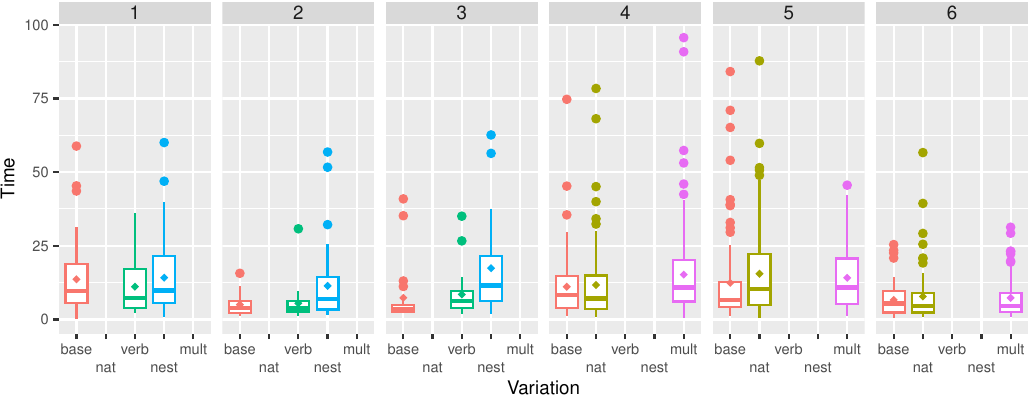}
	\caption{All anonymous sessions (including outliers) are taken into account. The
		outliers are represented as dots}
\end{figure}

\begin{figure}[H]
	\centering
	\includegraphics{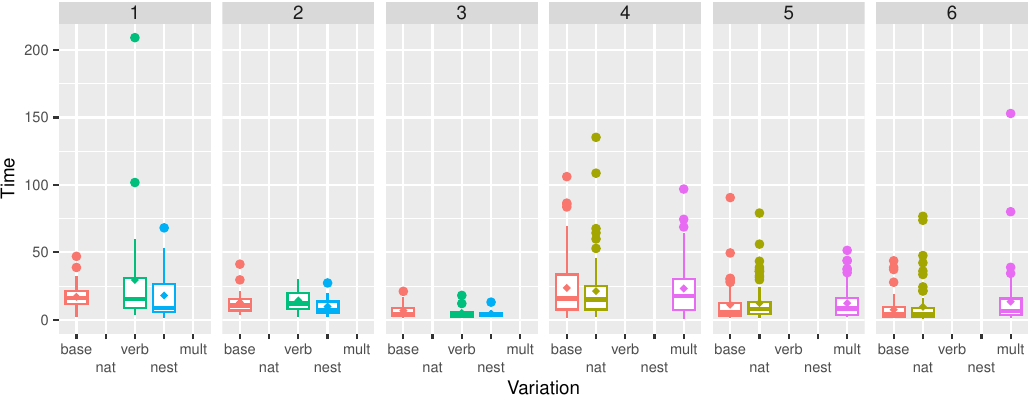}
	\caption{All anonymous sessions (including outliers) are taken into account. The
		outliers are represented as dots}
\end{figure}

\clearpage
	\subsubsection{Time without outliers}\hphantom{}

\begin{figure}[H]
	\centering
	\includegraphics{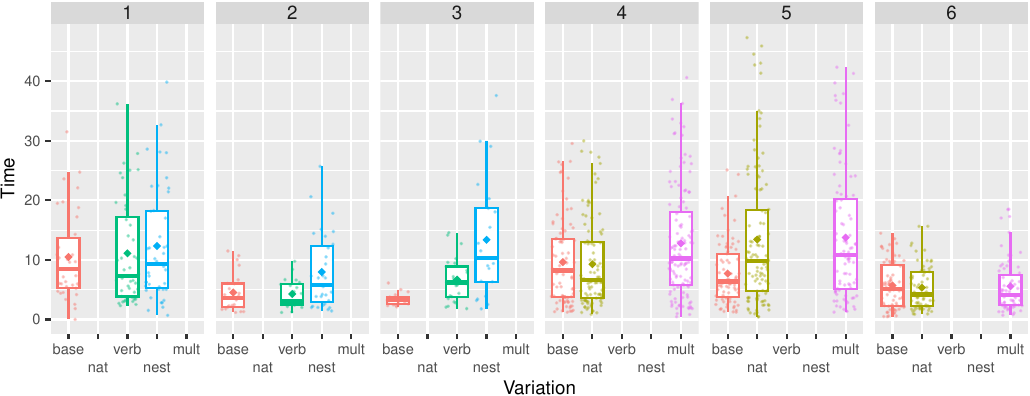}
	\caption{All anonymous sessions after the removal of outliers are
		visible. Additionally every participant is represented by a dot.}
\end{figure}

\begin{figure}[H]
	\centering
	\includegraphics{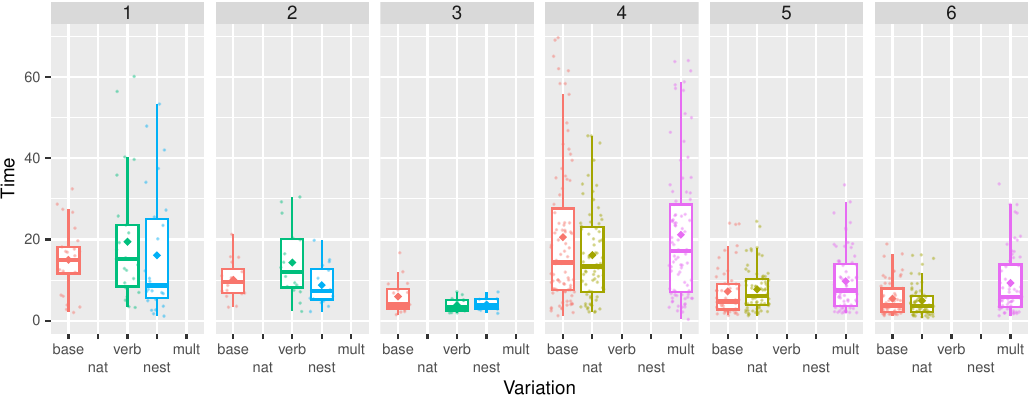}
	\caption{All anonymous sessions after the removal of outliers are
		visible. Additionally every participant is represented by a dot.}
\end{figure}
\clearpage

\subsection{Perceived
	difficulty}\label{app:sessions-perceived}
	Distribution of the perceived difficulty grouped by variation (base: baseline, verb: verbose set notation, nest: nesting, mult: multiplicities, nat: natural
	language). All anonymous sessions are taken into account. The number of observations is noted below the corresponding bar.

\subsubsection{Perceived interpretation difficulty}\hphantom{}
\begin{figure}[H]
	\centering
	{\includegraphics[keepaspectratio]{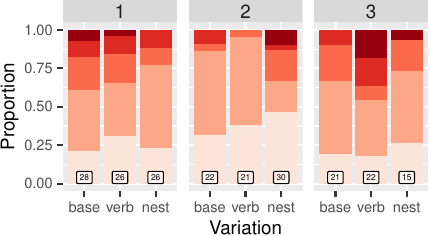}}
	\hfill
	{\includegraphics[keepaspectratio]{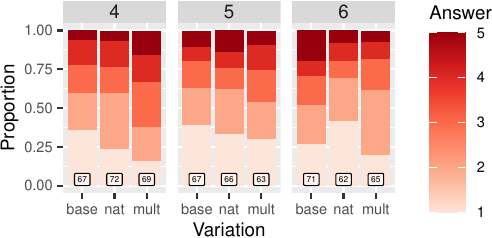}}
	\caption{Distribution of the perceived interpretation difficulty (1 = very easy to 5 = very difficult) for the tasks with a context free grammar as the
		target representation.}
\end{figure}

\begin{figure}[H]
	\centering
	{\includegraphics[keepaspectratio]{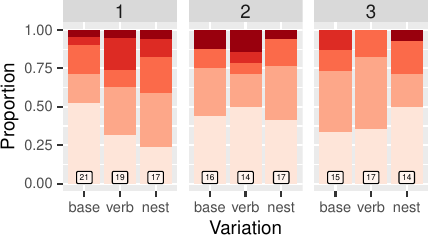}}
	\hfill
	{\includegraphics[keepaspectratio]{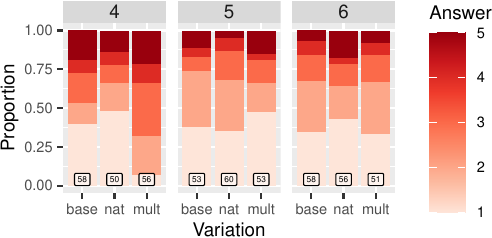}}
\caption{Distribution of the perceived interpretation difficulty (1 = very easy to 5 = very difficult) for the tasks with a pushdown automaton as the
	target representation.}
\end{figure}

\clearpage

\subsubsection{Perceived construction
	difficulty}\label{app:sessions-perceived-construction}\hphantom{}

\begin{figure}[H]
	\centering
	{\includegraphics[keepaspectratio]{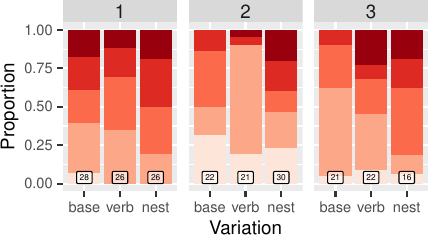}}
	\hfill
	{\includegraphics[keepaspectratio]{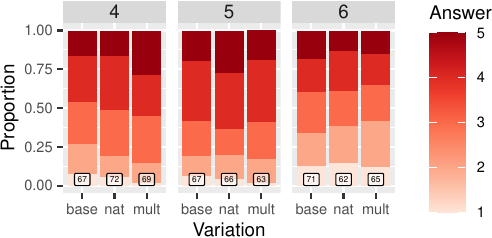}}
	\caption{Distribution of the perceived construction difficulty (1 = very easy to 5 = very difficult) for the tasks with a context free grammar as the
		target representation.}
\end{figure}

\begin{figure}[H]
	\centering
	{\includegraphics[keepaspectratio]{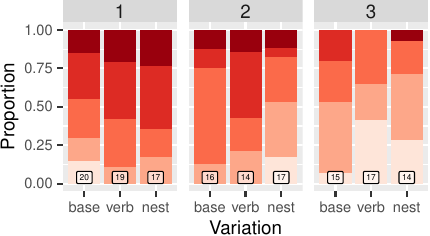}}
	\hfill
	{\includegraphics[keepaspectratio]{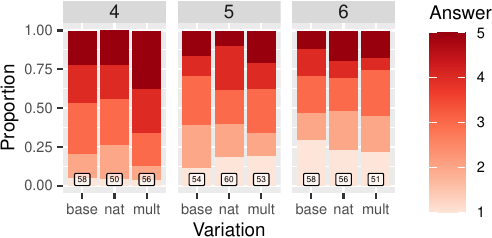}}
	\caption{Distribution of the perceived construction difficulty (1 = very easy to 5 = very difficult) for the tasks with a pushdown automaton as the
		target representation.}
\end{figure}

\clearpage
	\subsection{Task level linear OLS regression
		analysis}\label{app:sessions-tasklevel}

Results of the task level linear regression analysis. The rows present
the regression coefficients for each variation, i.e.~how much the
corresponding metric (column) differs due to the variation. Below
each coefficient the standard error is specified. Additionally to the
variations we included task-fixed effects to control for general
differences between tasks, which we omitted here for readability. For
each metric the number of included observations and the coefficient of
determination (\(R^2\)) are presented. The \(R^2\) value describes how
good our independent variables (variations) predict the dependent
outcome variable (measured metric).

\subsubsection{Context-free grammars as the target
		representation}\hphantom{}

\begin{table}[H]
	\caption{Task level linear regression on the full session data for Task group 1 regarding context-free grammars as the target representation.}
	\begin{center}
		\begin{tabular}{l c c c c c}
			\toprule
			& \parbox{1.5cm}{Number of attempts} & \parbox{1.5cm}{Time spent (min)} & \parbox{3.2cm}{Probability of success in first attempt (0--1)} & \parbox{3.2cm}{Perceived interpretation difficulty ($\times$SD)} & \parbox{3.2cm}{Perceived construction difficulty ($\times$SD)} \\
			\midrule
			(Intercept)          & $4.20^{***}$ & $10.48^{***}$ & $0.22^{**}$ & $0.06$   & $-0.04$  \\
			& $(0.60)$     & $(1.29)$      & $(0.07)$    & $(0.17)$ & $(0.19)$ \\
			Verbose set notation & $-0.01$      & $0.62$        & $-0.08$     & $-0.15$  & $-0.02$  \\
			& $(0.78)$     & $(1.85)$      & $(0.09)$    & $(0.24)$ & $(0.25)$ \\
			Nesting              & $0.68$       & $1.85$        & $-0.04$     & $-0.24$  & $0.31$   \\
			& $(0.80)$     & $(1.94)$      & $(0.09)$    & $(0.22)$ & $(0.25)$ \\
			\midrule
			R$^2$                & $0.01$       & $0.01$        & $0.01$      & $0.01$   & $0.03$   \\
			Num. obs.            & $120$        & $121$         & $126$       & $80$     & $80$     \\
			\bottomrule
			\multicolumn{6}{l}{\scriptsize{Note: *** p<0.001, ** p<0.01, * p<0.05, + p<0.1.}}
		\end{tabular}
		
	\end{center}
\end{table}

\begin{table}[H]
	\caption{Task level linear regression on the full session data for Task group 2 regarding context-free grammars as the target representation.}
	\begin{center}
		\begin{tabular}{l c c c c c}
			\toprule
			& \parbox{1.5cm}{Number of attempts} & \parbox{1.5cm}{Time spent (min)} & \parbox{3.2cm}{Probability of success in first attempt (0--1)} & \parbox{3.2cm}{Perceived interpretation difficulty ($\times$SD)} & \parbox{3.2cm}{Perceived construction difficulty ($\times$SD)} \\
			\midrule
			(Intercept)          & $1.60^{***}$ & $4.52^{***}$ & $0.52^{***}$ & $-0.34^{*}$ & $-0.67^{***}$ \\
			& $(0.20)$     & $(0.69)$     & $(0.11)$     & $(0.14)$    & $(0.18)$      \\
			Verbose set notation & $0.54$       & $-0.25$      & $-0.07$      & $-0.19$     & $-0.21$       \\
			& $(0.35)$     & $(0.86)$     & $(0.15)$     & $(0.17)$    & $(0.24)$      \\
			Nesting              & $1.73^{***}$ & $3.44^{*}$   & $-0.23^{+}$  & $0.15$      & $0.46$        \\
			& $(0.50)$     & $(1.33)$     & $(0.13)$     & $(0.23)$    & $(0.28)$      \\
			\midrule
			R$^2$                & $0.13$       & $0.13$       & $0.04$       & $0.03$      & $0.08$        \\
			Num. obs.            & $75$         & $74$         & $79$         & $73$        & $73$          \\
			\bottomrule
			\multicolumn{6}{l}{\scriptsize{Note: *** p<0.001, ** p<0.01, * p<0.05, + p<0.1.}}
		\end{tabular}
		
	\end{center}
\end{table}

\begin{table}[H]
	\caption{Task level linear regression on the full session data for Task group 3 regarding context-free grammars as the target representation.}
	\begin{center}
		\begin{tabular}{l c c c c c}
			\toprule
			& \parbox{1.5cm}{Number of attempts} & \parbox{1.5cm}{Time spent (min)} & \parbox{3.2cm}{Probability of success in first attempt (0--1)} & \parbox{3.2cm}{Perceived interpretation difficulty ($\times$SD)} & \parbox{3.2cm}{Perceived construction difficulty ($\times$SD)} \\
			\midrule
			(Intercept)          & $1.37^{***}$ & $3.38^{***}$  & $0.55^{***}$ & $-0.09$  & $-0.58^{***}$ \\
			& $(0.11)$     & $(0.25)$      & $(0.11)$     & $(0.15)$ & $(0.13)$      \\
			Verbose set notation & $1.25^{***}$ & $3.30^{***}$  & $-0.18$      & $0.44$   & $0.45^{+}$    \\
			& $(0.35)$     & $(0.77)$      & $(0.14)$     & $(0.28)$ & $(0.26)$      \\
			Nesting              & $4.95^{***}$ & $10.00^{***}$ & $-0.42^{**}$ & $-0.08$  & $0.70^{**}$   \\
			& $(1.23)$     & $(2.20)$      & $(0.13)$     & $(0.26)$ & $(0.26)$      \\
			\midrule
			R$^2$                & $0.27$       & $0.31$        & $0.12$       & $0.07$   & $0.10$        \\
			Num. obs.            & $67$         & $64$          & $72$         & $58$     & $59$          \\
			\bottomrule
			\multicolumn{6}{l}{\scriptsize{Note: *** p<0.001, ** p<0.01, * p<0.05, + p<0.1.}}
		\end{tabular}
		
	\end{center}
\end{table}

\begin{table}[H]
	\caption{Task level linear regression on the full session data for Task group 4 regarding context-free grammars as the target representation.}
	\begin{center}
		\begin{tabular}{l c c c c c}
			\toprule
			& \parbox{1.5cm}{Number of attempts} & \parbox{1.5cm}{Time spent (min)} & \parbox{3.2cm}{Probability of success in first attempt (0--1)} & \parbox{3.2cm}{Perceived interpretation difficulty ($\times$SD)} & \parbox{3.2cm}{Perceived construction difficulty ($\times$SD)} \\
			\midrule
			(Intercept)      & $4.70^{***}$ & $9.62^{***}$ & $0.17^{***}$ & $-0.02$     & $0.10$     \\
			& $(0.39)$     & $(0.80)$     & $(0.04)$     & $(0.12)$    & $(0.11)$   \\
			Natural language & $1.05^{+}$   & $-0.36$      & $-0.10^{*}$  & $0.11$      & $0.12$     \\
			& $(0.63)$     & $(1.10)$     & $(0.05)$     & $(0.16)$    & $(0.15)$   \\
			Multiplicities   & $1.71^{**}$  & $3.14^{**}$  & $-0.09^{+}$  & $0.48^{**}$ & $0.32^{*}$ \\
			& $(0.60)$     & $(1.16)$     & $(0.05)$     & $(0.17)$    & $(0.15)$   \\
			\midrule
			R$^2$            & $0.02$       & $0.04$       & $0.02$       & $0.04$      & $0.02$     \\
			Num. obs.        & $287$        & $297$        & $312$        & $208$       & $208$      \\
			\bottomrule
			\multicolumn{6}{l}{\scriptsize{Note: *** p<0.001, ** p<0.01, * p<0.05, + p<0.1.}}
		\end{tabular}
		
	\end{center}
\end{table}

\begin{table}[H]
	\caption{Task level linear regression on the full session data for Task group 5 regarding context-free grammars as the target representation.}
	\begin{center}
		\begin{tabular}{l c c c c c}
			\toprule
			& \parbox{1.5cm}{Number of attempts} & \parbox{1.5cm}{Time spent (min)} & \parbox{3.2cm}{Probability of success in first attempt (0--1)} & \parbox{3.2cm}{Perceived interpretation difficulty ($\times$SD)} & \parbox{3.2cm}{Perceived construction difficulty ($\times$SD)} \\
			\midrule
			(Intercept)      & $3.72^{***}$ & $7.70^{***}$ & $0.22^{***}$ & $-0.06$  & $0.29^{**}$ \\
			& $(0.33)$     & $(0.60)$     & $(0.04)$     & $(0.13)$ & $(0.11)$    \\
			Natural language & $0.10$       & $5.72^{***}$ & $-0.05$      & $0.11$   & $0.11$      \\
			& $(0.47)$     & $(1.27)$     & $(0.06)$     & $(0.18)$ & $(0.16)$    \\
			Multiplicities   & $0.75$       & $6.04^{***}$ & $-0.06$      & $0.17$   & $0.05$      \\
			& $(0.51)$     & $(1.30)$     & $(0.06)$     & $(0.18)$ & $(0.15)$    \\
			\midrule
			R$^2$            & $0.01$       & $0.07$       & $0.00$       & $0.00$   & $0.00$      \\
			Num. obs.        & $259$        & $271$        & $287$        & $196$    & $196$       \\
			\bottomrule
			\multicolumn{6}{l}{\scriptsize{Note: *** p<0.001, ** p<0.01, * p<0.05, + p<0.1.}}
		\end{tabular}
		
	\end{center}
\end{table}

\begin{table}[H]
	\caption{Task level linear regression on the full session data for Task group 6 regarding context-free grammars as the target representation.}
	\begin{center}
		\begin{tabular}{l c c c c c}
			\toprule
			& \parbox{1.5cm}{Number of attempts} & \parbox{1.5cm}{Time spent (min)} & \parbox{3.2cm}{Probability of success in first attempt (0--1)} & \parbox{3.2cm}{Perceived interpretation difficulty ($\times$SD)} & \parbox{3.2cm}{Perceived construction difficulty ($\times$SD)} \\
			\midrule
			(Intercept)      & $1.96^{***}$ & $5.69^{***}$ & $0.47^{***}$ & $0.27^{*}$  & $-0.04$  \\
			& $(0.16)$     & $(0.46)$     & $(0.06)$     & $(0.13)$    & $(0.12)$ \\
			Natural language & $0.74^{*}$   & $-0.36$      & $-0.13$      & $-0.42^{*}$ & $-0.10$  \\
			& $(0.31)$     & $(0.65)$     & $(0.08)$     & $(0.18)$    & $(0.18)$ \\
			Multiplicities   & $0.56^{*}$   & $-0.13$      & $-0.09$      & $-0.20$     & $-0.11$  \\
			& $(0.27)$     & $(0.70)$     & $(0.08)$     & $(0.17)$    & $(0.18)$ \\
			\midrule
			R$^2$            & $0.03$       & $0.00$       & $0.01$       & $0.03$      & $0.00$   \\
			Num. obs.        & $202$        & $202$        & $220$        & $198$       & $198$    \\
			\bottomrule
			\multicolumn{6}{l}{\scriptsize{Note: *** p<0.001, ** p<0.01, * p<0.05, + p<0.1.}}
		\end{tabular}
		
	\end{center}
\end{table}

	\subsubsection{Pushdown automata as the target
		representation}\hphantom{}

\begin{table}[H]
	\caption{Task level linear regression on the full session data for Task group 1 regarding pushdown automata as the target representation.}
	\begin{center}
		\begin{tabular}{l c c c c c}
			\toprule
			& \parbox{1.5cm}{Number of attempts} & \parbox{1.5cm}{Time spent (min)} & \parbox{3.2cm}{Probability of success in first attempt (0--1)} & \parbox{3.2cm}{Perceived interpretation difficulty ($\times$SD)} & \parbox{3.2cm}{Perceived construction difficulty ($\times$SD)} \\
			\midrule
			(Intercept)          & $4.88^{***}$ & $14.99^{***}$ & $0.15^{*}$ & $-0.35^{+}$ & $-0.01$  \\
			& $(0.82)$     & $(1.67)$      & $(0.07)$   & $(0.20)$    & $(0.23)$ \\
			Verbose set notation & $1.38$       & $4.45$        & $0.07$     & $0.36$      & $0.42$   \\
			& $(1.46)$     & $(3.55)$      & $(0.11)$   & $(0.30)$    & $(0.29)$ \\
			Nesting              & $-1.58$      & $1.11$        & $0.22^{+}$ & $0.39$      & $0.44$   \\
			& $(1.14)$     & $(3.46)$      & $(0.12)$   & $(0.29)$    & $(0.31)$ \\
			\midrule
			R$^2$                & $0.06$       & $0.02$        & $0.05$     & $0.04$      & $0.05$   \\
			Num. obs.            & $75$         & $76$          & $81$       & $57$        & $56$     \\
			\bottomrule
			\multicolumn{6}{l}{\scriptsize{Note: *** p<0.001, ** p<0.01, * p<0.05, + p<0.1.}}
		\end{tabular}
		
	\end{center}
\end{table}

\begin{table}[H]
	\caption{Task level linear regression on the full session data for Task group 2 regarding pushdown automata as the target representation.}
	\begin{center}
		\begin{tabular}{l c c c c c}
			\toprule
			& \parbox{1.5cm}{Number of attempts} & \parbox{1.5cm}{Time spent (min)} & \parbox{3.2cm}{Probability of success in first attempt (0--1)} & \parbox{3.2cm}{Perceived interpretation difficulty ($\times$SD)} & \parbox{3.2cm}{Perceived construction difficulty ($\times$SD)} \\
			\midrule
			(Intercept)          & $3.00^{***}$ & $10.06^{***}$ & $0.22^{*}$ & $-0.22$  & $0.07$      \\
			& $(0.38)$     & $(1.22)$      & $(0.10)$   & $(0.26)$ & $(0.17)$    \\
			Verbose set notation & $-0.71$      & $4.26^{+}$    & $0.22$     & $0.06$   & $0.20$      \\
			& $(0.58)$     & $(2.51)$      & $(0.16)$   & $(0.40)$ & $(0.27)$    \\
			Nesting              & $-1.12^{*}$  & $-1.30$       & $0.25$     & $-0.09$  & $-0.53^{+}$ \\
			& $(0.46)$     & $(1.74)$      & $(0.16)$   & $(0.33)$ & $(0.29)$    \\
			\midrule
			R$^2$                & $0.11$       & $0.13$        & $0.05$     & $0.00$   & $0.13$      \\
			Num. obs.            & $47$         & $48$          & $51$       & $47$     & $47$        \\
			\bottomrule
			\multicolumn{6}{l}{\scriptsize{Note: *** p<0.001, ** p<0.01, * p<0.05, + p<0.1.}}
		\end{tabular}
		
	\end{center}
\end{table}

\begin{table}[H]
	\caption{Task level linear regression on the full session data for Task group 3 regarding pushdown automata as the target representation.}
	\begin{center}
		\begin{tabular}{l c c c c c}
			\toprule
			& \parbox{1.5cm}{Number of attempts} & \parbox{1.5cm}{Time spent (min)} & \parbox{3.2cm}{Probability of success in first attempt (0--1)} & \parbox{3.2cm}{Perceived interpretation difficulty ($\times$SD)} & \parbox{3.2cm}{Perceived construction difficulty ($\times$SD)} \\
			\midrule
			(Intercept)          & $2.00^{***}$ & $5.93^{***}$ & $0.47^{***}$ & $-0.22$  & $-0.44^{*}$ \\
			& $(0.37)$     & $(1.05)$     & $(0.13)$     & $(0.20)$ & $(0.19)$    \\
			Verbose set notation & $-0.62$      & $-2.13^{+}$  & $0.18$       & $-0.19$  & $-0.52^{*}$ \\
			& $(0.40)$     & $(1.14)$     & $(0.17)$     & $(0.25)$ & $(0.26)$    \\
			Nesting              & $-0.43$      & $-1.98^{+}$  & $0.17$       & $-0.11$  & $-0.36$     \\
			& $(0.43)$     & $(1.13)$     & $(0.18)$     & $(0.32)$ & $(0.30)$    \\
			\midrule
			R$^2$                & $0.06$       & $0.12$       & $0.03$       & $0.01$   & $0.08$      \\
			Num. obs.            & $46$         & $44$         & $48$         & $46$     & $46$        \\
			\bottomrule
			\multicolumn{6}{l}{\scriptsize{Note: *** p<0.001, ** p<0.01, * p<0.05, + p<0.1.}}
		\end{tabular}
		
	\end{center}
\end{table}

\begin{table}[H]
	\caption{Task level linear regression on the full session data for Task group 4 regarding pushdown automata as the target representation.}
	\begin{center}
		\begin{tabular}{l c c c c c}
			\toprule
			& \parbox{1.5cm}{Number of attempts} & \parbox{1.5cm}{Time spent (min)} & \parbox{3.2cm}{Probability of success in first attempt (0--1)} & \parbox{3.2cm}{Perceived interpretation difficulty ($\times$SD)} & \parbox{3.2cm}{Perceived construction difficulty ($\times$SD)} \\
			\midrule
			(Intercept)      & $6.17^{***}$ & $20.57^{***}$ & $0.13^{***}$ & $0.14$     & $0.22^{+}$ \\
			& $(0.56)$     & $(1.91)$      & $(0.04)$     & $(0.16)$   & $(0.12)$   \\
			Natural language & $-0.40$      & $-4.43^{+}$   & $-0.04$      & $-0.24$    & $-0.06$    \\
			& $(0.77)$     & $(2.34)$      & $(0.05)$     & $(0.22)$   & $(0.18)$   \\
			Multiplicities   & $-0.33$      & $0.61$        & $0.03$       & $0.48^{*}$ & $0.35^{*}$ \\
			& $(0.79)$     & $(2.63)$      & $(0.05)$     & $(0.20)$   & $(0.17)$   \\
			\midrule
			R$^2$            & $0.00$       & $0.02$        & $0.01$       & $0.07$     & $0.04$     \\
			Num. obs.        & $238$        & $242$         & $255$        & $164$      & $164$      \\
			\bottomrule
			\multicolumn{6}{l}{\scriptsize{Note: *** p<0.001, ** p<0.01, * p<0.05, + p<0.1.}}
		\end{tabular}
		
	\end{center}
\end{table}

\begin{table}[H]
	\caption{Task level linear regression on the full session data for Task group 5 regarding pushdown automata as the target representation.}
	\begin{center}
		\begin{tabular}{l c c c c c}
			\toprule
			& \parbox{1.5cm}{Number of attempts} & \parbox{1.5cm}{Time spent (min)} & \parbox{3.2cm}{Probability of success in first attempt (0--1)} & \parbox{3.2cm}{Perceived interpretation difficulty ($\times$SD)} & \parbox{3.2cm}{Perceived construction difficulty ($\times$SD)} \\
			\midrule
			(Intercept)      & $3.06^{***}$ & $7.20^{***}$ & $0.19^{***}$ & $-0.14$  & $-0.15$  \\
			& $(0.29)$     & $(0.83)$     & $(0.05)$     & $(0.14)$ & $(0.13)$ \\
			Natural language & $0.39$       & $0.50$       & $0.17^{*}$   & $-0.02$  & $-0.05$  \\
			& $(0.45)$     & $(1.05)$     & $(0.07)$     & $(0.18)$ & $(0.19)$ \\
			Multiplicities   & $-0.22$      & $2.50^{*}$   & $0.22^{**}$  & $0.03$   & $0.07$   \\
			& $(0.43)$     & $(1.23)$     & $(0.08)$     & $(0.21)$ & $(0.20)$ \\
			\midrule
			R$^2$            & $0.01$       & $0.03$       & $0.04$       & $0.00$   & $0.00$   \\
			Num. obs.        & $195$        & $186$        & $210$        & $166$    & $167$    \\
			\bottomrule
			\multicolumn{6}{l}{\scriptsize{Note: *** p<0.001, ** p<0.01, * p<0.05, + p<0.1.}}
		\end{tabular}
		
	\end{center}
\end{table}

\begin{table}[H]
	\caption{Task level linear regression on the full session data for Task group 6 regarding pushdown automata as the target representation.}
	\begin{center}
		\begin{tabular}{l c c c c c}
			\toprule
			& \parbox{1.5cm}{Number of attempts} & \parbox{1.5cm}{Time spent (min)} & \parbox{3.2cm}{Probability of success in first attempt (0--1)} & \parbox{3.2cm}{Perceived interpretation difficulty ($\times$SD)} & \parbox{3.2cm}{Perceived construction difficulty ($\times$SD)} \\
			\midrule
			(Intercept)      & $2.00^{***}$ & $5.43^{***}$ & $0.46^{***}$ & $-0.11$  & $-0.40^{**}$ \\
			& $(0.19)$     & $(0.58)$     & $(0.06)$     & $(0.12)$ & $(0.14)$     \\
			Natural language & $-0.62^{**}$ & $-0.53$      & $0.03$       & $0.09$   & $0.10$       \\
			& $(0.21)$     & $(0.76)$     & $(0.09)$     & $(0.20)$ & $(0.21)$     \\
			Multiplicities   & $0.23$       & $3.86^{**}$  & $-0.13$      & $0.02$   & $0.09$       \\
			& $(0.28)$     & $(1.19)$     & $(0.09)$     & $(0.18)$ & $(0.21)$     \\
			\midrule
			R$^2$            & $0.08$       & $0.11$       & $0.02$       & $0.00$   & $0.00$       \\
			Num. obs.        & $166$        & $178$        & $195$        & $165$    & $165$        \\
			\bottomrule
			\multicolumn{6}{l}{\scriptsize{Note: *** p<0.001, ** p<0.01, * p<0.05, + p<0.1.}}
		\end{tabular}
		
	\end{center}
\end{table}

	\subsection{Full fixed-effects panel regression (including task dummy
		variables)}\label{app:sessions-panel}

Results of the fixed-effects panel regression analysis. The rows present
the regression coefficients for each variation, i.e.~how much the
corresponding metric (column) differs due to the variation. Below
each coefficient the standard error is specified. Additionally to the
variations we included task-fixed effects to control for general
differences between tasks. For each metric the number of included
observations and the coefficient of determination (\(R^2\)) are
presented. The \(R^2\) value describes how good our independent
variables (variations and task) predict the dependent outcome
variable (measured metric).

\begin{table}[H]
	\caption{Fixed-effects panel regression on the full session data regarding context-free grammars as the target representation.}
	\begin{center}
		\begin{tabular}{l c c c c c}
			\toprule
			& \parbox{1.5cm}{Number of attempts} & \parbox{1.5cm}{Time spent (min)} & \parbox{3.2cm}{Probability of success in first attempt (0--1)} & \parbox{3.2cm}{Perceived interpretation difficulty ($\times$SD)} & \parbox{3.2cm}{Perceived construction difficulty ($\times$SD)} \\
			\midrule
			Natural language     & $0.96^{**}$   & $3.13^{***}$  & $-0.10^{**}$ & $-0.06$     & $0.07$        \\
			& $(0.33)$      & $(0.70)$      & $(0.04)$     & $(0.07)$    & $(0.07)$      \\
			Verbose set notation & $-0.18$       & $0.46$        & $-0.05$      & $-0.01$     & $-0.01$       \\
			& $(0.44)$      & $(0.84)$      & $(0.07)$     & $(0.07)$    & $(0.11)$      \\
			Nesting              & $1.82^{**}$   & $4.35^{***}$  & $-0.16^{*}$  & $0.02$      & $0.49^{***}$  \\
			& $(0.59)$      & $(1.04)$      & $(0.07)$     & $(0.08)$    & $(0.11)$      \\
			Multiplicities       & $1.21^{***}$  & $3.45^{***}$  & $-0.09^{*}$  & $0.15^{*}$  & $0.10$        \\
			& $(0.30)$      & $(0.65)$      & $(0.04)$     & $(0.06)$    & $(0.07)$      \\
			Task group 2         & $-2.00^{***}$ & $-5.12^{***}$ & $0.23^{**}$  & $-0.21^{*}$ & $-0.65^{***}$ \\
			& $(0.45)$      & $(0.93)$      & $(0.07)$     & $(0.08)$    & $(0.10)$      \\
			Task group 3         & $-0.38$       & $-2.31^{*}$   & $0.15^{*}$   & $0.25^{**}$ & $-0.12$       \\
			& $(0.62)$      & $(1.17)$      & $(0.06)$     & $(0.09)$    & $(0.12)$      \\
			Task group 4         & $0.77$        & $-2.97^{*}$   & $-0.00$      & $0.03$      & $-0.09$       \\
			& $(0.69)$      & $(1.35)$      & $(0.08)$     & $(0.10)$    & $(0.16)$      \\
			Task group 5         & $-0.98$       & $-1.73$       & $0.04$       & $-0.13$     & $0.02$        \\
			& $(0.67)$      & $(1.28)$      & $(0.08)$     & $(0.09)$    & $(0.15)$      \\
			Task group 6         & $-2.77^{***}$ & $-7.86^{***}$ & $0.25^{**}$  & $-0.14$     & $-0.44^{**}$  \\
			& $(0.67)$      & $(1.32)$      & $(0.08)$     & $(0.10)$    & $(0.16)$      \\
			\midrule
			R$^2$                & $0.22$        & $0.22$        & $0.12$       & $0.07$      & $0.18$        \\
			Num. obs.            & $847$         & $862$         & $923$        & $752$       & $753$         \\
			\bottomrule
			\multicolumn{6}{l}{\scriptsize{Note: *** p<0.001, ** p<0.01, * p<0.05, + p<0.1.}}
		\end{tabular}
		
	\end{center}
\end{table}

\begin{table}[H]
	\caption{Fixed-effects panel regression on the full session data regarding pushdown automata as the target representation.}
	\begin{center}
		\begin{tabular}{l c c c c c}
			\toprule
			& \parbox{1.5cm}{Number of attempts} & \parbox{1.5cm}{Time spent (min)} & \parbox{3.2cm}{Probability of success in first attempt (0--1)} & \parbox{3.2cm}{Perceived interpretation difficulty ($\times$SD)} & \parbox{3.2cm}{Perceived construction difficulty ($\times$SD)} \\
			\midrule
			Natural language     & $0.08$        & $-1.56$        & $0.07^{+}$   & $-0.00$      & $0.00$        \\
			& $(0.36)$      & $(0.97)$       & $(0.04)$     & $(0.06)$     & $(0.07)$      \\
			Verbose set notation & $-0.59$       & $0.83$         & $0.20^{**}$  & $0.13$       & $0.04$        \\
			& $(0.55)$      & $(1.42)$       & $(0.07)$     & $(0.09)$     & $(0.14)$      \\
			Nesting              & $-1.17^{*}$   & $-1.65$        & $0.21^{**}$  & $0.10$       & $-0.07$       \\
			& $(0.45)$      & $(1.33)$       & $(0.08)$     & $(0.11)$     & $(0.12)$      \\
			Multiplicities       & $0.01$        & $1.62$         & $0.06$       & $0.23^{***}$ & $0.20^{**}$   \\
			& $(0.36)$      & $(1.15)$       & $(0.04)$     & $(0.06)$     & $(0.07)$      \\
			Task group 2         & $-1.97^{***}$ & $-3.28^{*}$    & $0.09$       & $-0.03$      & $-0.17$       \\
			& $(0.53)$      & $(1.66)$       & $(0.07)$     & $(0.10)$     & $(0.13)$      \\
			Task group 3         & $-2.69^{***}$ & $-9.25^{***}$  & $0.27^{***}$ & $-0.09$      & $-0.84^{***}$ \\
			& $(0.50)$      & $(1.34)$       & $(0.08)$     & $(0.12)$     & $(0.14)$      \\
			Task group 4         & $-0.49$       & $-1.55$        & $0.12$       & $0.05$       & $-0.48^{**}$  \\
			& $(0.75)$      & $(2.03)$       & $(0.07)$     & $(0.14)$     & $(0.18)$      \\
			Task group 5         & $-3.35^{***}$ & $-11.44^{***}$ & $0.33^{***}$ & $-0.36^{**}$ & $-1.00^{***}$ \\
			& $(0.70)$      & $(1.87)$       & $(0.07)$     & $(0.14)$     & $(0.18)$      \\
			Task group 6         & $-4.46^{***}$ & $-13.39^{***}$ & $0.40^{***}$ & $-0.30^{*}$  & $-1.17^{***}$ \\
			& $(0.72)$      & $(1.77)$       & $(0.08)$     & $(0.14)$     & $(0.18)$      \\
			\midrule
			R$^2$                & $0.28$        & $0.33$         & $0.13$       & $0.15$       & $0.31$        \\
			Num. obs.            & $647$         & $652$          & $709$        & $606$        & $607$         \\
			\bottomrule
			\multicolumn{6}{l}{\scriptsize{Note: *** p<0.001, ** p<0.01, * p<0.05, + p<0.1.}}
		\end{tabular}
		
	\end{center}
\end{table}

\clearpage
	\subsection{Full pooled linear regression (including task dummy
		variables)}\label{app:sessions-pooled}

Results of the pooled linear regression analysis. The rows present the
regression coefficients for each variation, i.e.~how much the
corresponding metric (column) differs due to the variation. Below
each coefficient the standard error is specified. Additionally to the
variations we included task-fixed effects to control for general
differences between tasks. For each metric the number of included
observations and the coefficient of determination (\(R^2\)) are
presented. The \(R^2\) value describes how good our independent
variables (variations and task) predict the dependent outcome
variable (measured metric).

\begin{table}[H]
	\caption{Pooled linear regression on the full session data regarding context-free grammars as the target representation.}
	\begin{center}
		\begin{tabular}{l c c c c c}
			\toprule
			& \parbox{1.5cm}{Number of attempts} & \parbox{1.5cm}{Time spent (min)} & \parbox{3.2cm}{Probability of success in first attempt (0--1)} & \parbox{3.2cm}{Perceived interpretation difficulty ($\times$SD)} & \parbox{3.2cm}{Perceived construction difficulty ($\times$SD)} \\
			\midrule
			(Intercept)          & $3.56^{***}$  & $9.47^{***}$  & $0.28^{***}$ & $-0.07$      & $-0.12$       \\
			& $(0.43)$      & $(0.91)$      & $(0.06)$     & $(0.11)$     & $(0.12)$      \\
			Natural language     & $0.60^{*}$    & $1.79^{**}$   & $-0.09^{**}$ & $-0.07$      & $0.04$        \\
			& $(0.28)$      & $(0.60)$      & $(0.03)$     & $(0.08)$     & $(0.07)$      \\
			Verbose set notation & $0.44$        & $1.05$        & $-0.10$      & $0.02$       & $0.06$        \\
			& $(0.40)$      & $(0.91)$      & $(0.06)$     & $(0.11)$     & $(0.11)$      \\
			Nesting              & $2.02^{***}$  & $4.24^{***}$  & $-0.19^{**}$ & $-0.04$      & $0.48^{***}$  \\
			& $(0.50)$      & $(1.04)$      & $(0.06)$     & $(0.10)$     & $(0.12)$      \\
			Multiplicities       & $1.06^{***}$  & $3.25^{***}$  & $-0.08^{*}$  & $0.15^{+}$   & $0.09$        \\
			& $(0.28)$      & $(0.59)$      & $(0.03)$     & $(0.08)$     & $(0.07)$      \\
			Task group 2         & $-2.06^{***}$ & $-5.66^{***}$ & $0.24^{***}$ & $-0.26^{**}$ & $-0.63^{***}$ \\
			& $(0.38)$      & $(0.88)$      & $(0.07)$     & $(0.09)$     & $(0.11)$      \\
			Task group 3         & $-0.92^{+}$   & $-3.32^{**}$  & $0.17^{**}$  & $0.13$       & $-0.25^{+}$   \\
			& $(0.55)$      & $(1.12)$      & $(0.06)$     & $(0.13)$     & $(0.13)$      \\
			Task group 4         & $1.54^{**}$   & $-0.57$       & $-0.11^{+}$  & $0.22^{+}$   & $0.33^{*}$    \\
			& $(0.52)$      & $(1.04)$      & $(0.06)$     & $(0.13)$     & $(0.14)$      \\
			Task group 5         & $-0.13$       & $0.67$        & $-0.04$      & $0.08$       & $0.43^{**}$   \\
			& $(0.49)$      & $(1.06)$      & $(0.06)$     & $(0.13)$     & $(0.13)$      \\
			Task group 6         & $-1.74^{***}$ & $-5.59^{***}$ & $0.17^{*}$   & $0.11$       & $-0.02$       \\
			& $(0.47)$      & $(0.98)$      & $(0.07)$     & $(0.13)$     & $(0.14)$      \\
			\midrule
			R$^2$                & $0.14$        & $0.12$        & $0.09$       & $0.03$       & $0.09$        \\
			Num. obs.            & $1010$        & $1029$        & $1096$       & $813$        & $814$         \\
			\bottomrule
			\multicolumn{6}{l}{\scriptsize{Note: *** p<0.001, ** p<0.01, * p<0.05, + p<0.1.}}
		\end{tabular}
		
	\end{center}
\end{table}

\begin{table}[H]
	\caption{Pooled linear regression on the full session data regarding pushdown automata as the target representation.}
	\begin{center}
		\begin{tabular}{l c c c c c}
			\toprule
			& \parbox{1.5cm}{Number of attempts} & \parbox{1.5cm}{Time spent (min)} & \parbox{3.2cm}{Probability of success in first attempt (0--1)} & \parbox{3.2cm}{Perceived interpretation difficulty ($\times$SD)} & \parbox{3.2cm}{Perceived construction difficulty ($\times$SD)} \\
			\midrule
			(Intercept)          & $5.13^{***}$  & $16.14^{***}$  & $0.13^{*}$   & $-0.17$    & $0.29^{+}$    \\
			& $(0.58)$      & $(1.40)$       & $(0.06)$     & $(0.15)$   & $(0.16)$      \\
			Natural language     & $-0.21$       & $-1.77^{+}$    & $0.05$       & $-0.04$    & $0.00$        \\
			& $(0.34)$      & $(1.00)$       & $(0.04)$     & $(0.08)$   & $(0.08)$      \\
			Verbose set notation & $0.27$        & $2.60$         & $0.14^{*}$   & $0.10$     & $0.05$        \\
			& $(0.66)$      & $(1.72)$       & $(0.07)$     & $(0.12)$   & $(0.14)$      \\
			Nesting              & $-1.13^{*}$   & $-0.46$        & $0.22^{**}$  & $0.09$     & $-0.12$       \\
			& $(0.52)$      & $(1.54)$       & $(0.07)$     & $(0.13)$   & $(0.13)$      \\
			Multiplicities       & $-0.17$       & $2.03^{+}$     & $0.04$       & $0.18^{*}$ & $0.18^{*}$    \\
			& $(0.35)$      & $(1.14)$       & $(0.04)$     & $(0.07)$   & $(0.08)$      \\
			Task group 2         & $-2.42^{***}$ & $-5.81^{***}$  & $0.13^{*}$   & $-0.13$    & $-0.32^{*}$   \\
			& $(0.56)$      & $(1.71)$       & $(0.07)$     & $(0.13)$   & $(0.14)$      \\
			Task group 3         & $-3.23^{***}$ & $-12.26^{***}$ & $0.34^{***}$ & $-0.22$    & $-1.02^{***}$ \\
			& $(0.56)$      & $(1.46)$       & $(0.07)$     & $(0.13)$   & $(0.14)$      \\
			Task group 4         & $0.92$        & $3.16^{+}$     & $-0.02$      & $0.35^{*}$ & $-0.03$       \\
			& $(0.70)$      & $(1.91)$       & $(0.06)$     & $(0.16)$   & $(0.17)$      \\
			Task group 5         & $-1.87^{**}$  & $-7.95^{***}$  & $0.17^{*}$   & $-0.01$    & $-0.50^{**}$  \\
			& $(0.65)$      & $(1.68)$       & $(0.07)$     & $(0.16)$   & $(0.17)$      \\
			Task group 6         & $-3.15^{***}$ & $-9.67^{***}$  & $0.27^{***}$ & $0.05$     & $-0.68^{***}$ \\
			& $(0.63)$      & $(1.58)$       & $(0.07)$     & $(0.16)$   & $(0.17)$      \\
			\midrule
			R$^2$                & $0.20$        & $0.24$         & $0.10$       & $0.04$     & $0.10$        \\
			Num. obs.            & $767$         & $774$          & $840$        & $645$      & $645$         \\
			\bottomrule
			\multicolumn{6}{l}{\scriptsize{Note: *** p<0.001, ** p<0.01, * p<0.05, + p<0.1.}}
		\end{tabular}
		
	\end{center}
\end{table}

\clearpage

	\section{BH-corrected regression tables for study
		participants}\label{app:bhCorrection}

To account for multiple testing, we report the results of a
Benjamini--Hochberg (BH) correction applied to all variation-related
\(p\)-values. While the main analyses rely on p-values
corrected only for heteroscedasticity and clustered standard
errors---consistent with our preregistration and theory-driven
hypotheses---we note that applying the BH procedure leads to a modest
attenuation of statistical significance in a handful of cases. Below are the regression tables with Benjamini--Hochberg--corrected
p-values.

	\subsection{Panel regression tables with
		BH-correction}\label{app:bhCorrection-panel}

\begin{table}[H]
	\caption{Fixed-effects panel regression on the linked data regarding context-free grammars as the target representation.}
	\begin{center}
		\begin{tabular}{l c c c c c}
			\toprule
			& \parbox{1.5cm}{Number of attempts} & \parbox{1.5cm}{Time spent (min)} & \parbox{3.2cm}{Probability of success in first attempt (0--1)} & \parbox{3.2cm}{Perceived interpretation difficulty ($\times$SD)} & \parbox{3.2cm}{Perceived construction difficulty ($\times$SD)} \\
			\midrule
			Natural language     & $1.83^{***}$  & $4.75^{***}$  & $-0.10^{+}$ & $-0.05$  & $0.11$        \\
			& $(0.49)$      & $(1.15)$      & $(0.05)$    & $(0.09)$ & $(0.08)$      \\
			Verbose set notation & $-0.25$       & $0.79$        & $-0.13$     & $0.07$   & $0.08$        \\
			& $(0.63)$      & $(1.09)$      & $(0.08)$    & $(0.12)$ & $(0.13)$      \\
			Nesting              & $1.53^{+}$    & $3.95^{*}$    & $-0.22^{*}$ & $-0.02$  & $0.61^{***}$  \\
			& $(0.80)$      & $(1.56)$      & $(0.08)$    & $(0.12)$ & $(0.16)$      \\
			Multiplicities       & $2.78^{***}$  & $5.39^{***}$  & $-0.09^{+}$ & $0.20$   & $0.15$        \\
			& $(0.56)$      & $(1.02)$      & $(0.05)$    & $(0.08)$ & $(0.09)$      \\
			Task group 2         & $-2.27^{**}$  & $-4.84^{***}$ & $0.27^{**}$ & $-0.30$  & $-0.73^{***}$ \\
			& $(0.67)$      & $(1.25)$      & $(0.09)$    & $(0.12)$ & $(0.14)$      \\
			Task group 3         & $-1.02$       & $-2.91^{+}$   & $0.17^{*}$  & $0.13$   & $-0.23$       \\
			& $(0.68)$      & $(1.54)$      & $(0.07)$    & $(0.15)$ & $(0.15)$      \\
			Task group 4         & $0.44$        & $-1.93$       & $-0.07$     & $-0.04$  & $-0.03$       \\
			& $(0.89)$      & $(1.73)$      & $(0.09)$    & $(0.16)$ & $(0.22)$      \\
			Task group 5         & $-1.42$       & $-2.04$       & $-0.02$     & $-0.24$  & $0.10$        \\
			& $(0.95)$      & $(1.79)$      & $(0.09)$    & $(0.15)$ & $(0.20)$      \\
			Task group 6         & $-4.18^{***}$ & $-9.74^{***}$ & $0.19^{+}$  & $-0.20$  & $-0.30$       \\
			& $(0.87)$      & $(1.76)$      & $(0.09)$    & $(0.15)$ & $(0.21)$      \\
			\midrule
			R$^2$                & $0.25$        & $0.24$        & $0.13$      & $0.08$   & $0.16$        \\
			Num. obs.            & $477$         & $487$         & $522$       & $476$    & $477$         \\
			\bottomrule
			\multicolumn{6}{l}{\scriptsize{Note: *** p<0.001, ** p<0.01, * p<0.05, + p<0.1.}}
		\end{tabular}
		
	\end{center}
\end{table}

\begin{table}[H]
	\caption{Fixed-effects panel regression on the linked data regarding pushdown automata as the target representation.}
	\begin{center}
		\begin{tabular}{l c c c c c}
			\toprule
			& \parbox{1.5cm}{Number of attempts} & \parbox{1.5cm}{Time spent (min)} & \parbox{3.2cm}{Probability of success in first attempt (0--1)} & \parbox{3.2cm}{Perceived interpretation difficulty ($\times$SD)} & \parbox{3.2cm}{Perceived construction difficulty ($\times$SD)} \\
			\midrule
			Natural language     & $0.67$        & $-0.10$        & $0.10^{+}$   & $-0.02$    & $0.09$        \\
			& $(0.54)$      & $(1.49)$       & $(0.05)$     & $(0.08)$   & $(0.09)$      \\
			Verbose set notation & $-0.49$       & $0.41$         & $0.24^{*}$   & $0.20$     & $-0.06$       \\
			& $(0.86)$      & $(1.99)$       & $(0.08)$     & $(0.12)$   & $(0.17)$      \\
			Nesting              & $-0.78$       & $-1.30$        & $0.20^{+}$   & $0.17$     & $-0.15$       \\
			& $(0.75)$      & $(2.01)$       & $(0.10)$     & $(0.15)$   & $(0.15)$      \\
			Multiplicities       & $0.28$        & $0.30$         & $0.11^{+}$   & $0.24^{+}$ & $0.24^{*}$    \\
			& $(0.59)$      & $(1.58)$       & $(0.05)$     & $(0.08)$   & $(0.08)$      \\
			Task group 2         & $-2.53^{*}$   & $-7.36^{**}$   & $0.16$       & $0.01$     & $-0.43^{*}$   \\
			& $(0.93)$      & $(2.31)$       & $(0.09)$     & $(0.13)$   & $(0.17)$      \\
			Task group 3         & $-3.23^{***}$ & $-11.16^{***}$ & $0.27^{*}$   & $-0.07$    & $-1.04^{***}$ \\
			& $(0.82)$      & $(1.97)$       & $(0.09)$     & $(0.15)$   & $(0.15)$      \\
			Task group 4         & $-0.38$       & $-1.85$        & $0.19^{+}$   & $0.20$     & $-0.83^{**}$  \\
			& $(1.10)$      & $(2.91)$       & $(0.09)$     & $(0.16)$   & $(0.22)$      \\
			Task group 5         & $-3.81^{***}$ & $-12.62^{***}$ & $0.38^{**}$  & $-0.23$    & $-1.24^{***}$ \\
			& $(1.00)$      & $(2.51)$       & $(0.10)$     & $(0.14)$   & $(0.21)$      \\
			Task group 6         & $-5.58^{***}$ & $-16.43^{***}$ & $0.44^{***}$ & $-0.12$    & $-1.45^{***}$ \\
			& $(1.05)$      & $(2.50)$       & $(0.10)$     & $(0.15)$   & $(0.22)$      \\
			\midrule
			R$^2$                & $0.28$        & $0.36$         & $0.15$       & $0.13$     & $0.33$        \\
			Num. obs.            & $395$         & $396$          & $435$        & $403$      & $402$         \\
			\bottomrule
			\multicolumn{6}{l}{\scriptsize{Note: *** p<0.001, ** p<0.01, * p<0.05, + p<0.1.}}
		\end{tabular}
		
	\end{center}
\end{table}

	\subsection{Pooled OLS regression tables with
		BH-correction}\label{app:bhCorrection-pooled}

\begin{table}[H]
	\caption{Pooled linear regression on the BH‐corrected participant data for context‐free grammars.}
	\begin{center}
		\begin{tabular}{l c c c c c}
			\toprule
			& \parbox{1.5cm}{Number of attempts} & \parbox{1.5cm}{Time spent (min)} & \parbox{3.2cm}{Probability of success in first attempt (0--1)} & \parbox{3.2cm}{Perceived interpretation difficulty ($\times$SD)} & \parbox{3.2cm}{Perceived construction difficulty ($\times$SD)} \\
			\midrule
			(Intercept)          & $4.47^{***}$  & $10.87^{***}$ & $0.27^{**}$ & $-0.01$     & $-0.17$       \\
			& $(0.68)$      & $(1.40)$      & $(0.07)$    & $(0.12)$    & $(0.16)$      \\
			Natural language     & $1.73^{**}$   & $4.70^{***}$  & $-0.09^{+}$ & $-0.06$     & $0.09$        \\
			& $(0.49)$      & $(1.11)$      & $(0.05)$    & $(0.09)$    & $(0.08)$      \\
			Verbose set notation & $0.07$        & $1.09$        & $-0.14$     & $0.04$      & $0.03$        \\
			& $(0.57)$      & $(1.25)$      & $(0.08)$    & $(0.13)$    & $(0.14)$      \\
			Nesting              & $1.85^{*}$    & $4.01^{**}$   & $-0.23^{*}$ & $-0.08$     & $0.56^{**}$   \\
			& $(0.73)$      & $(1.44)$      & $(0.08)$    & $(0.12)$    & $(0.16)$      \\
			Multiplicities       & $2.54^{***}$  & $5.16^{***}$  & $-0.09^{+}$ & $0.18$      & $0.15$        \\
			& $(0.55)$      & $(0.99)$      & $(0.04)$    & $(0.09)$    & $(0.09)$      \\
			Task group 2         & $-2.93^{***}$ & $-6.47^{***}$ & $0.30^{**}$ & $-0.32^{+}$ & $-0.64^{***}$ \\
			& $(0.60)$      & $(1.20)$      & $(0.08)$    & $(0.12)$    & $(0.14)$      \\
			Task group 3         & $-1.80^{*}$   & $-5.00^{**}$  & $0.25^{**}$ & $0.11$      & $-0.27$       \\
			& $(0.65)$      & $(1.55)$      & $(0.07)$    & $(0.15)$    & $(0.15)$      \\
			Task group 4         & $1.48$        & $-0.48$       & $-0.11$     & $0.14$      & $0.31$        \\
			& $(0.84)$      & $(1.62)$      & $(0.08)$    & $(0.16)$    & $(0.19)$      \\
			Task group 5         & $-0.48$       & $-0.47$       & $-0.06$     & $-0.06$     & $0.39^{+}$    \\
			& $(0.85)$      & $(1.73)$      & $(0.08)$    & $(0.15)$    & $(0.18)$      \\
			Task group 6         & $-3.36^{***}$ & $-8.25^{***}$ & $0.15$      & $-0.01$     & $0.00$        \\
			& $(0.75)$      & $(1.53)$      & $(0.08)$    & $(0.15)$    & $(0.18)$      \\
			\midrule
			R$^2$                & $0.21$        & $0.17$        & $0.11$      & $0.03$      & $0.10$        \\
			Num. obs.            & $484$         & $494$         & $529$       & $479$       & $480$         \\
			\bottomrule
			\multicolumn{6}{l}{\scriptsize{Note: *** p<0.001, ** p<0.01, * p<0.05, + p<0.1.}}
		\end{tabular}
		
	\end{center}
\end{table}

\begin{table}[H]
	\caption{Pooled linear regression on the BH‐corrected participant data for pushdown automata.}
	\begin{center}
		\begin{tabular}{l c c c c c}
			\toprule
			& \parbox{1.5cm}{Number of attempts} & \parbox{1.5cm}{Time spent (min)} & \parbox{3.2cm}{Probability of success in first attempt (0--1)} & \parbox{3.2cm}{Perceived interpretation difficulty ($\times$SD)} & \parbox{3.2cm}{Perceived construction difficulty ($\times$SD)} \\
			\midrule
			(Intercept)          & $5.87^{***}$  & $18.40^{***}$  & $0.06$      & $-0.26$    & $0.49^{*}$    \\
			& $(0.96)$      & $(2.22)$       & $(0.08)$    & $(0.17)$   & $(0.19)$      \\
			Natural language     & $0.29$        & $0.91$         & $0.10$      & $-0.09$    & $0.04$        \\
			& $(0.62)$      & $(1.75)$       & $(0.05)$    & $(0.09)$   & $(0.09)$      \\
			Verbose set notation & $-0.15$       & $0.95$         & $0.22^{*}$  & $0.20$     & $-0.06$       \\
			& $(0.90)$      & $(2.17)$       & $(0.08)$    & $(0.15)$   & $(0.18)$      \\
			Nesting              & $-0.21$       & $-1.06$        & $0.21^{*}$  & $0.14$     & $-0.19$       \\
			& $(0.76)$      & $(2.04)$       & $(0.09)$    & $(0.15)$   & $(0.15)$      \\
			Multiplicities       & $-0.15$       & $1.04$         & $0.11^{+}$  & $0.14$     & $0.15$        \\
			& $(0.63)$      & $(1.62)$       & $(0.05)$    & $(0.09)$   & $(0.08)$      \\
			Task group 2         & $-3.14^{**}$  & $-7.80^{**}$   & $0.20^{*}$  & $-0.02$    & $-0.47^{*}$   \\
			& $(0.92)$      & $(2.29)$       & $(0.08)$    & $(0.17)$   & $(0.18)$      \\
			Task group 3         & $-3.84^{***}$ & $-12.94^{***}$ & $0.34^{**}$ & $-0.15$    & $-1.11^{***}$ \\
			& $(0.81)$      & $(1.98)$       & $(0.09)$    & $(0.16)$   & $(0.15)$      \\
			Task group 4         & $1.72$        & $3.40$         & $0.02$      & $0.51^{+}$ & $-0.26$       \\
			& $(1.19)$      & $(3.17)$       & $(0.09)$    & $(0.19)$   & $(0.21)$      \\
			Task group 5         & $-2.01^{+}$   & $-9.45^{***}$  & $0.22^{*}$  & $0.14$     & $-0.61^{*}$   \\
			& $(1.03)$      & $(2.57)$       & $(0.09)$    & $(0.19)$   & $(0.21)$      \\
			Task group 6         & $-3.71^{**}$  & $-13.09^{***}$ & $0.28^{*}$  & $0.23$     & $-0.86^{***}$ \\
			& $(1.03)$      & $(2.46)$       & $(0.09)$    & $(0.19)$   & $(0.22)$      \\
			\midrule
			R$^2$                & $0.20$        & $0.25$         & $0.09$      & $0.03$     & $0.10$        \\
			Num. obs.            & $399$         & $400$          & $440$       & $405$      & $404$         \\
			\bottomrule
			\multicolumn{6}{l}{\scriptsize{Note: *** p<0.001, ** p<0.01, * p<0.05, + p<0.1.}}
		\end{tabular}
		
	\end{center}
\end{table}

	\subsection{Comparing p-values}\label{app:bhCorrection-comparison}

To account for multiple testing, we applied a Benjamini--Hochberg (BH)
correction to every variation-related \(p\) in our five focal
outcomes---number of attempts, time needed, probability of success on
the first try, perceived interpretation difficulty, and perceived
construction difficulty---in both the CFG and PDA panels (a total of
\(5 \text{ outcomes} \times 4 \text{ coefficients} \times 2 \text{ settings} =40\)
tests. From these 19, 12 for CFG and 7 for PDA, (at least marginally)
significant results, a total of 16, 10 for CFG and 6 for PDA, remained
(at least marginally) significant. More specific

\begin{itemize}
	\item
	\textbf{CFG models}:
	
	\begin{itemize}
		\item
		\textbf{2} variation coefficients (the multiplicities effect on
		interpretation and on construction difficulty) are pushed fully
		above the \(p = .10\) threshold.
		\item
		\textbf{2} additional coefficients (the natural-language and
		multiplicities variations in the success-on-first-try model) are
		attenuated from \(p < .05\) to the marginal range
		\(.05 \leq p < .10\), so they remain marginally significant.
	\end{itemize}
	\item
	\textbf{PDA models}:
	
	\begin{itemize}
		\item
		\textbf{1} variation coefficient (the verbose-set-notation effect
		on interpretation difficulty) is pushed above \(p = .10\).
		\item
		\textbf{4} further coefficients (all three variations in the
		success-on-first-try model, plus the multiplicities effect on
		interpretation difficulty) move from \(p < .05\) to
		\(.05 \leq p < .10\), staying marginally significant.
	\end{itemize}
\end{itemize}

In other words, only a small minority of our 40 variation tests (3,
or 7.5\%) lose even marginal significance, and only 6 (15\%) move into
the .05--.10 range; the vast majority (over 75\%) stay at or below
\(p = .05\). Thus, none of our core theoretical conclusions are
overturned by the BH correction, and the regressions reported above---using only heteroscedasticity- and cluster-robust standard
errors---can be interpreted with confidence.

\clearpage 
\begin{longtable}[]{@{}llrrr@{}}
	\caption{Raw vs BH-adjusted p-values and diff for CFG}\tabularnewline
	\toprule
	Metric & Variation & Raw p value & Adjusted p value & Difference of p
	values\tabularnewline
	\midrule
	\endfirsthead
	\toprule
	Metric & Variation & Raw p value & Adjusted p value & Difference of p
	values\tabularnewline
	\midrule
	\endhead
	Number of attempts & Natural language & 0.000 & 0.001 &
	0.000\tabularnewline
	Number of attempts & Verbose set notation & 0.690 & 0.759 &
	0.069\tabularnewline
	Number of attempts & Nesting & 0.057 & 0.087 & 0.030\tabularnewline
	Number of attempts & Multiplicities & 0.000 & 0.000 &
	0.000\tabularnewline
	Time spent & Natural language & 0.000 & 0.000 & 0.000\tabularnewline
	Time spent & Verbose set notation & 0.470 & 0.571 & 0.101\tabularnewline
	Time spent & Nesting & 0.012 & 0.019 & 0.008\tabularnewline
	Time spent & Multiplicities & 0.000 & 0.000 & 0.000\tabularnewline
	Probability of success in first try & Natural language & 0.042 & 0.078 &
	0.037\tabularnewline
	Probability of success in first try & Verbose set notation & 0.107 &
	0.161 & 0.054\tabularnewline
	Probability of success in first try & Nesting & 0.006 & 0.018 &
	0.012\tabularnewline
	Probability of success in first try & Multiplicities & 0.039 & 0.078 &
	0.039\tabularnewline
	Perceived interpretation difficulty & Natural language & 0.584 & 0.746 &
	0.162\tabularnewline
	Perceived interpretation difficulty & Verbose set notation & 0.549 &
	0.713 & 0.164\tabularnewline
	Perceived interpretation difficulty & Nesting & 0.868 & 0.942 &
	0.074\tabularnewline
	Perceived interpretation difficulty & Multiplicities & 0.016 & 0.142 &
	0.126\tabularnewline
	Perceived construction difficulty & Natural language & 0.185 & 0.315 &
	0.130\tabularnewline
	Perceived construction difficulty & Verbose set notation & 0.577 & 0.690
	& 0.114\tabularnewline
	Perceived construction difficulty & Nesting & 0.000 & 0.001 &
	0.001\tabularnewline
	Perceived construction difficulty & Multiplicities & 0.078 & 0.168 &
	0.090\tabularnewline
	\bottomrule
\end{longtable}

\begin{longtable}[]{@{}llrrr@{}}
	\caption{Raw vs BH-adjusted p-values and diff for PDA}\tabularnewline
	\toprule
	Metric & Variation & Raw p value & Adjusted p value & Difference of p
	values\tabularnewline
	\midrule
	\endfirsthead
	\toprule
	Metric & Variation & Raw p value & Adjusted p value & Difference of p
	values\tabularnewline
	\midrule
	\endhead
	Number of attempts & Natural language & 0.218 & 0.288 &
	0.070\tabularnewline
	Number of attempts & Verbose set notation & 0.573 & 0.672 &
	0.100\tabularnewline
	Number of attempts & Nesting & 0.302 & 0.387 & 0.085\tabularnewline
	Number of attempts & Multiplicities & 0.639 & 0.726 &
	0.087\tabularnewline
	Time spent & Natural language & 0.949 & 0.960 & 0.012\tabularnewline
	Time spent & Verbose set notation & 0.837 & 0.880 & 0.043\tabularnewline
	Time spent & Nesting & 0.517 & 0.618 & 0.101\tabularnewline
	Time spent & Multiplicities & 0.849 & 0.889 & 0.040\tabularnewline
	Probability of success in first try & Natural language & 0.046 & 0.080 &
	0.035\tabularnewline
	Probability of success in first try & Verbose set notation & 0.004 &
	0.015 & 0.011\tabularnewline
	Probability of success in first try & Nesting & 0.042 & 0.078 &
	0.036\tabularnewline
	Probability of success in first try & Multiplicities & 0.034 & 0.073 &
	0.039\tabularnewline
	Perceived interpretation difficulty & Natural language & 0.823 & 0.905 &
	0.082\tabularnewline
	Perceived interpretation difficulty & Verbose set notation & 0.096 &
	0.357 & 0.261\tabularnewline
	Perceived interpretation difficulty & Nesting & 0.256 & 0.556 &
	0.300\tabularnewline
	Perceived interpretation difficulty & Multiplicities & 0.002 & 0.053 &
	0.051\tabularnewline
	Perceived construction difficulty & Natural language & 0.297 & 0.420 &
	0.123\tabularnewline
	Perceived construction difficulty & Verbose set notation & 0.717 & 0.806
	& 0.090\tabularnewline
	Perceived construction difficulty & Nesting & 0.327 & 0.447 &
	0.119\tabularnewline
	Perceived construction difficulty & Multiplicities & 0.003 & 0.011 &
	0.008\tabularnewline
	\bottomrule
\end{longtable}

\clearpage
	\section{Languages that were used in the
		lecture}\label{app:lecture}

	\subsection{For context-free grammars as the target
		representation}\label{app:lecture-cfg}

\begin{itemize}
	\item Palindromic strings over $\{a,b\}$ (4.1)
	\item Arithmetic expressions (4.1)
	\item Strings over $\{a,b\}$ with an equal number of $a$ and $b$ (4.1)
	\item $\{a^ib^jc^k\mid i=j \vee j=k\}$ (4.1)
	\item Strings over $\{a,b\}$, where the first half is unequal to the second half (4.2)
\end{itemize}

	\subsection{For pushdown automata as the target
		representation}\label{app:lecture-pda}

\begin{itemize}
	\item Correctly tagged expressions with \texttt{<a>} and \texttt{<b>} (5.1)
	\item $\{a^ib^j\mid i\ge j\}$ (5.1)
	\item Palindromic strings of even length over $\{a,b\}$ (5.1)
	\item $\{a^nb^n\mid n\ge0\}$ (5.2)
\end{itemize}

\end{document}